\documentclass[12pt,preprint]{aastex}
\usepackage{rotating}
\usepackage{bm}

\newcommand{\threej}[6]
{ \left(\begin{array}{ccc}
#1&#2&#3\\
#4&#5&#6
\end{array}\right) }

\begin{document}
\title{Advanced Forward Modeling and Inversion of 
Stokes Profiles Resulting from the Joint Action of the Hanle and Zeeman Effects}

\author{A. Asensio Ramos, J. Trujillo Bueno\altaffilmark{1}}
\affil{Instituto de Astrof\'{\i}sica de Canarias, V\'\i a L\'actea s/n, E-38205 La Laguna, Tenerife, Spain}
\and
\author{E. Landi Degl'Innocenti}
\affil{Universit\`a degli Studi di Firenze, 
Dipartimento di Astronomia e Scienza dello Spazio, Largo Enrico Fermi 2, I-50125 Florence, Italy}
\altaffiltext{1}{Consejo Superior de Investigaciones Cient\'{\i}ficas (Spain)}
\email{aasensio@iac.es,jtb@iac.es,landie@arcetri.astro.it}

\begin{abstract}
A big challenge in solar and stellar physics in the coming years will be
to 
decipher the magnetism of the solar outer atmosphere (chromosphere and corona)
along with 
its dynamic coupling with the magnetic fields of the underlying photosphere. To
this end, it 
is important to develop rigorous diagnostic tools for the physical
interpretation of  
spectropolarimetric observations in suitably chosen spectral lines. Here we
present a 
computer program for the synthesis and inversion of Stokes profiles caused by
the joint 
action of atomic level polarization and the Hanle and Zeeman effects in some
spectral lines of diagnostic interest, such 
as those of the He {\sc i} 10830 \AA\ and 5876 \AA\ (or D$_3$) multiplets. It is
based
on the quantum theory of spectral line polarization, which takes into account in
a rigorous
way all the relevant physical mechanisms and ingredients 
(optical pumping, atomic level polarization, level crossings and repulsions,
Zeeman, Paschen-Back and Hanle effects). The influence of radiative transfer on
the emergent spectral line radiation is taken into account through a suitable 
slab model. The user can either calculate the emergent intensity and polarization
for any given magnetic field vector or infer the dynamical and magnetic
properties from the observed Stokes profiles via an efficient inversion
algorithm based on 
global optimization methods. The reliability of the forward modeling and
inversion code presented here is 
demonstrated through several applications, which range from the inference of the
magnetic field vector in solar active regions to determining whether or not it is
canopy-like in quiet chromospheric regions. This user-friendly diagnostic tool 
called ``HAZEL" (from HAnle and ZEeman Light) is offered to
the astrophysical community, with the hope that it will facilitate new advances in solar and
stellar physics.
\end{abstract}


\section{Introduction} 

The present paper describes a computer program for the synthesis and inversion
of Stokes profiles resulting from the joint 
action of the Hanle and Zeeman effects in some spectral lines of diagnostic
interest, such as those pertaining to the \ion{He}{1} 
10830 \AA\ and 5876 \AA\ (or D$_3$) multiplets. The effects of radiative
transfer on the emergent spectral
line radiation are taken into account through a suitable slab model. 
Our aim is to provide the solar and stellar physics communities with a robust
but user-friendly tool for understanding 
and interpreting spectropolarimetric observations, with the hope that this will
facilitate new advances in solar and stellar
physics.

In particular, the lines of neutral helium at 10830 \AA\ are of great interest
for empirical investigations of the 
dynamic and magnetic properties of plasma structures in the solar chromosphere
and corona, such as active 
regions \citep[e.g.,][]{harvey_hall71,ruedi96,Lagg04,centeno06}, filaments 
\citep[e.g.,][]{lin98,trujillo_nature02}, prominences \citep[e.g.,][]{merenda06}
and 
spicules \citep[e.g.,][]{trujillo_merenda05,socas_elmore05}. The same applies to
the lines of the 
\ion{He}{1} D$_3$ multiplet at 5876 \AA\, which have been used for investigating
the magnetic field vector in solar 
prominences and spicules
\citep[e.g.,][]{landi_d3_82,querfeld85,bommier94,casini03,lopezariste_casini05,ramelli06_2,
ramelli06}

Such helium lines result from transitions between terms of the triplet system of
helium (ortho-helium), whose 
respective $J$-levels (with $J$ the level's total angular momentum) 
are far less populated than the ground level of helium (that is, than the
singlet level $^1$S$_0$), except
perhaps in flaring regions. On the other 
hand, the lower term (2s$^3$S$_1$) of the \ion{He}{1} 10830 \AA\ multiplet is the
ground level of ortho-helium, while 
its upper term (2p$^3$P$_{2,1,0}$) is the lower one of 5876 \AA\ (whose upper term
is 3d$^3$D$_{3,2,1}$). Therefore, the 
significant difference in the ensuing optical thicknesses of the observed solar
plasma structure implies that when 
the radiation in these spectral lines is observed on the solar disk it is much
easier to see structures in
10830 \AA\ than in 5876 \AA, while both lines are clearly seen in emission when
observing off-limb structures 
such as prominences and spicules. The additional fact that the Hanle effect in
forward scattering creates 
measurable linear polarization signals in the lines of the \ion{He}{1} 10830
\AA\ multiplet when the magnetic 
field is inclined with respect to the local solar vertical direction
\citep{trujillo_nature02}, and that 
there is a nearby photospheric line of Si {\sc i}, makes the 10830 \AA\ spectral
region very suitable for 
investigating the coupling between the photosphere and the corona. 

While the Stokes $I$ profiles of the 10830 \AA\ and 5876 \AA\ helium lines
depend mainly on the 
distribution of the populations of their respective upper ($J_u$) and lower
($J_l$) levels along the 
line-of-sight (LOS), their Stokes $Q$, $U$ and $V$ profiles depend on the
strengths 
and wavelength positions of the $\pi$ ($\Delta{M}=M_u-M_l=0$),
$\sigma_{\rm blue}$ ($\Delta{M}=+1$) 
and $\sigma_{\rm
red}$ ($\Delta{M}=-1$) transitions, which can only be calculated correctly
within the framework of the 
Paschen-Back effect theory. Moreover, the $Q$, $U$ and $V$ profiles are also
affected by the atomic level polarization
induced by anisotropic pumping 
processes \citep[e.g.,][]{landi_landolfi04}. An atomic level of
total angular 
momentum $J$ is said to be polarized when its magnetic sublevels are unequally
populated and/or when there 
are quantum coherences between them. The radiative transitions induced by the 
anisotropic illumination of the helium atoms in the solar atmosphere 
are able to create a significant amount of atomic polarization in the helium
levels, even in the metastable 
lower level of the \ion{He}{1} 10830 \AA\ multiplet
\citep{trujillo_nature02}. 
If the net circular polarization of the incident radiation at the wavelengths of
the helium transitions is 
negligible, as it uses to be the case, 
the radiatively induced atomic level polarization is such that the 
populations of substates with different values of $|M|$ are different (non-zero
atomic alignment), while substates 
with magnetic quantum numbers $M$ and $-M$ are equally populated (zero atomic
orientation). On the other 
hand, elastic collisions with the neutral hydrogen atoms of the solar
chromospheric and coronal structures 
are unable to destroy the atomic polarization of the He {\sc i} levels. 
As a result, even in the
absence of magnetic fields,
linearly polarized spectral 
line radiation would be produced, simply because the population imbalances among
the magnetic sublevels imply 
more or fewer $\pi$-transitions, per unit volume and time, than $\sigma$
transitions. The atomic polarization 
of the upper level of the line transition under consideration is thus
responsible of a {\em selective emission} 
of polarization components, while that of the lower level may give rise to a
{\em selective absorption} of 
polarization components (``zero-field'' dichroism). In order for this type of
dichroism to produce a measurable 
contribution to the emergent linear polarization it is necessary to have a
substantial line-center optical 
thickness along the LOS or, assuming a small but non-negligible optical
thickness, that the plasma structure under 
consideration is observed against the bright background of the solar disk
\citep{trujillo_nature02,trujillo_asensio07}. Therefore, the observable effects
of dichroism are easier to detect in \ion{He}{1} 10830 \AA\ than in 5876 \AA.

In the presence of a magnetic field the emergent polarization changes because of
the following two reasons. 
First, because a magnetic field modifies the atomic level polarization, not only
by producing the Hanle-effect 
relaxation of the quantum coherences pertaining to each individual $J$-level,
but also through possible 
interferences between the magnetic sublevels pertaining to different $J$-levels,
which give rise to a variety of 
remarkable effects such as the transfer of atomic alignment to atomic
orientation in the $J$-levels of the 
upper term of the \ion{He}{1} D$_3$ multiplet \citep{landi_d3_82} or the
enhancement of the scattering 
polarization in the D$_2$ line of \ion{Na}{1} by a vertical magnetic field
\citep{trujillo_casini02}.
Second, because the magnetic 
splitting of the atomic energy levels give rise to significant wavelength shifts
between the $\pi$ and 
$\sigma$ transitions (as compared with the spectral line width) and,
consequently, to the generation of 
measurable linear and/or circular polarization (i.e., the familiar transverse
and longitudinal Zeeman effects, 
respectively). Obviously, a correct modeling of the spectral line polarization
that results from the joint 
action of the Hanle and Zeeman effects requires 
the application of the 
quantum theory of spectral line polarization \citep[see the monograph by][]{landi_landolfi04}, as done by several researchers for interpreting spectropolarimetric observations of solar plasma structures in the \ion{He}{1} D$_3$ multiplet \citep[e.g.,][]{landi_d3_82,bommier94,casini03,lopezariste_casini05} and in the \ion{He}{1} 10830 \AA\ triplet \citep[e.g.,][]{trujillo_nature02,trujillo_merenda05,merenda06}.

Over the last few years new computer programs for the synthesis and inversion of Stokes profiles induced by the joint action of atomic level polarization and the Hanle 
and Zeeman effects have been developed and applied to the interpretation of spectropolarimetric observations.
For example, \cite{landi_landolfi04} have developed some forward modeling codes with which they have calculated several Hanle effect diagrams and theoretical Stokes profiles of lines from complex atomic models, while the computer program of 
\cite{casini_manso05} can treat even the case of a hyperfine structured multiplet taking into account the quantum interferences between the $F$ levels belonging to the $J$ levels of different terms. Concerning Stokes inversion techniques we should 
mention that \cite{casini03} applied an inversion code based on principal component analysis \citep{arturo_casini02} to spectropolarimetric observations of solar prominences in the \ion{He}{1} D$_3$ multiplet, providing two-dimensional maps of the magnetic field vector and further evidence for strengths significantly larger than average. On the other hand, 
\cite{merenda06} opted for an inversion strategy for the \ion{He}{1} 10830 \AA\ multiplet 
in which the longitudinal component of the magnetic field 
vector is obtained from the measured Stokes $V$ profiles (which are dominated by
the Zeeman effect for the lines of the \ion{He}{1} 10830 \AA\ multiplet), and its
orientation from the observed Stokes $Q$ and $U$ profiles (which in solar prominences are due to the presence of atomic level 
polarization). In the just mentioned computer programs and 
applications the optically thin 
assumption was used, which was a suitable approximation for the particular
prominences observed, but an 
unreliable one in general (and especially for the interpretation of observations
in the \ion{He}{1} 10830 \AA\ multiplet). 

A simple but suitable model for taking into account radiative transfer effects is the constant-property 
slab model used by \cite{trujillo_nature02,trujillo_merenda05}
for the interpretation of spectropolarimetric observations in solar 
filaments and spicules, 
which has been recently extended by \cite{trujillo_asensio07} to include
magneto-optical effects. 
These authors have applied this ``cloud" model for the interpretation of
spectropolarimetric observations in order to point out that the atomic
polarization of the helium levels 
may have an important impact on the emergent linear polarization of the
\ion{He}{1} 10830 \AA\ multiplet, even 
for magnetic field strengths as large as 1000 G. Therefore, inversion codes that
neglect the influence of 
atomic level polarization, such as the Milne-Eddington codes of \cite{Lagg04} and \cite{socas_trujillo04},
should ideally be used 
only for the inversion of Stokes profiles emerging from strongly magnetized
regions (with $B>1000$ G), or 
when the observed Stokes $Q$ and $U$ profiles turn out to be dominated by the
transverse Zeeman effect 
\citep[e.g., as it happens with some active regions filaments as a result of the
particular illumination conditions explained in][]{trujillo_asensio07}. 

It is also necessary to mention that \cite{manso_trujillo03a} developed a general 
radiative transfer computer program for solving  the so-called non-LTE problem of the 
2nd kind --that is, the multilevel scattering polarization problem including the Hanle 
effect of a weak magnetic field. These authors considered the multilevel atom 
approximation (which neglects quantum interferences between the sublevels 
of {\em different} $J$-levels), but took fully into account the effects of radiative 
transfer in realistic atmospheric models and the role of elastic and inelastic 
collisions in addition to all the relevant optical pumping mechanisms. An interesting 
application using a semi-empirical model of the solar atmosphere can be found in 
\cite{manso_trujillo03b}. The recent advances in the development of efficient numerical 
methods for the solution of non-LTE polarization transfer 
problems \citep[e.g., the review by][]{trujillo03} and in computer technology 
make now possible even the performing of three-dimensional radiative transfer 
simulations of the Hanle effect in convective atmospheres \citep[e.g.,][]{trujillo_shchukina07}.

The previous introductory paragraphs strongly suggest that it would be of great
interest to develop 
a robust but user-friendly computer program for the synthesis and  inversion of
Stokes profiles resulting from 
the joint action of atomic level polarization and the Hanle and Zeeman
effects. We have done this 
by implementing an efficient global optimization method for the solution of the
inversion problem and 
by calculating at each iterative step the emergent spectral line polarization
through the solution of 
the Stokes-vector transfer equation in a slab of constant physical properties in
which the 
radiatively-induced atomic level polarization is assumed to be dominated by the
photospheric continuum 
radiation. At each point of the observed field of view the slab's optical
thickness is chosen to fit 
the Stokes $I$ profile, which is a strategy that accounts implicitly for the
true physical mechanisms  
that populate the triplet levels of \ion{He}{1} 
\citep[e.g., the photoionization-recombination mechanism, as shown
by][]{centeno08b}. The observed Stokes $Q$, $U$ and $V$ profiles are then used 
to infer the magnetic field vector along with a few extra physical quantities.

The outline of this paper is as follows. The formulation of the problem is
presented in \S2, where 
we review the relevant equations 
within the framework of the quantum theory of spectral line polarization. The
forward modelling option of 
our computer program is described in \S3, including some interesting
examples of possible 
applications. \S4 deals with a detailed description of the Stokes
inversion option, while \S5 
considers a variety of applications aimed at an in-depth testing of this
diagnostic tool. The 
important issue of the possible ambiguities and degeneracies is addressed in
\S6, with emphasis 
on the Van Vleck ambiguity and on the possibility of inferring the atmospheric
height at which the 
observed on-disk plasma structure is located. Finally, in \S7 we summarize
the main conclusions and  
comment on some ongoing developments.

\section{Formulation of the problem}

We consider a constant-property slab of \ion{He}{1} atoms, located at a height
$h$ above the 
visible solar ``surface", in the presence of a deterministic magnetic field of
arbitrary strength $B$, 
inclination $\theta_B$ and azimuth $\chi_B$ (see Fig. 1). The slab's optical
thickness at the wavelength 
and line of sight under consideration is $\tau$.
We assume that
all the \ion{He}{1} atoms inside this slab are illuminated from below by the
photospheric solar continuum radiation field, whose center-to-limb variation has
been tabulated by \cite{pierce00}. The ensuing anisotropic radiation pumping
produces population imbalances and quantum coherences between pairs of magnetic
sublevels, even among those pertaining to the different $J$-levels of the
adopted \ion{He}{1} 
atomic model. This atomic level polarization and the Zeeman-induced
wavelength shifts between the $\pi$ ($\Delta{M}=M_u-M_l=0$), $\sigma_{\rm blue}$
($\Delta{M}=+1$) and $\sigma_{\rm red}$ ($\Delta{M}=-1$) transitions produce
polarization in the emergent spectral line radiation.

In order to facilitate the reading and understanding of this paper, in 
the following subsections we summarize the basic equations which allow us to 
calculate the spectral line polarization taking rigorously into account the 
joint action of the Hanle and Zeeman effects. To this end, we have applied the 
quantum theory of spectral line polarization, which is described in great detail 
in the monograph by \cite{landi_landolfi04}. We have also applied several methods 
of solution of the Stokes-vector transfer equation, some of which can be 
considered as particular cases of the two general methods explained in \S6 of \cite{trujillo03}.

\subsection{The radiative transfer approach}
\label{sec:radiative_transfer}
The emergent Stokes vector $\mathbf{I}(\nu,\mathbf{\Omega})=(I,Q,U,V)^{\dag}$
(with $\dag$=transpose, $\nu$ the frequency and $\mathbf{\Omega}$ the unit vector indicating the direction of propagation of the ray) is obtained by solving the radiative transfer equation

\begin{equation}
\frac{d}{ds}\mathbf{I}(\nu,\mathbf{\Omega}) =
\bm{\epsilon}(\nu,\mathbf{\Omega}) - \mathbf{K}(\nu,\mathbf{\Omega}) 
\mathbf{I}(\nu,\mathbf{\Omega}),
\label{eq:rad_transfer}
\end{equation}
where $s$ is the geometrical distance along the ray under consideration,
$\bm{\epsilon}(\nu,\mathbf{\Omega})=({\epsilon}_I,{\epsilon}_Q,{\epsilon
}_U,{\epsilon}_V)^{\dag}$ is the emission vector and
\begin{equation}
\mathbf{K} = \left( \begin{array}{cccc}
\eta_I & \eta_Q & \eta_U & \eta_V \\
\eta_Q & \eta_I & \rho_V & -\rho_U \\
\eta_U & -\rho_V & \eta_I & \rho_Q \\
\eta_V & \rho_U & -\rho_Q & \eta_I
\end{array} \right)
\label{eq:propagation}
\end{equation}
is the propagation matrix. Alternatively, introducing the optical distance along the ray,  
${\rm d}{\tau}=-{\eta_I}{\rm d}s$, one can write the Stokes-vector transfer Eq. (\ref{eq:rad_transfer}) in the following two ways:

\begin{itemize}

\item The first one, whose formal solution requires the use of the evolution operator introduced by \cite{landi_landi_85}, is 
\begin{equation}
{{d}\over{d{\tau}}}{\bf I}\,=\,{\bf K}^{*}
{\bf I}\,-\,{\bf S}, 
\label{eq:rad_transfer_peo}
\end{equation}
where ${\bf K}^{*}={\bf K}/{\eta_I}$ and ${\bf S}=\bm{\epsilon}/{\eta_I}$. 
The formal solution of this equation can be seen in eq. (23) of \cite{trujillo03}.

\item The second one, whose formal solution does not require the use of the above-mentioned evolution operator is \citep[e.g.,][]{rees89}
\begin{equation}
{{d}\over{d{\tau}}}{\bf I}\,=\,{\bf I}\,-\,{\bf S}_{\rm eff},  
\label{eq:rad_transfer_delo}
\end{equation}
where the effective source-function vector
$\,{\bf S}_{\rm eff}\,=\,{\bf S}\,-\,
{\bf K}^{'}{\bf I},\,\,\,$ being $\,{\bf K}^{'}={\bf K}^{*}-{\bf 1}$
(with $\bf 1$ the unit matrix). The formal solution of this equation can be seen in eq. (26) of \cite{trujillo03}.

\end{itemize}

Once
the coefficients $\epsilon_I$ and $\epsilon_X$ (with
$X=Q,U,V$) of the emission vector 
and the coefficients $\eta_I$, $\eta_X$, and
$\rho_X$ of the $4\times4$ propagation matrix are known 
at each point within the medium it is possible to solve formally Eq.
(\ref{eq:rad_transfer_peo}) or Eq.
(\ref{eq:rad_transfer_delo}) for
obtaining the emergent Stokes profiles for any desired line of sight.
Our computer program considers the following levels of sophistication for the solution of the radiative transfer equation: 

\begin{itemize}

\item {\em Numerical Solutions}.
The most general case, where the properties of the slab vary
along the ray path, has to be solved numerically. To this
end, two efficient and accurate methods of solution of 
the Stokes-vector transfer equation are those proposed by \cite{trujillo03} (see his eqs. (24) and (27), respectively). The starting points for the development of these two numerical methods were Eq. (\ref{eq:rad_transfer_peo}) and Eq. (\ref{eq:rad_transfer_delo}), respectively. Both methods can be considered as generalizations, to the Stokes-vector transfer case, of the well-known short characteristics method for the solution of the standard (scalar) transfer equation. 

\item {\em Exact analytical solution of the problem of a constant-property slab including the magneto-optical terms of the propagation matrix}. For the general case of a constant-property slab of arbitrary optical thickness we actually have the following analytical solution, which can be easily obtained as a particular case of eq. (24) of \cite{trujillo03}:

\begin{equation}
{\bf I}={\rm e}^{-{\mathbf{K}^{*}}\tau}\,{\bf I}_{\rm sun}\,+\,\left[{\mathbf{K}^{*}}\right]^{-1}\,
\left( \mathbf{1} - {\rm e}^{-{\mathbf{K}^{*}}\tau} \right) \,\mathbf{S},
\label{eq:slab_peo}
\end{equation}
where $\mathbf{I}_{\rm sun}$ is the Stokes
vector that illuminates the slab's boundary that is most distant from the
observer. We point out that the exponential of the propagation 
matrix ${\mathbf{K}^{*}}$ has an analytical expression similar to eq. (8.23) in \cite{landi_landolfi04}.

\item {\em Approximate analytical solution of the problem of a constant-property slab including the magneto-optical terms of the propagation matrix}. An approximate analytical solution to the constant-property slab problem can be easily obtained as a particular case of eq. (27) of \cite{trujillo03}:

\begin{equation}
\mathbf{I} = \left[ \mathbf{1}+\Psi_0 \mathbf{K}' \right]^{-1} \left[ \left(
e^{-\tau} \mathbf{1} - \Psi_M \mathbf{K}' \right) \mathbf{I}_{\rm sun} +
(\Psi_M+\Psi_0) \mathbf{S} \right],
\label{eq:slab_delo}
\end{equation}
where the coefficients $\Psi_M$ and $\Psi_0$ depend only on the optical thickness of the slab at the frequency and line-of-sight under consideration, since their expressions are:
\begin{eqnarray}
\Psi_M&=& \frac{1-e^{-\tau}}{\tau} - e^{-\tau},\nonumber \\
\Psi_0 &=&1-\frac{1-e^{-\tau}}{\tau}.
\end{eqnarray}

Note that Eq. (\ref{eq:slab_delo}) for the emergent Stokes vector is the one used by \cite{trujillo_asensio07} for 
investigating the impact of atomic level polarization on the Stokes profiles of the He {\sc i} 10830 \AA\ multiplet. 
We point out that, strictly speaking, it can be considered only as the exact analytical solution of the optically-thin 
constant-property slab problem\footnote{More precisely, when the optical thickness of the slab is small in comparison 
with the eigenvalues of the matrix $\mathbf{K}'$.}. The reason why Eq. (\ref{eq:slab_delo}) is, in general, an approximate 
expression for calculating the 
emergent Stokes vector is because  
its derivation assumes that the Stokes vector within the slab varies linearly with the optical distance. However, it provides 
a fairly good approximation to the emergent Stokes profiles (at least for all the problems we have investigated in this paper). 
Moreover, the results of fig. 2 of \cite{trujillo_asensio07} remain also virtually the same when using instead the exact 
Eq. (\ref{eq:slab_peo}), which from a computational viewpoint is significantly less efficient than the approximate Eq. (\ref{eq:slab_delo}).

\item {\em Exact analytical solution of the problem of a constant-property slab when neglecting the second-order terms of the Stokes-vector transfer equation}. Simplified expressions for the emergent Stokes vector can be obtained when 
$\epsilon_I{\gg}\epsilon_X$ and $\eta_I{\gg}(\eta_X,\rho_X)$, which justifies to neglect the second-order terms of Eq. (\ref{eq:rad_transfer}). The resulting approximate formulae for the emergent Stokes parameters are given by eqs. (9) and (10) of \cite{trujillo_asensio07}, which are identical to those used by \cite{trujillo_merenda05} for modeling the Stokes profiles observed in solar chromospheric spicules. We point out that there is a typing error in the sentence that introduces such eqs. (9) and (10) in \cite{trujillo_asensio07}, since they are obtained only when the above-mentioned second-order terms are neglected in Eq. (\ref{eq:rad_transfer}), although it is true that there are no magneto-optical terms in the resulting equations. 

\item {\em Optically thin limit}. Finally, the most simple solution 
is obtained when taking the optically thin limit ($\tau{\ll}1$) in the equations reported in the previous point, which lead to the equations (11) and (12) of \cite{trujillo_asensio07}. Note that if $\mathbf{I}_{\rm sun}=0$ (i.e., $I_0=X_0=0$), then such optically thin equations imply that ${X/I}\,{\approx}\,{\epsilon_X}/{\epsilon_I}$. 

\end{itemize}

The coefficients of the emission vector and of the propagation matrix
depend on the multipolar components, $\rho^K_Q(J,J^{'})$, of the atomic density
matrix. Let us recall now the meaning of these physical quantities and how to
calculate them in the presence of an arbitrary magnetic field under given
illumination conditions.  

\subsection{The multipolar components of the atomic density matrix}

We quantify the atomic polarization of the \ion{He}{1} levels using the multipolar components of the atomic density matrix. The
\ion{He}{1} atom can be correctly described under 
the framework of the $L$-$S$ coupling \citep[e.g.,][]{condon_shortley35}. As
illustrated in Fig. 2, the  
different $J$-levels are grouped in terms with well defined values of the
electronic 
angular momentum $L$ and the spin $S$. Since the $^4$He atoms are devoid of
nuclear 
angular momentum, we do not have to consider hyperfine structure\footnote{See
\cite*{belluzzi07} for
the formulation and solution of an interesting problem where hyperfine structure
is important.}. The energy
separation between the $J$-levels pertaining to 
each term is very small in comparison with the energy difference between
different terms. For example, the energy separation between the $J=3$ and
$J=2$ levels of the term 3d$^3$D (the upper term of the D$_3$ multiplet) is of
the order of 
$0.0025\,{\rm cm}^{-1}$, which is $\sim$2$\times$10$^{5}$ times smaller than
the separation between the 3d$^3$D and 3p$^3$P terms. On the other hand,
although the energy separations between the $J$-levels of the upper terms of the
10830 \AA\ and D$_3$ multiplets are much larger than their natural widths, such
$J$-levels suffer
crossings and repulsions for the typical magnetic strengths encountered in the
solar atmospheric plasma (e.g., the $J=2$ and $J=1$ levels of the upper term of
the \ion{He}{1} 10830 \AA\ multiplet cross for magnetic strengths between 400 G
and 1600 G, while the $J=3$ and $J=2$ levels of the upper term of the D$_3$
multiplet cross 
for strengths between 10 G and 100 G, approximately). This can be seen clearly in Fig. \ref{fig:splitting}.
Therefore, it turns out to be fundamental to allow for coherences between
different 
$J$-levels pertaining to the same term but not between the $J$-levels
pertaining to 
different terms. As a result, we can represent the atom under the 
formalism of the multi-term atom discussed by \cite{landi_landolfi04}.

In the absence of magnetic fields the energy eigenvectors can be written using
Dirac's notation as $|\beta L S J M\rangle$, where
$\beta$ indicates a set of inner quantum numbers specifying the electronic
configuration. In general, if a magnetic field of 
arbitrary strength is present, the vectors $|\beta L S J M\rangle$ are no longer
eigenfunctions of the total Hamiltonian and $J$ is no longer a good quantum
number. In this
case, the eigenfunctions of the full Hamiltonian can be written as the following
linear combination:
\begin{equation}
\label{eq:eigenfunctions_total_hamiltonian}
|\beta L S j M\rangle = \sum_J C_J^j(\beta L S, M) |\beta L S J M\rangle,
\end{equation}
where $j$ is a pseudo-quantum number which is used for labeling the energy
eigenstates belonging to the subspace corresponding to assigned values of the
quantum numbers $\beta$, $L$, $S$, and $M$, and where the coefficients $C_J^j$
can be chosen to be real. 

In the presence of a magnetic field sufficiently weak so that the magnetic
energy is much smaller than the energy intervals between the $J$-levels, the energy eigenvectors are still
of the form $|\beta L S J M\rangle$ ($C_J^j(\beta L S, M) \approx \delta_{Jj}$), and the
splitting of the magnetic sublevels pertaining to each $J$-level is linear with the magnetic field strength. For stronger magnetic fields, we enter the incomplete Paschen-Back effect regime in which the energy eigenvectors are
of the general form given by Eq. (\ref{eq:eigenfunctions_total_hamiltonian}),
and the splitting among the various $M$-sublevels is no longer linear with the
magnetic strength. This regime is reached for magnetic strengths of the order of
10 G for the \ion{He}{1} D$_3$ multiplet and of the order of 100 G for the 10830
\AA\ multiplet (see Fig. \ref{fig:splitting}). If the magnetic field strength is further increased we
eventually reach the so-called complete Paschen-Back effect regime, where the
energy eigenvectors are of the form $|L S M_L M_S\rangle$ and each $L$-$S$ term
splits into a number of components, each of which corresponding to particular
values of ($M_L+2M_S$). 

Within the framework of the multi-term atom model the atomic polarization of the
energy levels is described with the
aid of the density matrix elements
\begin{equation}
\rho^{\beta L S}(jM,j'M') = \langle \beta L S j M | \rho | \beta L S j' M'\rangle,
\end{equation}
where $\rho$ is the atomic density matrix operator. Using the expression of the
eigenfunctions of the
total Hamiltonian given by Eq. (\ref{eq:eigenfunctions_total_hamiltonian}), the
density matrix 
elements can be rewritten as:
\begin{equation}
\rho^{\beta L S}(jM,j'M') = \sum_{JJ'} C_J^j(\beta L S, M) C_{J'}^{j'}(\beta L
S, M') \rho^{\beta L S}(JM,J'M'),
\end{equation}
where $\rho^{\beta L S}(JM,J'M')$ are the density matrix elements on the basis of
the eigenvectors $| \beta L S J M\rangle$.

Following \cite{landi_landolfi04}, it is helpful to use the spherical
statistical tensor 
representation, which is related to the previous one by the following linear
combination:
\begin{eqnarray}
{^{\beta LS}\rho^K_Q(J,J')} &=& \sum_{jj'MM'} C_J^j(\beta L S, M)
C_{J'}^{j'}(\beta L S, M') \nonumber \\
&\times& (-1)^{J-M} \sqrt{2K+1} \threej{J}{J'}{K}{M}{-M'}{-Q} 
\rho^{\beta L S}(jM,j'M'),
\end{eqnarray}
where the 3-j symbol is defined as indicated by any
suitable textbook on Racah algebra.
This alternative representation has some advantages. Firstly, the well-known
results 
obtained when atomic polarization effects are disregarded are easily
recovered by 
considering only the elements of the density matrix with $K=0$ and $Q=0$.
Secondly, 
the transformation law under rotations is much simpler because it involves only
one rotation matrix. Finally, each $\rho^K_Q$ element 
has a clear physical interpretation: the $\rho^2_Q$ elements (with $Q=0,{\pm}1, {\pm}2$) are 
called the atomic alignment components, while the $\rho^1_Q$ elements (with $Q=0,{\pm}1$) are 
the atomic orientation components. The 
${^{\beta LS}\rho^K_Q(J,J')}$ elements are, in general, complex quantities
(except for the elements with
$Q=0$ and $J=J'$, that are real quantities), so that, taking into account that 
the density matrix is an hermitian operator, the number of 
real quantities required to describe the
atomic polarization properties of a given $L$-$S$ term is $(2S+1)^2(2L+1)^2$. This makes a total of 405 real quantities to describe the atomic polarization of the 5-term model atom shown in Fig. \ref{fig:helium_atom}.

\subsection{Statistical equilibrium equations}
In order to obtain the
${^{\beta LS}\rho^K_Q(J,J')}$ elements we have to solve the  
statistical equilibrium equations. These equations, written in a reference
system in which 
the quantization axis ($Z$) 
is directed along the
magnetic field vector and
neglecting the 
influence of collisions, can be written as \citep{landi_landolfi04}:
\begin{eqnarray}
\frac{d}{dt} {^{\beta LS}\rho^K_Q(J,J')} &=& -2\pi \mathrm{i} \sum_{K' Q'}
\sum_{J'' J'''} N_{\beta LS}(KQJJ',K'Q'J''J''') {^{\beta LS}\rho^{K'}_{Q'}(J'',J''')}
\nonumber \\
&+& \sum_{\beta_\ell L_\ell K_\ell Q_\ell J_\ell J_\ell'} {^{\beta_\ell L_\ell
S}\rho^{K_\ell}_{Q_\ell}(J_\ell,J_\ell')} 
\mathbb{T}_A(\beta L S K Q J J', \beta_\ell L_\ell S K_\ell Q_\ell J_\ell
J_\ell') \nonumber \\
&+& \sum_{\beta_u L_u K_u Q_u J_u J_u'} {^{\beta_u L_u
S}\rho^{K_u}_{Q_u}(J_u,J_u')} 
\Big[ \mathbb{T}_E(\beta L S K Q J J', \beta_u L_u S K_u Q_u J_u J_u') \nonumber \\
& &\qquad \qquad \qquad \qquad \qquad + \mathbb{T}_S(\beta L S K Q
J J', \beta_u L_u S K_u Q_u J_u J_u') \Big] \nonumber \\
&-& \sum_{K' Q' J'' J'''} {^{\beta L S}\rho^{K'}_{Q'}(J'',J''') } \Big[
\mathbb{R}_A(\beta L S K Q J J' K' Q' J'' J''') \nonumber \\
& & + \mathbb{R}_E(\beta L S K Q J J' K'
Q' J'' J''') + \mathbb{R}_S(\beta L S K Q J J' K' Q' J'' J''') \Big].
\label{eq:see}
\end{eqnarray}
The first term in the right hand side of Eq. (\ref{eq:see}) takes into account
the 
influence of the magnetic field on the atomic level polarization. This term has 
its simplest expression in the chosen magnetic field
reference frame \citep[see eq. 7.41 of][]{landi_landolfi04}. 
In any other reference system, a more complicated expression
arises.
The second, third and fourth terms account, respectively, for coherence transfer due
to 
absorption from lower levels ($\mathbb{T}_A$), spontaneous emission from upper
levels 
($\mathbb{T}_E$) and stimulated emission from upper levels ($\mathbb{T}_S$).
The remaining terms account for the relaxation of coherences due to absorption to
upper 
levels ($\mathbb{R}_A$), spontaneous emission to lower levels ($\mathbb{R}_E$) 
and stimulated emission to lower levels ($\mathbb{R}_S$), respectively. 

The stimulated emission and absorption transfer and relaxation rates depend explicitly on 
the radiation field properties \citep[see eqs. 7.45 and 7.46 of][]{landi_landolfi04}.
The symmetry properties of the
radiation 
field are accounted for by the spherical components of the radiation field
tensor:

\begin{equation}
J^K_Q(\nu) = \oint \frac{d\Omega}{4\pi} \sum_{i=0}^3
\mathcal{T}^K_Q(i,\mathbf{\Omega}) S_i(\nu,\mathbf{\Omega}).
\label{eq:jkq}
\end{equation}
The quantities $\mathcal{T}^K_Q(i,\mathbf{\Omega})$ are spherical tensors that
depend
on the reference frame and on the  
ray direction $\mathbf{\Omega}$. They are given by
\begin{equation}
\mathcal{T}^K_Q(i,\mathbf{\Omega}) = \sum_P t^K_P(i) \mathcal{D}^K_{PQ}(R'),
\label{eq:tkq}
\end{equation}
where $R'$ is the rotation that carries the reference system defined by
the line-of-sight $\mathbf{\Omega}$ and by the polarization unit vectors $\mathbf{e}_1$ and
$\mathbf{e}_2$ into the reference system of the magnetic field, 
while $\mathcal{D}^K_{PQ}(R')$
is the usual rotation matrix \citep[e.g.,][]{edmonds60}.
Table 5.6 in \cite{landi_landolfi04} gives the
$\mathcal{T}^K_Q(i,\mathbf{\Omega})$ values for each Stokes parameter $S_i$ (with $S_0=I$, $S_1=Q$, $S_2=U$ and $S_3=V$).

\subsection{Emission and absorption coefficients}
Once the multipolar components ${^{\beta L S}\rho^{K}_{Q}(J,J') }$ are known, the
coefficients $\epsilon_I$ and $\epsilon_X$ (with
$X=Q,U,V$) of the emission vector 
and the coefficients $\eta_I$, $\eta_X$, and
$\rho_X$ of the propagation matrix
for a given transition between an upper 
term $(\beta L_u S)$ and an lower term $(\beta L_\ell S)$ can be
calculated with the expressions of \S7.6.b in  
\cite{landi_landolfi04}. 
These radiative transfer coefficients are proportional to the number density of \ion{He}{1} atoms, $\mathcal{N}$. Their defining expressions contain also the Voigt profile and the Faraday-Voigt profile \citep[see \S5.4 in][]{landi_landolfi04}, which involve the following parameters: $a$ (i.e., the reduced damping constant), $v_\mathrm{th}$ (i.e., the velocity  that characterizes the thermal motions, which
broaden the line profiles), and $v_\mathrm{mac}$ (i.e., the velocity of possible bulk motions in the plasma, which produce a Doppler shift). 

It is important to emphasize that the expressions for 
the emission and absorption coefficients and those of the statistical
equilibrium equations are written in the reference system whose quantization
axis is parallel to the magnetic field. The following equation indicates how to
obtain the density matrix elements in a new reference system:
\begin{equation}
\left[ {^{\beta L S}\rho^{K}_{Q}(J,J') } \right]_\mathrm{new} = \sum_{Q'} \left[
{^{\beta L S}\rho^{K}_{Q'}(J,J') } \right]_\mathrm{old}
\mathcal{D}^K_{Q' Q}(R)^*,
\end{equation}
where $\mathcal{D}^K_{Q' Q}(R)^*$ is the complex conjugate of the rotation matrix for the rotation $R$ that carries the 
old reference system into the new one.


\subsection{The atomic model}

The atomic model we have used in our calculations 
includes the following five terms of the 
triplet system of neutral helium: 2s$^3$S, 3s$^3$S, 2p$^3$P, 3s$^3$P and 3d$^3$D
(see Fig. \ref{fig:helium_atom}). 
It has been concluded that the inclusion of these five terms is sufficient for a
reliable calculation of the atomic polarization that the anisotropic pumping of
the photospheric continuum radiation produces in the lower and upper $J$-levels
of the D$_3$ multiplet \citep{bommier80}. Since the only level (with $J=1$) of
the lower term 2s$^3$S is metastable, the adopted atomic model should be also
satisfactory for the 10830 \AA\ multiplet. In order to check this 
we have compared the results obtained using two different atomic models. The
first one is that of  
Fig. \ref{fig:helium_atom}. The second one is 
a simplified version in which only the 2s$^3$S, 2p$^3$P and 3d$^3$D terms are taken into account.
We have verified, for a large number of possible configurations, that
differences
in the resulting values of the multipolar components of the atomic density
matrix
are never larger than $\sim$5\%
for the terms involved in the 10830 \AA\ transitions.
Although the selection rules
allow six transitions among the 5 terms, we have only included the four
transitions indicated in Fig. \ref{fig:helium_atom} (i.e., 
we have neglected the influence of the infrared transitions 3d$^3$D-3p$^3$P
and 3p$^3$P-3s$^3$S). The Einstein 
coefficients for the four included transitions shown in Table
\ref{tab:tab_einstein} and the energy of the levels shown in Fig.
\ref{fig:helium_atom} 
have been obtained from the NIST
database\footnote{\texttt{http://physics.nist.gov/PhysRefData/ASD/index.html}}
\citep[see also][]{wiese_nist66,drake_helium98}. Table \ref{tab:tab_einstein} also indicates the
value of the critical magnetic field strength for the operation of the
upper-level Hanle effect, which results from equating the Zeeman splitting of
the level with its natural width:

\begin{equation}
B_\mathrm{critical} \approx 1.137 \times 10^{-7} / (t_\mathrm{life} g_L), 
\label{eq:critical_field}
\end{equation}
where $t_\mathrm{life}$ is the level's lifetime in seconds, $g_L$ is its Land\'e factor and
$B_\mathrm{critical}$ is given in gauss.
Note that for obtaining the $B_\mathrm{critical}^\mathrm{upper}$ values of Table
\ref{tab:tab_einstein} we have used $t_\mathrm{life}{\approx}1/A_{ul}$, while
an estimation of the critical magnetic field strength for the operation of
the lower-level Hanle effect requires using
$t_\mathrm{life} \approx 1/(B_{lu}J^0_0)$ (which for the metastable lower-level
of the He {\sc i} 10830 \AA\ multiplet gives
$B_\mathrm{critical}^\mathrm{lower} \approx 0.1$ G).

We point out that in our modeling we are not explicitly taking into account the 
radiative mechanism that is thought to be responsible of the overpopulation of
the triplet levels of \ion{He}{1} required to 
produce the
absorption or emission features observed in the spectral lines of such two
multiplets --that
is, ionizations from the singlet states of \ion{He}{1} caused by EUV ionizing
radiation coming downwards from the corona followed by recombinations towards
both the singlet and triplet states \citep[e.g.,][]{avrett94}. The fact that
most of the ionizations take place from the singlet states suggests that the
atomic polarization of the triplet states should be little affected by such EUV
coronal irradiation. This expectation is reinforced by the fact that the number
of photoionizations 
per unit volume and time from the triplet levels of \ion{He}{1} is way smaller
than the number of bound-bound 
transitions \citep[see Fig. 7 in][]{centeno08b}, which suggests that the atomic
polarization of the $J$-levels 
of the 10830 \AA\ and D$_3$ multiplets is indeed dominated by optical pumping in
the line transitions 
themselves. Following our approach, the key mechanism responsible of the
absorption or emission observed 
in such lines of neutral helium, be it the above-mentioned
photoionization-recombination mechanism 
and/or collisional excitation in regions with T$>20000$ K
\citep[cf.,][]{andretta_jones97}, is 
unimportant in our forward modeling and inversion approach. The reason is that
they are 
implicitly accounted for via the definition of the optical thickness of the slab
which, being 
a free parameter in our modeling, is used to fit the observed intensity
profiles. This point has been clarified by \cite{centeno08b}.

\subsection{The incident radiation field}

As mentioned before, we consider a slab of \ion{He}{1} atoms
anisotropically
illuminated from below by the photospheric continuum radiation, which we assume
to have axial symmetry around the solar local vertical direction. Since the
illumination conditions are assumed to be known a priori, the radiation field
tensors can be calculated
directly from the given incident radiation. 

For symmetry reasons, it is advantageous to calculate the statistical tensors 
of the radiation field in a reference frame in which the $Z$-axis is along the 
vertical of the atmosphere. Since the incoming radiation is assumed to be
unpolarized, the only 
non-zero spherical tensor components of the radiation field are $J^0_0$ and
$J^2_0$. They 
quantify the mean intensity and the ``degree of anisotropy'' of the radiation 
field, respectively. For the case of our plane-parallel slab model, their
expressions 
reduce to:
\begin{eqnarray}
\label{eq:j00_j20}
J^0_0 &=& \frac{1}{2} \int_{-1}^{1} I(\mu) \mathrm{d}\mu \\
J^2_0 &=& \frac{1}{2\sqrt{2}} \int_{-1}^{1}  (3\mu^2-1) I(\mu) \mathrm{d}\mu,
\end{eqnarray}
where $\mu=\cos \theta$ is the cosine of the heliocentric angle $\theta$.

It is convenient to parameterize the radiation
field
in
terms of the number of photons per mode $\bar n$ and the anisotropy factor $w$:
\begin{equation}
\bar n = \frac{c^2}{2 h \nu^3} J^0_0, \qquad w = \sqrt{2} \frac{J^2_0}{J^0_0},
\label{eq:nbar_omega}
\end{equation}
where $c$ is the light speed, $h$ is the Planck constant and $\nu$ is the 
frequency of the transition.  
The anisotropy
factor 
fulfills $-1/2 \leq w \leq 1$. It reaches $w=1$ when the radiation is
unidirectional 
along the polar axis of the reference system, while $w=-1/2$ when the
radiation field
is azimuth-independent and confined to the plane perpendicular to the
quantization axis.
We have calculated the tensors of the radiation field by using the
center-to-limb variation and the wavelength dependence of the solar continuum
radiation
field 
tabulated by \cite{pierce00}. Since the \ion{He}{1} lines originate in the 
outer regions of the solar/stellar atmosphere, it is necessary to take into
account
the geometrical effect produced by the non-negligible height $h$ above the
surface of
the star. To this end, we have applied the strategy outlined in \S12.3 of
\cite{landi_landolfi04}.
Figure \ref{fig:nbar_omega} shows the sensitivity of $w$ and $\bar n$ to the
height in the solar atmosphere at which our slab is assumed to be located. The
number of photons 
per mode decreases, while the anisotropy factor rapidly increases. This effect
is of a purely geometrical nature.

It is important to note that the theory of spectral line
polarization 
presented in the monograph of \cite{landi_landolfi04} is only valid if the
radiation 
field illuminating the atomic system is spectrally flat (independent of
frequency) over a frequency interval larger than the Bohr frequencies connecting
the
levels that present quantum coherences. Fortunately, this is the case for all the lines
included in the atomic model of Fig. \ref{fig:helium_atom}, except for the
3p$^3$P-2s$^3$S transition at 3888.6 \AA\ which
is situated in a quite crowded region of the spectrum. 
Fortunately, this line is of 
secondary importance in setting the statistical equilibrium, and according to
\cite{landi_landolfi04} the statistical tensors can be calculated by simply reducing the
photospheric continuum radiation intensity at that wavelength by a factor 5. In any case,
its inclusion
has a rather negligible impact on the
final results.

\section{The forward modeling code}

By forward modeling we mean the calculation 
of the emergent Stokes profiles for 
given values of the height $h$ at which the slab 
is located above the visible solar ``surface",
of the slab's optical thickness and of the magnetic field vector. We have 
written this option of our computer program in a way such that it performs the 
calculation at various levels of realism. We point out that similar type of 
forward-modeling calculations can be carried out also with some of the computer 
programs mentioned in \S1, which are likewise based on the density-matrix theory 
of spectral line polarization. The main difference with ours is that we have taken 
into account radiative transfer effects (without neglecting the magneto-optical 
terms of the Stokes-vector transfer equation) and that we have developed a 
user-friendly interface to facilitate the performing of numerical experiments 
(see Fig. \ref{fig:front-end}). In what follows we first describe the various 
options of our computer program and then show some illustrative examples of possible applications.

\subsection{Description of the computing options}

The solution of the statistical equilibrium equations, using the radiation field
tensors calculated from the given illumination conditions, provides the
mutipolar components of the atomic density matrix. The computer program
calculates such $\rho^K_Q(J,J^{'})$ elements in both the magnetic field
reference frame (if a deterministic
magnetic field is present) and in a reference system where the quantization
axis lies along the solar local vertical
direction (hereafter, ``vertical frame"). This option of the forward modeling
code helps to understand what's going on at the atomic level when an atomic
system is subjected to anisotropic radiative pumping processes in the absence
and in the presence of a magnetic field. Some illustrative examples for the 
$J$-levels involved in the 10830 \AA\ and D$_3$ transitions are shown in \S3.2
below.

The values of the multipolar components of the density
matrix is all we need for calculating the coefficients of the emission
vector and of the propagation matrix, which enter the Stokes vector transfer
equation (\ref{eq:rad_transfer}). Our forward modeling code can solve this
equation at the various
levels of approximation explained in \S\ref{sec:radiative_transfer}. Moreover,
it can also compute
the emergent Stokes profiles calculating the wavelength positions of the $\pi$
and $\sigma$ components assuming the linear Zeeman-effect regime (instead of
using the general Paschen-Back effect theory), incorporating or discarding the
influence of atomic level polarization. 

\subsection{Atomic level polarization in two reference systems}

Figure \ref{fig:coherences} shows an illustrative
example of the population imbalances [$\rho^2_0(J)$ and $\rho^1_0(J)$]
and quantum coherences [$\rho^2_Q(J)$ and $\rho^1_Q(J)$, with $Q{\ne}0$] induced
by optical pumping processes in the levels of the \ion{He}{1} 10830 \AA\
multiplet that can carry atomic polarization (i.e., the lower level, with
$J_{l}=1$, and the upper levels with $J_{u}=2$ and $J_{u}=1$). Like in
\cite{trujillo_asensio07}, we
assume a slab with $\Delta{\tau}_{\rm red}=1$, where the label ``red'' indicates
that the optical 
thickness of the slab along its normal direction 
is measured at the frequency where the peak of the red blended 
component of the 10830 \AA\ multiplet is located. The slab is assumed to be at a height
of only 3 arcseconds and in the presence of a horizontal magnetic field whose strength we can vary at will. 

Consider first the results for $J_{l}=1$ and $B=0$ G in a reference system with the quantization axis along the local vertical. 
Since the pumping radiation is assumed to be unpolarized
and with axial symmetry around the vertical, for $B \approx 0$ G we see only population imbalances of the form $\rho^2_0(J)$.
Note that in the absence of magnetic fields
the $\rho^1_0(J_l) $ atomic orientation value is zero because there is no net
circular polarization in the incident radiation field. 
The bottom right panel of
Fig. \ref{fig:coherences} shows that, for $B=0$ G, we have lower-level quantum
coherences of the form $\rho^2_Q(J)$ (with $Q{\ne}0$) 
in the magnetic reference system, whose $Z$-axis is inclined with respect to the
symmetry axis of the pumping radiation field.  
In fact, for given illumination conditions in the
absence of a magnetic field, whether or not we have such quantum coherences
depends only on the reference system.
In each
of the two bottom panels it is shown what happens with the population imbalances
and the
coherences of the $J_{l}=1$ lower level as the strength of the assumed
horizontal field is increased. Consider, for instance, the right panel results
corresponding to the magnetic field reference frame. Note that the lower-level
Hanle effect starts to operate for field strengths as low as 0.01 G, and that
for field intensities of the order or larger than 1 G all the lower-level coherences have been
relaxed. This happens because such a lower level 
is metastable, which implies that its critical
Hanle field is only $0.1$ G (see Eq. \ref{eq:critical_field}).  
It is also of interest to point out that  
if the $J$-levels of our multiterm atomic model were isolated levels then the
$\rho^2_0(J)$ population imbalances would be constant in the magnetic field
reference frame, and the $\rho^1_0(J)$ atomic orientation value would be zero.
However, since the $J$-levels
suffer crossings and repulsions
non-zero $\rho^1_0(J)$ values appear through the alignment-to-orientation conversion
mechanism \citep[cf.][]{landi_d3_82}, which for the 10830 \AA\ triplet has
however a negligible impact on the emergent Stokes $V$ profiles. In addition, as
shown in the bottom right panel, the $\rho^2_0(J_l)$ alignment coefficient itself
starts to be modified as soon as the field strength is sensibly larger than 100
G. Although this lower level does not suffer any crossings with the other
$J$-levels of the model atom of Fig. \ref{fig:helium_atom}, its atomic polarization is modified
because it sensitively depends on that of the upper levels of
the 10830 \AA\ multiplet.

Consider now the
results for the upper levels with $J_{u}=1$ (central panels of Fig.
\ref{fig:coherences}) and $J_{u}=2$ (top panels).
Obviously, we only see population imbalances of the form $\rho^2_0(J)$ at
zero field in the vertical frame. Around $B=10^{-2}$ G, the density matrix
elements start to be 
affected by the magnetic field.
This modification is due to the
feedback effect that the alteration of the lower-level polarization has on the
upper levels.
It is also possible to note in the central and top panels of Fig.
\ref{fig:coherences}
the action of the upper level Hanle effect on the $J_{u}=1$ and $J_{u}=2$
levels. In fact, there is a hint of a small plateau just above 0.1 G, 
as expected from the fact that the critical upper-level Hanle field value is 0.8
G for the 10830 \AA\ multiplet (see Table \ref{tab:tab_einstein}). Similar
features can be seen in the corresponding right panels of Fig.
\ref{fig:coherences}.
Probably, the most notable conclusion to highlight 
from this figure is that between approximately 10 and 100 G we only have
population imbalances of the form $\rho^2_0(J)$ in the magnetic field reference
frame (i.e., we can consider that between 10 and 100 G the He
{\sc i} 10830 \AA\ multiplet is in the saturation regime of the Hanle effect,
where the coherences are negligible and the atomic alignment values of the lower
and upper levels are insensitive to the strength of the magnetic field). 

Finally, in Fig. \ref{fig:coherencesD3} we show similar results but for the upper levels of the He
{\sc i} D$_3$ multiplet. The situation now is much more complicated, as
evidenced by the fact that it is impossible to find a range of solar magnetic
field strength values between which the coherences are zero in the magnetic
field reference frame and, at the same time, the population imbalances are
insensitive to the magnetic strength.

\subsection{How to investigate empirically the possibility of magnetic canopies
in the quiet solar chromosphere?}
\label{sec:canopies}
In the presence of an \emph{inclined} magnetic field forward scattering
processes
can produce linear polarization signatures in spectral lines 
\citep[e.g., the review by][]{trujillo01}. In this geometry, the polarization signal
is \emph{created} by the Hanle effect, a physical phenomenon that has been
clearly demonstrated via spectropolarimetry of solar coronal filaments in the
\ion{He}{1} 10830 \AA\ multiplet \citep{trujillo_nature02}.

Fig. \ref{fig:canopies} shows theoretical examples of the emergent fractional
linear
polarization in the lines of the \ion{He}{1} 10830 \AA\ multiplet
assuming a constant-property slab of helium atoms permeated by a horizontal
magnetic field of 10 G. As expected, the smaller the optical thickness of the
assumed plasma structure the smaller the fractional polarization amplitude. In
principle, the Tenerife Infrared
Polarimeter \citep[TIP;][]{martinez_pillet99,collados_tipII07} mounted on the
Vacuum Tower Telescope (VTT) of the Observatorio del Teide allows the detection
of very low
amplitude polarization signals, such as those corresponding to the
$\Delta{\tau}_{\rm red}=0.1$ case of Fig. \ref{fig:canopies}. However, in order to be able
to achieve
this goal without having to sacrifice the spatial and/or temporal resolution we
need a larger aperture solar telescope.

\begin{figure}
\plotone{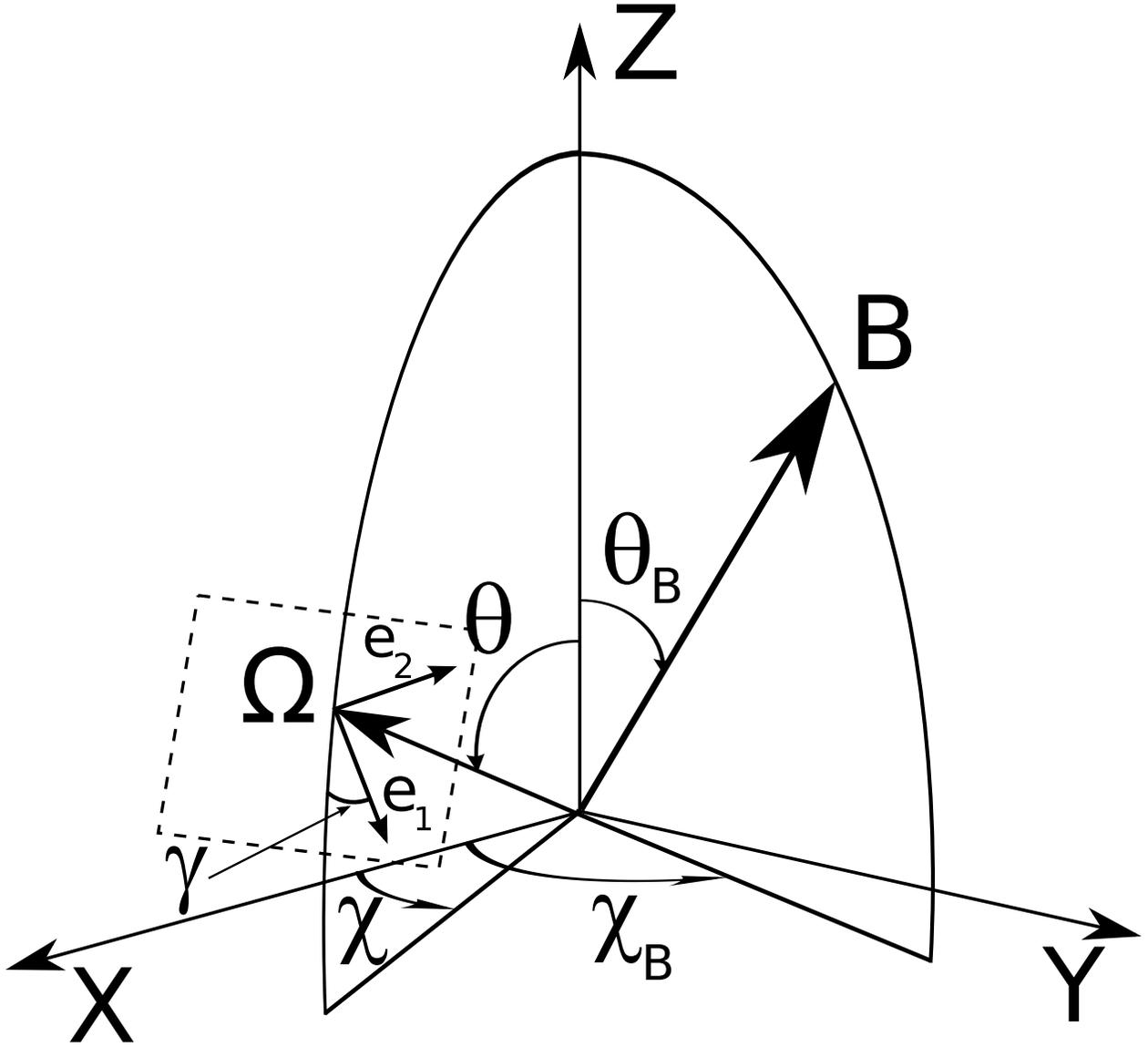}
\caption{The geometry for the scattering event. The $Z$-axis is placed along the vertical
to the solar atmosphere. The magnetic field vector,
$\mathbf{B}$,
is characterized by its modulus $B$, the inclination angle $\theta_B$ and
the azimuth $\chi_B$. The line-of-sight, indicated by the unit vector
$\mathbf{\Omega}$, 
is characterized by the two angles $\theta$ and $\chi$. 
The reference direction for Stokes $Q$ is defined by the vector $\mathbf{e}_1$
on the plane
perpendicular to the line-of-sight. This vector makes an angle $\gamma$ with
respect to the plane formed by
the vertical and the line-of-sight. In the figures showing examples of the
emergent Stokes profiles, our
choice for the positive reference direction for Stokes $Q$ is $\gamma=90^\circ$, unless otherwise stated.
For off-limb observations, we have $\theta=90^\circ$, while for observations 
on the solar disk, we have $\theta<90^\circ$. Note also that $\chi$ is generally taken to be $0^\circ$.
\label{fig:geometry}}
\end{figure}

\begin{figure}
\plotone{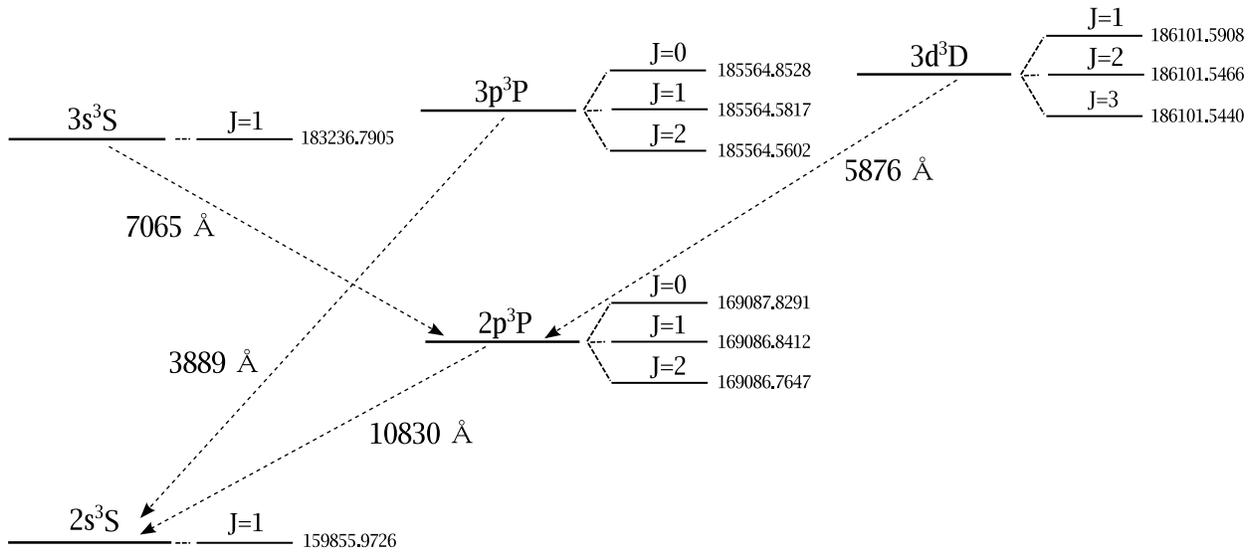}
\caption{Model atom of the triplet system of \ion{He}{1} used in this
investigation. This work focuses on 
the polarization properties of the 10830 \AA\ multiplet
between the 2p$^3$P and 2s$^3$S terms and on the D$_3$ multiplet between the
3d$^3D$ and 
2p$^3$P terms. The energy of each $J$-level is taken from
\cite{drake_helium98} and it 
is given in cm$^{-1}$ above the fundamental energy level (1s$^2 \, ^1S_0$). Note
that the separation
between the $J$-levels pertaining to each term is not drawn to scale.
\label{fig:helium_atom}}
\end{figure}

\begin{figure*}
\plottwo{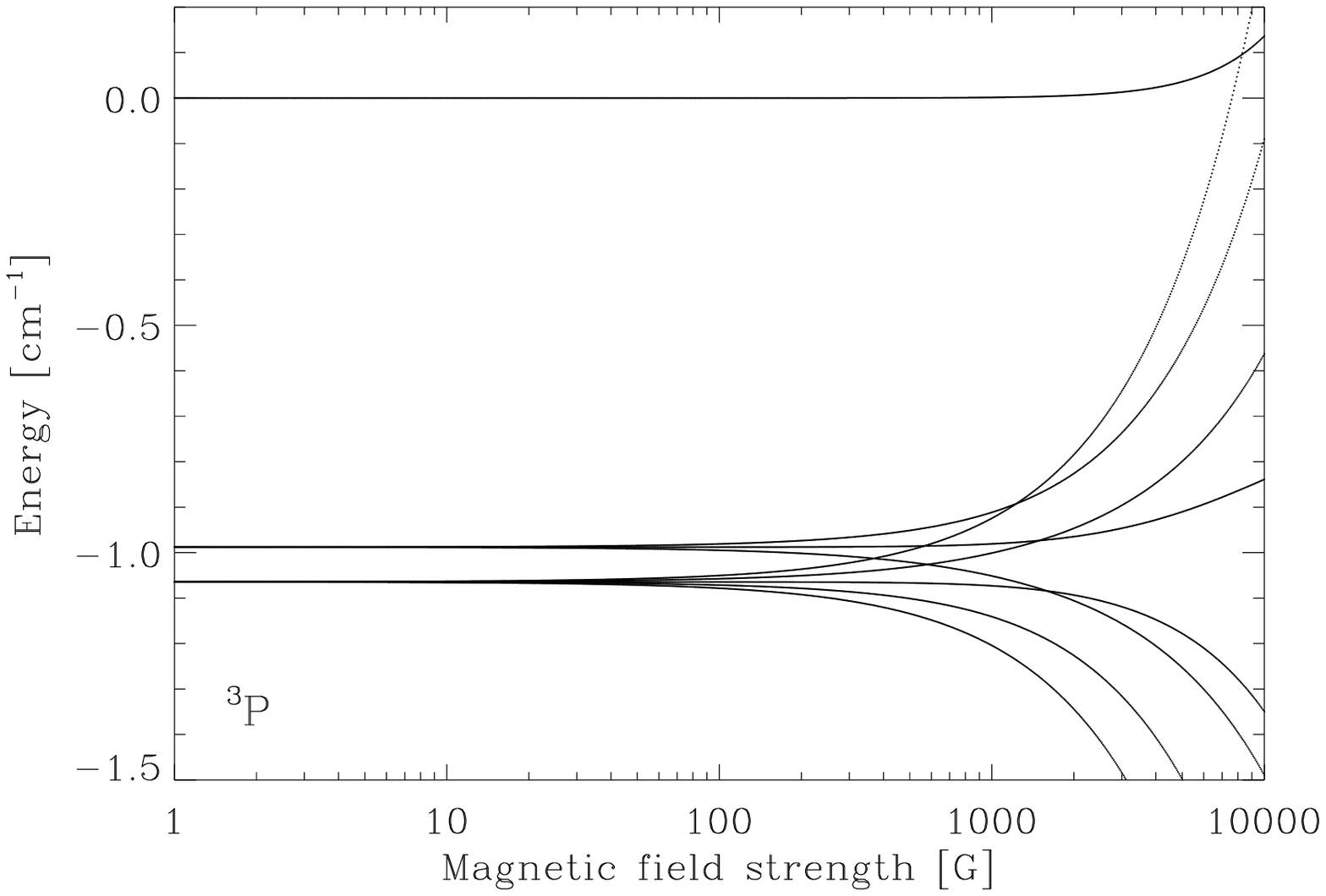}{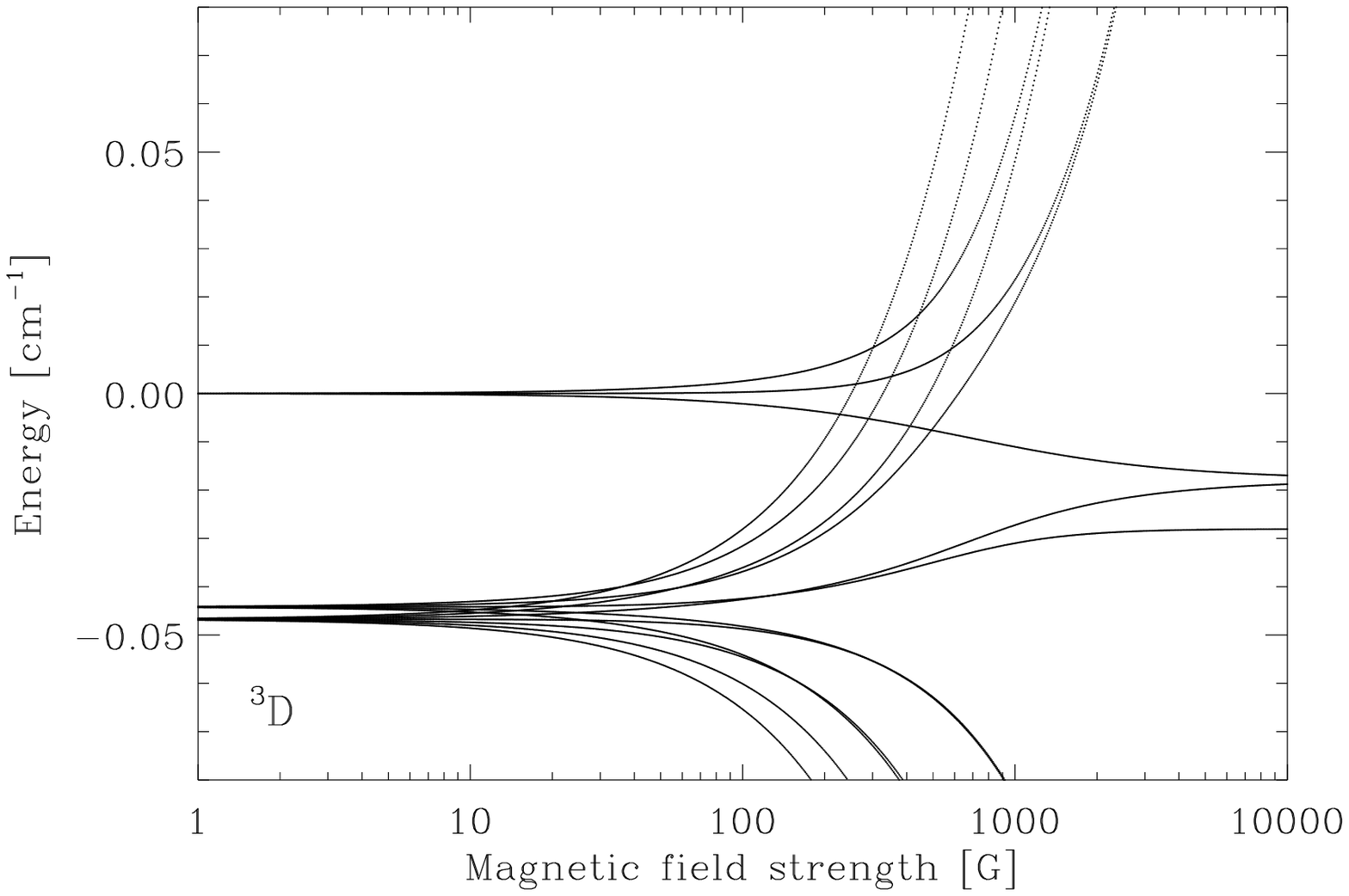}
\plottwo{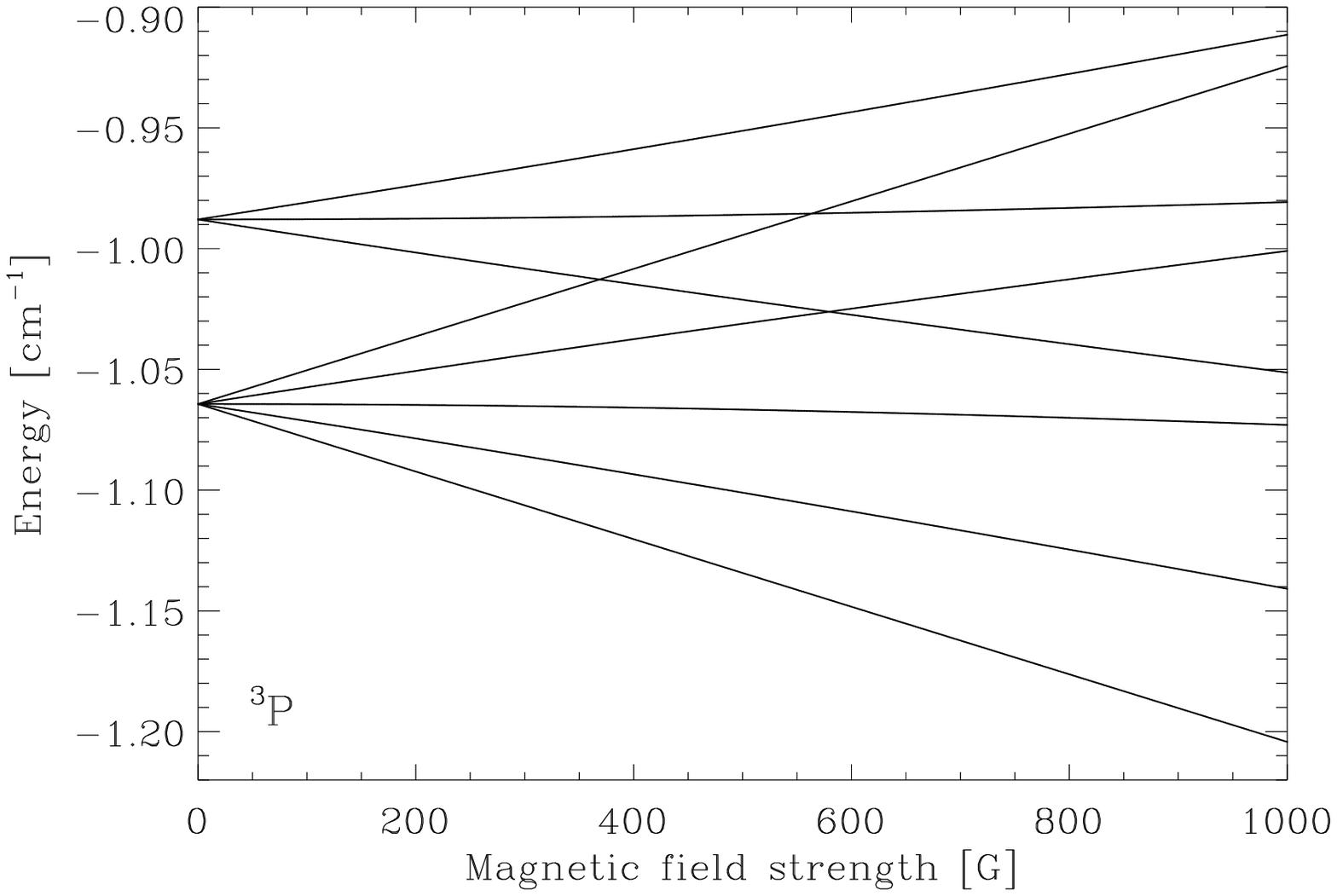}{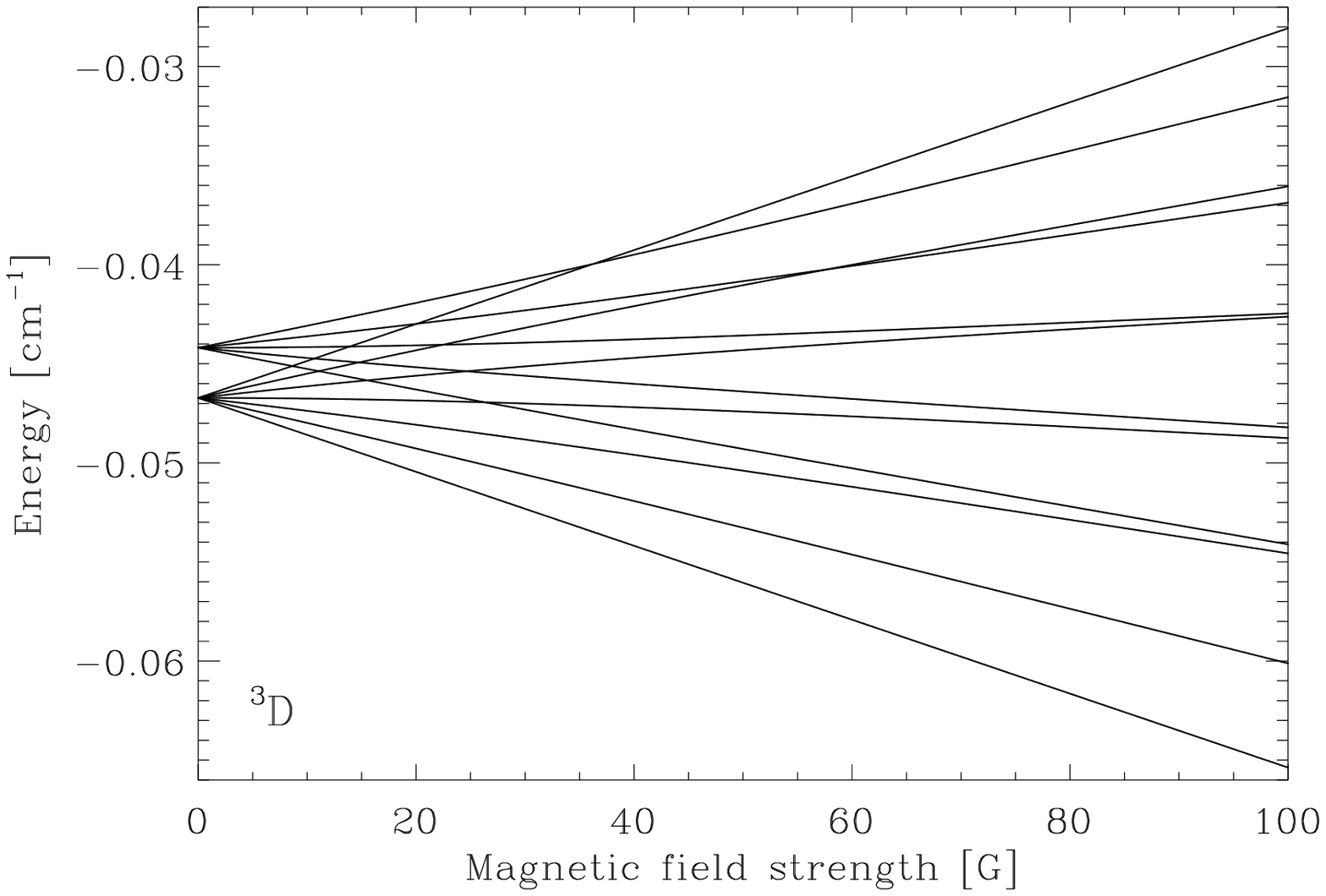}
\caption{Energy splitting due to the presence of a magnetic field for the upper
levels of the 10830 \AA\ multiplet
(left panels) and for the upper levels of the D$_3$ multiplet at 5876 \AA\
(right panels). Note that the 
regimes where level
crossings and repulsions occur are different for the two terms. They can be
better identified in the lower panels. The energy of each level is referred to
the 
energy of the level with the smallest value of $J$ at zero magnetic field (i.e.,
$J=0$ for the upper levels of the
10830 \AA\ multiplet and $J=1$ for the upper levels of the D$_3$ multiplet). The
energy separation at $B=0$ G
is obtained from the information presented in Fig. \ref{fig:helium_atom}.
\label{fig:splitting}}
\end{figure*}

\begin{figure*}
\plottwo{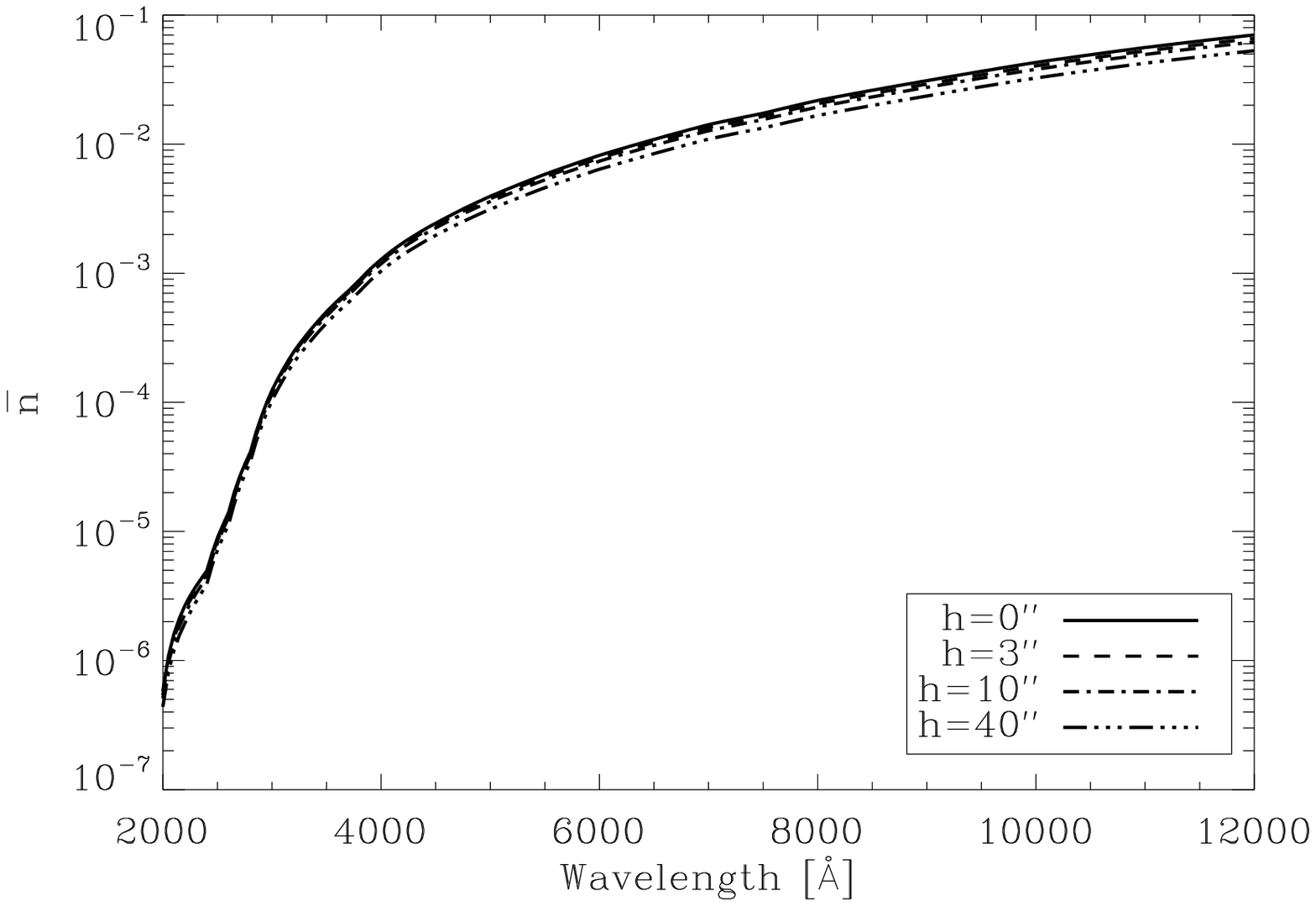}{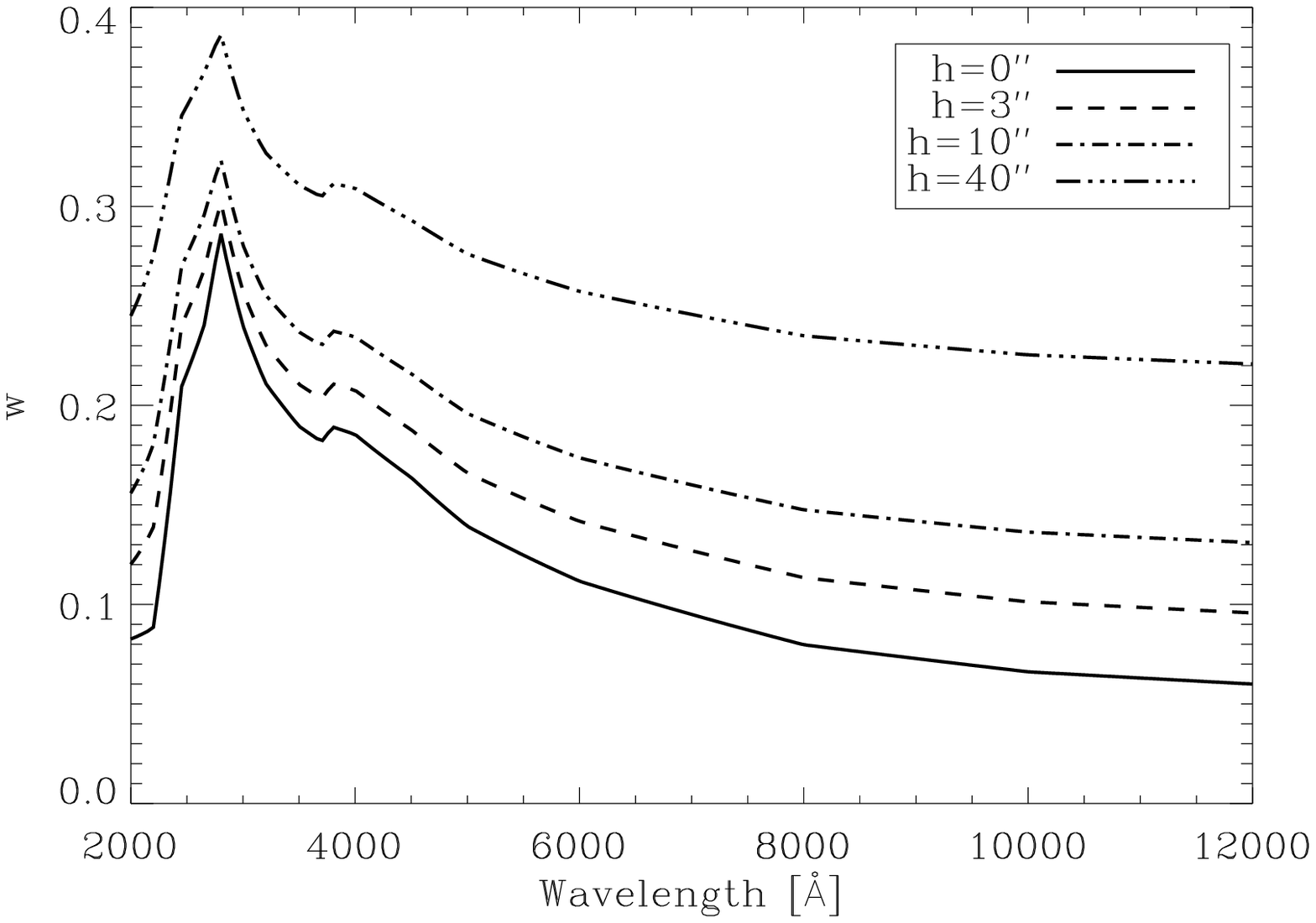}
\caption{These two quantities (see Eqs. \ref{eq:nbar_omega}) characterize the radiation field that produces optical
pumping processes in the \ion{He}{1} atoms. 
The number of photons per mode $\bar n$
(\emph{left panel}) is proportional to the mean intensity, J$^0_0$, while the
anisotropy factor 
$w$ (\emph{right panel}) is proportional to the $J^2_0$ tensor of 
the radiation field. Both quantities have been obtained using the 
limb-darkening data tabulated
by \cite{pierce00}. The figures show also the variation of these quantities with
the
atmospheric height at which the slab of helium atoms is assumed to be located.
Note that, while the number of photons per
mode is almost insensitive to $h$, the geometrical effects produce a 
significant variation on the anisotropy factor.
\label{fig:nbar_omega}}
\end{figure*}

We consider now the question of whether it is safer to interpret disk-center
observations or to opt for a different scattering geometry. To this end, we have
investigated how is the variation of $Q/I$ and $U/I$ at the central wavelengths
of the red and blue components
of the 10830 \AA\ multiplet for different inclinations ($\theta_B$) of the
magnetic field vector and for different line-of-sights, assuming
$\Delta{\tau}_{\rm red}=0.1$ and $B=10$ G (which implies that we are very close
to the saturation regime of the upper level Hanle effect). Fig. \ref{fig:canopies_qi_peak} shows
only the case of magnetic field vectors with azimuth $\chi_B=90^{\circ}$ (i.e.,
contained in the Z-Y plane of Fig. \ref{fig:geometry}) and for line-of-sights with $\chi=0$
(i.e., contained in the Z-X plane of Fig. \ref{fig:geometry}). It is interesting to note that for
the case of a horizontal magnetic field (i.e., $\theta_B=90^{\circ}$, which
implies that the magnetic field vector forms always an angle of $90^{\circ}$
with respect to any of the line-of-sights of Fig. \ref{fig:canopies_qi_peak}) 
Stokes $U=0$ and the
Stokes $Q$ amplitudes of the emergent spectral line radiation are identical 
for all such line-of-sights. This is easy to understand by using 
the following approximate expressions for $\epsilon_Q/\epsilon_I$ and
$\eta_Q/\eta_I$, which for the case of a deterministic magnetic field provide a
suitable approximation if we are in the saturation regime of the upper-level
Hanle effect \citep{trujillo03}:

\begin{equation}
{\frac{\epsilon_Q}{\epsilon_I}}\,{\approx}\,{\frac{3}{2\sqrt{2}}}\,({\mu}_B^2 -
1)\,{\cal W}\,[\sigma^2_0(J_u)]_B\, ,
\label{eq:QIa}
\end{equation}

\begin{equation}
{\frac{\eta_Q}{\eta_I}}\,{\approx}\,{\frac{3}{2\sqrt{2}}}\,({\mu}_B^2 -
1)\,{\cal Z}\,[\sigma^2_0(J_l)]_B\, ,
\label{eq:QIb}
\end{equation}
where $[\sigma^2_0]_B=[\rho^2_0]_B/\rho^0_0$ quantifies the 
fractional atomic alignment in the magnetic
field reference frame, while ${\cal W}$ and ${\cal Z}$ are numerical
coefficients that depend on the $J_l$ and $J_u$
values\footnote{Actually, ${\cal W}=w^{(2)}_{J_uJ_l}$ and ${\cal
Z}=w^{(2)}_{J_lJ_u}$, with $w^{(2)}_{JJ^{'}}$ given by Eq. (10.12) of 
\cite{landi_landolfi04}. For instance, ${\cal W}=0$ and ${\cal Z}=1$ for a
line transition with $J_l=1$ and $J_u=0$, such as that of the blue component of
the 10830 \AA\ triplet.}. It is very important to note that in Eqs.
(\ref{eq:QIa}) and (\ref{eq:QIb}) ${\mu}_B$ is {\em the cosine of the angle
between the magnetic field vector and the line of sight}. Note also that in these
expressions for $\epsilon_Q/\epsilon_I$ and $\eta_Q/\eta_I$, that are valid in the
magnetic field reference frame, we have chosen the positive reference
direction for Stokes $Q$ parallel to the projection of the magnetic field vector
onto the plane perpendicular to the LOS, while in the similar Eqs. (16) and 
(17) of \cite{trujillo_asensio07} we chose it along the perpendicular direction. In both cases, we have $\epsilon_U=\eta_U=0$ (if we are really in the above-mentioned Hanle-effect saturation regime).
Obviously, the observed $Q/I$ amplitude depends on the fractional atomic
alignment of the upper and lower levels of the 10830 \AA\ line transitions
calculated in the magnetic field reference frame, but their values are independent of the LOS.
Actually, in the weak anisotropy limit they are given by

\begin{equation}
[\sigma^2_0]_{B}=[\sigma^2_0]_{V}{\,} \frac{1}{2} (3 \cos^2{\theta_B}-1),
\end{equation}
where $[\sigma^2_0]_{V}$ is the fractional atomic alignment in the vertical
frame for the zero field case. Therefore, the dependency of the emergent Stokes $Q$ signal on the scattering angle is only through the factor $({\mu_B}^2 - 1)$, with 
${\mu_B}=0$ for all the line-of-sights of Fig. \ref{fig:canopies_qi_peak} if 
$\theta_B=90^{\circ}$.

As seen in Fig. \ref{fig:canopies_qi_peak}, for non-horizontal fields there are
notable 
differences between the curves corresponding to each LOS, mainly for the cases
with an 
inclination ($\theta_B$) of the magnetic field smaller than the Van-Vleck angle 
($\theta_{VV}=54.73^{\circ}$, which corresponds to ${\rm
cos}^{2}(\theta_{VV})=1/3$). For the forward-scattering case of a disk-center
observation (the $\mu=1$ curves of 
Fig. \ref{fig:canopies_qi_peak}) Stokes $U \approx 0$ and Stokes $Q$ admits only
one 
solution for $\theta_B > \theta_{VV}$ (with the exception of the well-known $180^{\circ}$ ambiguity of the Hanle effect)\footnote{The reason why Stokes $U$ in forward scattering geometry is
not exactly zero for the red component is because for B=10 G some of the
coherences are not completely insignificant.}. 
However, for $\theta_B < \theta_{VV}$ we may have two different magnetic 
field inclinations producing the same Stokes $Q$ values, each of them having 
its corresponding $180^{\circ}$ ambiguity of the Hanle effect. As seen in the figure, the range of 
$\theta_B$ values where two such solutions for Stokes $Q$ are possible depends
on 
the $\mu$-value of the LOS. 
Stokes $U$ is clearly non-zero for $\mu < 1$, but the fact that
$U \approx 0$ for the case of magnetic field with fixed inclination but with
random-azimuth within the 
spatio-temporal resolution element of the observation, leads us to conclude
that 
the best one can do for a reliable empirical investigation of the possibility
of 
canopy-like fields in the quiet solar chromosphere is to interpret
spectropolarimetric 
observations at the solar disk center, such as those considered in
\S\ref{sec:internetwork_regions}.

\section{The Inversion Code}
\label{sec:inversion}

Our inversion strategy is based on the minimization of a merit function 
that quantifies how well the Stokes profiles calculated in our atmospheric model
reproduce the observed Stokes
profiles. To this end, we have 
chosen the standard $\chi^2$--function, defined as:
\begin{equation}
\chi^2 = \frac{1}{4N_\lambda} \sum_{i=1}^4 \sum_{j=1}^{N_\lambda} 
\frac{\left[S_i^\mathrm{syn}(\lambda_j)-S_i^\mathrm{obs}(\lambda_j) \right]^2}{
\sigma_i^2(\lambda_j)} ,
\end{equation}
where $N_\lambda$ is the number of wavelength points and $\sigma_i^2(\lambda_j)$ is the
variance associated to the $j$-th wavelength point of the $i$-th Stokes profiles. The minimization 
algorithm tries to find the value of the parameters of our model that lead to
synthetic Stokes profiles $S_i^\mathrm{syn}$ with the best possible fit to the 
observations. 
For our slab model, the number of
parameters (number of dimensions of the $\chi^2$ hypersurface) lies between 5
and
7, the maximum value corresponding to the optically thick case (see Table
\ref{tab:parameters}). 
The magnetic field vector 
($B$, $\theta_B$ and $\chi_B$), the thermal velocity ($v_\mathrm{th}$) and the
macroscopic velocity ($v_\mathrm{mac}$) are always required. This set of
parameters is enough 
for the case of an optically thin slab. In order to account for radiative
transfer 
effects, we need to define the optical depth of the slab along its normal
direction and at a suitable
reference wavelength (e.g., the central wavelength of the red blended component
for the \ion{He}{1} 10830 \AA\ multiplet). In addition, 
we may additionally need to include the damping parameter ($a$) of the Voigt profile if
the wings of the observed Stokes profiles cannot be fitted 
using Gaussian line profiles.

A critical problem in any inversion code is to identify possible degeneracies
among different parameters of the model. When two or more parameters 
produce similar effects on the emergent Stokes profiles, the inversion algorithm
is unable to distinguish between them. As a result, the emergent Stokes profiles
corresponding to different combinations of the model parameters are
indistinguishable within the noise level. Concerning this critical point, we
investigate in detail in \S\ref{sec:invert_height} the possibility of
using the observed Stokes profiles of the 10830 \AA\ triplet to obtain the value
of the height $h$ at which the observed plasma structure is located. 

Ambiguities in the determination of the model's parameters can also result from
the presence of degeneracies.
However, this type of ambiguities occur only for a finite number of combinations
of some
parameters. Although the problem is complicated, it is possible to develop
techniques that can help selecting one of the combinations as the most 
plausible. The well-known 180$^\circ$ ambiguity of the Hanle effect 
adds to the unfamiliar  
Van Vleck ambiguity \citep{house77,casini_judge99,casini05,merenda06}. The role
of these ambiguities in the inferred model's parameters will be explored in \S\ref{sec:van_vleck}, 
although some hints have been already given in
\S\ref{sec:canopies}.

It is also important to point out the interest of developing methods capable of
providing reliable confidence intervals for all the inferred parameters. In a
future development, we plan to implement Bayesian inference techniques 
\citep{asensio_martinez_rubino07}.

\subsection{Levenberg-Marquardt}
Probably, the most well-known 
procedure for the minimization of the $\chi^2$-function is the 
Levenberg-Marquardt (LM) method \citep[e.g.,][]{numerical_recipes86}. 
The minimization strategy uses the Hessian method when the 
parameters are close to the minimum of the $\chi^2$-function 
(a quadratic form approximately describes this function
around the minimum) and the steepest descent method when the 
parameters are far from the minimum. The 
transition between both methods is done in an adaptive manner.
Its main drawback 
(which applies also to the majority of the standard numerical methods of function minimization) is that it can easily get trapped in local minima of the
$\chi^2$-function. Some alternatives are available to overcome this difficulty. The 
most straightforward but time consuming one is to restart the minimization
process 
at different values of the initial parameters. If the obtained minimum is 
systematically the same, the probability that this is the global optimum is 
high. However, when the $\chi^2$ parameter hypersurface is complicated, this 
technique does not give any confidence on the validity of the global minimum. 
Other possibilities rely on the application of some kind of ``inertia'' to the
method, so that the LM method can overcome such a local minimum problem when 
moving on the parameter hypersurface. Again, these methods do not guarantee the 
success of getting the global minimum of the function.
On the contrary, the LM method turns out to be one of the fastest and simplest
options when the initial estimate of the parameters is close to the 
absolute minimum.

\subsection{Global Optimization techniques}
In order to avoid the possibility of getting trapped in a local minimum of the
$\chi^2$ hypersurface, global 
optimization methods have to be used. Several optimization methods have been 
developed to obtain the global minimum of a function 
\citep[e.g.,][]{global_optimization95}. The majority of them are based
on stochastic optimization techniques. The essential philosophy of 
these methods is to sample efficiently the whole space of parameters 
to find the global minimum of a given function. 
One of the most promising methods is genetic optimization (inspired by
the fact of biological evolution), in 
which the parameters of the merit function are encoded in a gene. 
Although no mathematical proof of the convergence properties of these 
algorithms exists, recent advances suggest that the probability of 
convergence is very high \citep{gutowsky04}. Actually, they perform quite well
in practice 
for the optimization of very hard problems. In solar spectropolarimetry, 
genetic optimization methods have been recently applied by \cite{Lagg04} to the 
inversion of Stokes profiles induced by the Zeeman effect in the \ion{He}{1} 
10830 \AA\ triplet, neglecting the influence of atomic level polarization. The main
disadvantage is that the computing 
time needed to reach convergence increases dramatically (by a factor $\sim$10
with 
respect to standard methods based on the gradient descent like LM). 

Another different approach is based on deterministic algorithms 
\citep{global_optimization95}. Typically, these algorithms rely on a strong
mathematical basis, so that their convergence properties are well known.
We have chosen the DIRECT algorithm 
\citep{Jones_DIRECT93}, whose name derives from one of its main 
features: \emph{di}viding \emph{rect}angles. The idea is to recursively sample 
parts of the space of parameters, improving in each iteration the location of
the part of the space 
where the global minimum is potentially located. The decision algorithm is based
on the assumption that the function is Lipschitz continuous \citep[see][for details]{Jones_DIRECT93}.
The method works very well in practice and can indeed find the minimum in 
functions that do not fulfill the condition of Lipschitz continuity. The reason
is that the DIRECT algorithm does not require the explicit calculation of the 
Lipschitz constant but it uses all possible values of such a constant to determine
if 
a region of the parameter space should be broken into subregions because of
its potential interest \citep[see][for details]{Jones_DIRECT93}. A schematic
illustration of the subdivision process for
a function of two parameters is shown in Fig. \ref{fig:direct_method}.

\subsection{Convergence}
\label{sec:convergence}
Taking into account that the dimension of our space of parameters is 
between 5 and 7, it seems unreasonable to use an algorithm like DIRECT to 
obtain a precise determination of the values of the model's parameters at the
global minimum. The reason is that the precision 
in the values of the parameters decreases with the size of
the hyperrectangles. Therefore, we would need to perform many divisions 
to end up with a reasonable precision. What we do is to let the DIRECT algorithm
locate
the global minimum in a region whose hypervolume is $V$. This hypervolume is 
obtained as the product of the length $d_i$ of each dimension associated with
each of the $N$ parameters:
\begin{equation}
V = \prod_i^N d_i.
\end{equation}
When the hypervolume decreases by a factor $f$ after the DIRECT algorithm
has discarded some of the hyperrectangles, its size along each dimension is
approximately decreased by a factor $f^{1/N}$. 
In order to end up with a small region
where the global minimum is located, many subdivisions are 
necessary, thus requiring many function evaluations. 
For this reason, it has been observed that although
the DIRECT algorithm rapidly finds the region where the  
global minimum is located, its local convergence properties are rather poor 
\citep[see, e.g., ][for applications in the extremely hard problems of the
design of 
high-speed civil transport, aircrafts and 
bioinformatics]{cox01,bartholomew02,ljungberg04}. In summary, the DIRECT 
method is an ideal candidate for its application as an estimator of the region 
where the global minimum is located, but not for determining it. 

The most time consuming part of any optimization procedure is the evaluation of
the merit function. The DIRECT algorithm needs only a reduced number of evaluations 
of the merit function to find
the region where the global minimum is located. For this reason, we have
chosen it as the initialization part of the LM method. Since the initialization
point is close to the global minimum, the LM method, thanks to its quadratic behavior,
rapidly converges to the minimum.

\subsection{Stopping criterium}
A critical and fundamental problem in the optimization of functions (either
local or global) is to identify when the method has converged to the solution. 
We have used two stopping criteria for the
DIRECT algorithm. The first one is stopping when the ratio between the
hypervolume where
the global minimum is located and the original hypervolume is smaller than a
given threshold.
This method has been chosen when using the DIRECT 
algorithm as an initialization for the LM method, giving very good results. The
other good
option, suggested by \cite{Jones_DIRECT93}, is to stop after a fixed number of
evaluations of the merit function.  

Since the 
intensity profile is not very sensitive to
the presence of a magnetic field (at least for magnetic field 
strengths of the order of or smaller than 1000 G), we have decided to estimate
the optical
thickness of the slab, the thermal and the macroscopic velocity of the
plasma and the damping constant by using only the Stokes $I$ profile, and then to determine the magnetic
field
vector by using the polarization profiles. 
The full inversion scheme, shown schematically in Table \ref{tab:inversion},
begins by applying the DIRECT method to obtain a first estimation of the
indicated four
parameters by using only Stokes $I$.  Afterwards, 
some LM iterations are carried out to refine the initial values of the  
model's parameters obtained in the previous step. Once the LM method 
has converged, the inferred values of $v_\mathrm{th}$, $v_\mathrm{mac}$ 
(together with $a$ and $\Delta \tau$, when these are parameters of the model)
are kept fixed in the next steps, 
in which the DIRECT method is used again for obtaining an initial approximation
of 
the magnetic field vector 
($B$,$\theta_B$,$\chi_B$). 
According to our experience,
the first estimate of the magnetic field vector given by the DIRECT algorithm 
is typically very close to the final solution. Nevertheless, some iterations of
the LM method are performed to refine the value of the magnetic field strength,
inclination and azimuth.
In any case, although we have found very good results with this procedure, the
specific inversion scheme
is fully configurable and can be tuned for specific problems.

\section{Applications}
The main aim of this section is to illustrate the application of our inversion
code to some selected spectropolarimetric
observations in the \ion{He}{1} 10830 \AA\ multiplet, showing that it gives
results that are 
in agreement with the published ones. In addition, in
\S\ref{sec:internetwork_regions} we show a new application aimed 
at determining the strength and inclination of the magnetic field vector in the chromosphere above an 
internetwork region observed at solar disk center. Note that in the following 
applications $\Delta{\tau}_{\rm red}$ will continue denoting the optical thickness 
of the constant-property slab, along its normal direction, measured at the center 
of the red blended component of the \ion{He}{1} 10830 \AA\ multiplet.

\subsection{Prominences}
The first application is for the case of solar prominences, which are relatively
cool and dense plasma structures embedded in the $T\sim 10^6$ K solar corona. In
these objects the 
observed Stokes $Q$ and $U$ profiles of the \ion{He}{1} 10830 \AA\ multiplet are
dominated by the presence of atomic level 
polarization, while the Stokes $V$ profile
is dominated by the Zeeman effect. Recently, 
\cite{merenda06} have shown how to infer the 
magnetic field that confines the plasma of solar prominences via the inversion
of the Stokes profiles induced by scattering processes and the Hanle and Zeeman
effects in 
the lines of the \ion{He}{1} 10830 \AA\ multiplet. They analyzed in detail
spectropolarimetric
observations of the \ion{He}{1} 10830 \AA\ multiplet  
in a polar crown prominence and concluded that if the observed prominence 
was located in the plane of the sky
the
magnetic field had to be relatively strong ($B \approx 30 $ G) and inclined by only
$25^{\circ}$ with respect to the local vertical.

We have applied our inversion code to the
spectropolarimetric observations 
shown in Fig. \ref{fig:prominence}, taken from Fig. 9 of \cite{merenda06}. We
have assumed that 
the observed plasma structure was optically thin
and that the prominence
is located in the plane of the sky. The inversion code was
used to infer the value of the thermal velocity $v_\mathrm{th}$, the
macroscopic 
velocity shift $v_\mathrm{mac}$ (to allow for a shift in the wavelength
calibration) and 
the magnetic field vector $(B,\theta_B,\chi_B)$. The atmospheric height was
fixed to
$h=20"$,
the same value used by \cite{merenda06}.
After the four steps summarized in 
Table \ref{tab:inversion}, we end up with a thermal velocity  
$v_\mathrm{th}=7.97$ km s$^{-1}$, a bulk velocity that is compatible with zero 
and a magnetic field vector characterized by $B=26.8$ G,
$\theta_B=25.5^\circ$ and  
$\chi_B=161.0^\circ$. These values are in very good agreement with the results
of \cite{merenda06}, namely $B=26$ G, $\theta_B=25^\circ$ and 
$\chi_B=160.5^\circ$. Note that since the prominence plasma was assumed to lie in the 
plane of the sky, the following solutions are also valid:  
$\theta_B^{*}=180^\circ-\theta_B$ and $\chi_B^{*}=-\chi_B$ (i.e., the 
well-known $180^{\circ}$ ambiguity of the Hanle effect).

We point out that the total number of evaluations of the merit function was 132. For the inversion of the Stokes profiles corresponding to other points of the field of view, one can initialize the inversion using the model's parameters corresponding to the
previous point. Using a few LM iterations, one should be able
to reach the global minimum. In case this procedure does not work properly, one should return to the four-steps 
inversion scheme already presented in Table \ref{tab:inversion}.

\subsection{Spicules}
\label{sec:spicules}
Another interesting problem is
the determination of the magnetic field vector in solar 
chromospheric spicules. \cite{trujillo_merenda05} interpreted 
spectropolarimetric observations of spicules in the 
\ion{He}{1} 10830 \AA\ multiplet and concluded that the magnetic
field of spicules in quiet regions of the solar chromosphere has a strength 
of the order of 10 G and is inclined by about 35$^\circ$ with respect to the 
local vertical. Their conclusion that the 
typical magnetic field strength is $\sim 10$ G required to obtain the 
longitudinal component of the magnetic field vector via
some careful measurements of the Stokes
$V$ profiles, such as that shown in Fig. 13 of \cite{trujillo_esa05}.  
This was needed because for field strengths larger than a few gauss
the \ion{He}{1} 10830 \AA\ multiplet enters the saturation regime of the
upper-level Hanle effect and the observed Stokes $Q$ and $U$ profiles provide only 
information on the orientation of the magnetic field vector.
In fact, the application of our inversion code to the observed Stokes profiles 
shown in Fig. \ref{fig:spicules} (where the Stokes $V$ profile is at the noise level), 
assuming that the spicular material
is located in the plane of the sky, provides several different magnetic field vectors that lead to equally
reliable fits. One of them, given by $B=10$ G, $\theta_B=37^\circ$ and
$\chi_B=172^\circ$ is similar 
to the one chosen by \cite{trujillo_merenda05}. Another possible fit is 
the one illustrated in Fig. \ref{fig:spicules}, which corresponds to $B=2.6$ G, $\theta_B=37^\circ$ and $\chi_B=35^\circ$.

Fig. \ref{fig:spicules_chi2} gives the values of the $\chi^2$ function for
all possible magnetic field inclinations and azimuths corresponding to 
the cases $B=10$ G (left panel) and $B=2.6$ G (right panel). In each of these 
panels we have indicated with white dots and numbers the four solutions that
correspond to equally good best fits to the observed Stokes profiles of Fig. \ref{fig:spicules}. 
The pair of solutions $1$ and $2$ correspond to the Van-Vleck ambiguity\footnote{Information on this ambiguity,
typical of the Hanle-effect saturation regime, can be found in \cite{casini05}, in \cite{merenda06} and in \S\ref{sec:van_vleck} below.}. 
The same applies to the $1'$ and $2'$ solutions.
On the contrary, the pair of solutions
$1$ and $1'$ or the $2$ and $2'$ are not strictly equivalent. The inversion code considers 
such pairs of solutions as equivalent because the observed Stokes $V$ profile is at 
the noise level and it is not able to differentiate between the two cases.
Note that these pairs of solutions give exactly the same Stokes $Q$ and $U$ profiles, but their corresponding 
Stokes $V$ profiles have opposite signs.
Concerning each pair of solutions in Fig. \ref{fig:spicules_chi2}, it is
possible to verify that the projections on the plane of the sky of the magnetic fields
corresponding to solutions $1$ and $2$ form an angle close to to 90$^\circ$, which is typical of the Van-Vleck ambiguity. 
The same happens for the magnetic fields corresponding to
solutions $1'$ and $2'$. This holds for both cases, $B=10$ G and $B=2.6$ G. As pointed out above, when the observed Stokes $V$ signal is very small, it is 
very hard (or impossible) to differentiate between the two
possibilities having azimuths $\chi_B$ and $180^\circ-\chi_B$. The  
magnetic field vectors $1$ and $1'$ (or those corresponding to the $2$ and $2'$ 
solutions of Fig. \ref{fig:spicules_chi2}) have the same projection on the
line of sight, except for a sign change. Therefore, the detection of Stokes $V$ turns out to be fundamental to
determine which is the
correct one \citep{trujillo_merenda05,merenda06}. Apart from the considered solutions, which are
restricted to the interval $0^\circ < \theta_B < 180^\circ$, one has also 
to take into account the well-known ambiguity of the Hanle effect, which applies when 
the emitting plasma is located in the plane of the sky. In this case, we have to add to the possible 
set of solutions all the combinations fulfilling $\theta_B^{*}=180^\circ-\theta_B$ and $\chi_B^{*}=-\chi_B$ since both 
pairs produce the same Stokes profiles.\footnote{We point out that this  ambiguity of the Hanle effect applies only to some particular scattering geometries (i.e., to those of 90$^\circ$ and of forward scattering). If the observed plasma structure is not located in the plane of the sky (which implies scattering processes at an angle different from 90$^\circ$), or if it is not located at the solar disk center, then one has a sort of quasi-degeneracy which can disappear when the angle $\theta$ of Fig. 1 is considerably different from 90$^\circ$ or from 0$^\circ$. This fact has been exploited by \cite{landi_bommier93} to propose a method for removing the azimuth ambiguity intrinsically present in vector magnetograms.}.

\subsection{Filaments}
We have also considered the inversion of the Stokes profiles presented in
\cite{trujillo_nature02}, which were observed
in a solar coronal filament at the solar disk center. Such profiles, which are
reproduced in Fig. \ref{fig:filament_observation}, were used by those authors to
demonstrate the presence of atomic polarization in the lower level of the 10830
\AA\ multiplet
and that the Hanle effect due to an inclined magnetic field creates 
linearly polarized radiation in forward scattering geometry. Note that the
Stokes $Q$, $U$
and $V$ profiles are normalized to the maximum depression in Stokes $I$ 
(which is $0.4\,I_c$, approximately). The
application of our inversion code using $h=40"$ confirms the conclusions of
\cite{trujillo_nature02}, yielding the following values for the model's
parameters: $\Delta \tau=0.86$, $v_\mathrm{th}=6.6$ km s$^{-1}$, $a=0.19$ and a
magnetic field vector characterized by $B=18$ G and $\theta_B=105^\circ$.

\subsection{Inter-network regions}
\label{sec:internetwork_regions}
As discussed in \S\ref{sec:canopies}, the investigation of the possibility of horizontal magnetic canopies in the quiet solar chromosphere above internetwork regions is feasible
with TIP, especially when interpreting measurements of the polarization of the \ion{He}{1} 10830 \AA\ multiplet in forward scattering at the solar disk center\footnote{For a preliminary interpretation of an observation of a quiet region located at $\mu=0.5$ 
see \cite{Lagg07}.}. 
We present in Fig. \ref{fig:canopy_observation} an observation carried
out with TIP very close
to the solar disk center ($\mu=0.98$). The slit was crossing an enhanced network
region of circular shape. 
A time series of 50 steps with an integration time of 3 seconds was performed.
The resulting polarimetric 
sensitivity after averaging over the 50 time steps and along a small spatial
interval within the observed 
internetwork region is close to 6$\times$10$^{-5}$ in units of the continuum
intensity. According to the results
of the right panel of Fig. \ref{fig:canopies}, this is sufficient for detecting the 
linear polarization signal of a horizontal magnetic field provided the optical
depth at the wavelength of the red 
component of the 10830 \AA\ multiplet is larger than $\sim 0.01$. The reference
system has been rotated until
Stokes $U$ is minimized. Since the inferred magnetic field 
strength implies that the \ion{He}{1} 10830 \AA\ is in the saturation 
regime of the upper-level Hanle effect, the resulting
reference direction for Stokes $Q$ lies either
along the projection of the magnetic
field vector on the solar disk, or along the direction perpendicular to such a projection.
We have applied our inversion
code to the above-mentioned observed profiles assuming a 
slab located at a
height of 3 arcsec and we have obtained the following 
results: $\Delta \tau=0.19$, $v_\mathrm{th}=9.2$ km s$^{-1}$, $a=0.62$ and a
magnetic field vector characterized by
$B=35$ G, $\theta_B=21^\circ$ and $\chi_B=0^\circ$. However, other possible
solutions
can be found with a similar goodness of the fit (e.g.,  
$B=47$ G, $\theta_B=47^\circ$ and $\chi_B=0^\circ$).
These results obtained from a solar disk center observation 
suggest the presence of magnetic fields inclined by no more than
$50^{\circ}$ in the observed quiet Sun chromospheric region. 

\subsection{Emerging magnetic flux regions}
As pointed out by \citet{trujillo_asensio07}, the modeling of the emergent 
Stokes $Q$ and $U$ profiles of the \ion{He}{1} 10830 \AA\ multiplet should be done by 
taking into account the possible presence of 
atomic level polarization, even for magnetic field strengths 
as large as 1000 G. 
An example of a spectropolarimetric observation of
an emerging magnetic flux region is shown by the circles of Fig.
\ref{fig:lagg_emerging}.
The solid lines show the best theoretical fit to these 
observations of \cite{Lagg04}. Here, in addition to the Zeeman effect, we 
took into account the influence of atomic level polarization. The 
dotted lines neglect the atomic level polarization that is induced by
anisotropic radiation pumping
in the solar atmosphere. Our results 
indicate the presence of atomic level polarization
in a relatively strong field region (${\sim}$1000 G). However, it may be
tranquilizing to point out that both inversions of the observed profiles yield, at least
for this case, a
similar 
magnetic field vector, in spite of
the fact that the corresponding theoretical fit is much better for the case that
includes atomic
level polarization. 

\section{Ambiguity and degeneracies}

In the previous subsections we have shown how our inversion code can be used for
recovering the parameters of the 
assumed slab atmospheric model from the Stokes profiles observed in different
solar plasma structures. The aim of this section is to discuss the Van Vleck
ambiguity and to investigate 
whether we can infer the height of the
observed plasma structure directly through the inversion approach.

\subsection{Van Vleck Ambiguity}
\label{sec:van_vleck}
In general, the solution to any inversion problem is not unique --that is, it is often possible
to detect several solutions which are compatible with the observations
\cite[e.g.,][]{asensio_martinez_rubino07}.
Some of the unicity problems are associated with the physics of the polarization
phenomena (e.g., the 180$^\circ$ ambiguity of the Hanle effect for plasma structures located in the plane of the sky or the Van Vleck ambiguity).
However, as seen in \S\ref{sec:spicules},
in addition to this type of ambiguities, other degeneracies can appear because of the presence of 
noise in the observed Stokes profiles. 

The Van Vleck ambiguity occurs only for some
combinations 
of the inclinations and azimuths. Moreover, it occurs mainly in the saturation regime
of the Hanle effect. For example, Fig. 6 of \cite{merenda06} 
shows the region of parameters for which the Van
Vleck ambiguity occurs in the Hanle-effect
saturation regime of the \ion{He}{1} 10830 \AA\ triplet. 
Since two different 
magnetic field vectors give rise to exactly the same emergent Stokes 
profiles, it is impossible to distinguish between them 
using only the 10830 \AA\ multiplet
(or four solutions, if the 180$^\circ$ ambiguity of the Hanle effect also applies). However, 
if more information is introduced
in the inversion procedure (for instance, by using simultaneous observations in
the 10830 \AA\ 
and D$_3$ multiplets), it might be possible to distinguish between the two possible solutions.

Unfortunately, it is not easy to determine
the range of parameters in which we may have 
the Van Vleck ambiguity. One possibility \citep[used by][]{merenda06} is to 
consider the theoretical Hanle diagram of the red line of the He {\sc i} 10830 \AA\ multiplet and detect if the 
observed profiles fall in the ambiguity region. We propose another 
method based on the DIRECT algorithm implemented in our inversion code. 
The DIRECT algorithm can rapidly detect regions of the space of parameters
where the global minimum may be located. Therefore, we can take advantage of
this property to identify the two (or more) points in the space of parameters
$(\theta_B,\chi_B)$ that 
produce the same emergent Stokes profiles for a given magnetic field strength.

To this end, we have calculated the synthetic emergent Stokes profiles of the
\ion{He}{1} 10830 \AA\ line from an optically thin prominence, located in the plane of
the sky, with $v_\mathrm{th}=8$ km s$^{-1}$, $h=20"$, $B=25$ G, $\theta_B=40^\circ$ and 
$\chi_B=19^\circ$.
According to the Hanle diagram shown by \cite{merenda06}, these profiles are 
indistinguishable from the ones given by the combination $B=22$ G, 
$\theta_B=100^\circ$ and $\chi_B=46^\circ$. To these two combinations,
we have to add those corresponding to the 180$^\circ$ ambiguity: ($B=25$ G,
$\theta_B=140^\circ$, 
$\chi_B=-19^\circ$) and ($B=22$ G, $\theta_B=80^\circ$, $\chi_B=-46^\circ$).
Using 
the standard four steps inversion procedure explained in
\S \ref{sec:inversion},
the global minimum is rapidly located at position $B=22$ G,
$\theta_B=100^\circ$ and 
$\chi_B=46^\circ$. Keeping fixed the value of all the thermodynamical
properties 
and the field strength, the DIRECT algorithm is used to recover the inclination
and 
azimuth of the magnetic field vector. We show in Fig \ref{fig:vanvleck_ambiguity} the position in the
$(\theta_B,\chi_B)$ space of parameters of the $N$ evaluations performed by the 
DIRECT method. It has been possible to detect
the two combinations of parameters that give the same emergent Stokes profiles, 
as stated above. With only $N=100$ evaluations of the merit function, the 
DIRECT algorithm has located and refined the position of the global minimum. It
has also identified the second possible solution. 
When the 
number of function evaluations increases (even with only $N=200$), the DIRECT 
algorithm rapidly locates the two minima. For $N>200$, we face a
degradation in the convergence rate as discussed in \S\ref{sec:convergence}.

As already discussed, an interesting property of the DIRECT method is that no
hyper-rectangle is ever discarded 
from the search. Therefore, a rectangle that in one iteration is not considered
to be potentially interesting, can be chosen for division in a later iteration\footnote{This
proves to be fundamental to demonstrate that the global minimum will always be found 
\citep{Jones_DIRECT93}.}. 
This behavior is shown in Fig. 
\ref{fig:vanvleck_ambiguity}. When $N=100$, only a part of the space of
parameters 
has been sampled, with clear gaps for inclinations above 110$^\circ$. In spite
of these gaps, the two global minima have been already
found. However, when the number of function evaluation is increased, the
numerical scheme finally evaluates the function in those regions with the 
aim of discarding the presence of additional global minima. 

\subsection{Can we infer the height of the observed plasma structure?}
\label{sec:invert_height}
In this section we briefly discuss the possibility of determining the height at 
which the \ion{He}{1} atoms are located by only using the information contained
in 
the Stokes profiles of the 10830 \AA\ multiplet. For simplicity, we consider
first the optically thin limit, the case of off-limb observations (i.e., $90^{\circ}$ 
scattering geometry) and a magnetic field with a fixed
azimuth and strength. The
synthetic 
profiles correspond to the case $v_\mathrm{th}=8$ km s$^{-1}$ and 
$h=20"$, with the magnetic field vector given by $B=25$ G,
$\theta_B=40^\circ$ and  
$\chi_B=19^\circ$. The aim of this experiment is to infer the inclination
$\theta_B$ and height $h$ from synthetic Stokes profiles without noise. 
The positions where the merit 
function has been evaluated by the DIRECT algorithm are presented in the upper panels
and in the bottom left panel of Fig. 
\ref{fig:height_degeneration}. 
The $\chi^2$ surface is shown in the bottom right
panel of Fig. \ref{fig:height_degeneration}. The presence of the vertical strip where the
minimum is located makes it
very difficult to converge to the minimum using gradient-based methods like the 
LM method. The derivatives cannot be correctly approximated when the $\chi^2$ 
function has a large variation in such a small region of the space of
parameters.
This shallow strip is produced by the quasi-degeneracy of the problem in both
parameters. An infinity of combinations of both parameters give Stokes
profiles
that can approximately reproduce the observations. The difference in the
$\chi^2$ merit function between 
these spurious cases and the exact one is very small. Two reasons produce 
such a behavior. On the one hand, the linear polarization signal is 
enhanced when the height is increased because the anisotropy of the radiation
field 
increases (see Fig. \ref{fig:nbar_omega}, right panel). 
On the other hand, the Hanle effect turns 
out to be particularly efficient in reducing the atomic polarization when the magnetic field is significantly inclined with respect to the symmetry axis of the radiation field (the vertical
direction).
In a 
realistic case, the problem is much more complicated due to the presence of
other additional
parameters and the noise contamination.

When the observed structure is off the limb, imaging techniques can be used to estimate $h$.
On the contrary, the case of on-disk observations is much more complicated since no 
straightforward technique for estimating the height is available. One
possibility is to follow the observed active region until it approaches the limb. The height 
can then be estimated if we assume that the height of the plasma structure has not changed
between both observations. An even less precise procedure is to assume a given $h$ value  
based on the typical height of the solar structure type under study. Obviously, the ideal situation
would be the one where $h$ could be inferred directly from the observed Stokes profiles.
In order
to investigate this possibility, we have performed an experiment in which the DIRECT method is used
with disk-center ($\theta=0^\circ$) synthetic Stokes profiles. The emergent profiles have been calculated  
with $v_\mathrm{th}=8$ km s$^{-1}$, $\Delta \tau=0.8$, $h=20"$, $B=25$ G, $\theta_B=40^\circ$ and $\chi_B=19^\circ$, 
taking into account the effects of radiative transfer in the slab. We keep fixed 
all the parameters except for the inclination of the magnetic field $\theta_B$
and the height $h$. The upper panels and the bottom left panel of Figure
\ref{fig:height_diskcenter} present the points at which the 
DIRECT algorithm has evaluated the merit function, showing that it is possible
to infer the height of the observation by only using the Stokes profiles. The shape
of the $\chi^2$ surface is shown in the bottom right panel of Fig.
\ref{fig:height_diskcenter}.
In comparison with the off-limb case shown in Fig.
\ref{fig:height_degeneration}, the minimum is located in a much less
complicated region of the $\chi^2$ surface.
The quasi-degeneracy present in 
the off-limb case is not present in the on-disk case. 
This is associated with the fact that the blue 
component gives no signal in the optically thin limit, while it does if an inclined field is 
present for the disk center case \citep{trujillo_nature02}.

Interestingly, if one wants to infer the magnetic field vector and the height simultaneously
from the observations, the code is unable to get a suitable global minimum, even in the noiseless 
case. However, an easily accessible global minimum exists when one of the parameters is kept 
fixed, thus inferring only the following combinations of 
parameters: ($B$, $\theta_B$, $h$), ($B$, $\chi_B$, $h$) and ($\theta_B$,
$\chi_B$, $h$).

\section{Conclusions}
\label{sec:conclusions}

The physical interpretation of spectropolarimetric observations of lines of
neutral helium, such as those of the 10830 \AA\ and D$_3$  multiplets,
represents a very important diagnostic window for investigating the dynamical
behavior and the magnetic field of plasma structures in the solar chromosphere
and corona, such as spicules, filaments, regions of emerging magnetic flux,
network and internetwork regions, sunspots, flaring regions, etc. In order to
facilitate this type of investigations we have developed a
powerful forward modeling and inversion code that permits either to 
calculate the emergent spectral line intensity and polarization for any given
magnetic field vector or to infer the dynamical and magnetic properties from the
observed Stokes profiles. This diagnostic tool is based on the quantum theory of
spectral line polarization \citep[see][]{landi_landolfi04}, which self-consistently
accounts for the presence of atomic level polarization and the Hanle and Zeeman
effects in the most general situation of the incomplete Paschen-Back effect regime.  It is
also of interest to mention that the same computer program can be easily applied
to other chemical species apart from \ion{He}{1} (e.g., in order to 
investigate the magnetic sensitivity of the polarization caused by the joint
action of the Hanle and Zeeman effects in many other spectral lines of
diagnostic interest, both in the solar atmosphere and in other astrophysical plasmas).

The influence of radiative transfer on the emergent spectral line radiation is
taken into account by solving the Stokes-vector transfer equation in a slab of
constant physical properties, including the magneto-optical terms of the
propagation matrix. Although this ``cloud" model for the interpretation of
polarimetric observations in such lines of \ion{He}{1} is suitable for inferring
the magnetic field vector of plasma structures embedded in the solar
chromosphere and corona, there are several interesting improvements and
generalizations on which we are presently working on. The first one will be
useful for improving the modeling of the Stokes profiles observed in 
low-lying optically-thick plasma structures embedded in the solar chromosphere,
such as those of active region filaments. It consists in taking into account
that in optically-thick plasma structures located at low atmospheric heights,
the atomic level polarization is not going to be necessarily dominated by the
anisotropic continuum radiation coming from the underlying solar photosphere (as
we have assumed here), given that the radiation field generated by the
optically-thick structure itself will tend to reduce the anisotropy factor of
the true radiation field that pumps the helium atoms of the plasma structure
under consideration \citep[see][]{trujillo_asensio07}. The second additional 
development consists in considering a Milne-Eddington atmospheric
model, but determining consistently the height-dependent atomic level
polarization induced by the anisotropic radiation field within the
atmosphere model that provides the best fit to the observed
Stokes profiles. Since the anisotropy factor is very sensitive to
the source-function gradient \citep[e.g., Fig. 4 in][]{trujillo01} the solution of these type of problems in stratified model atmospheres may be facilitated by the application of efficient  
iterative schemes, such as those used by 
\cite{manso_trujillo03a,manso_trujillo03b} for developing a general multilevel radiative 
transfer program for modeling scattering line polarization and the Hanle effect in weakly 
magnetized stellar atmospheres.

For the solution of the Stokes inversion problem we have applied an efficient
algorithm based on global optimization methods, which permits a fast and
reliable determination of the global minimum and facilitates the determination
of the solutions corresponding to the unfamiliar Van-Vleck ambiguity. Our
inversion approach is based
on the application of the Levenberg-Marquardt (LM) method for locating
the minimum of the merit function that quantifies the goodness of the fit between the observed and synthetic
Stokes profiles. However, gradient-based methods suffer from convergence problems when the initial
value of the parameters is not close to the minimum. In order to improve the convergence 
properties of the LM method, we have introduced a novel initialization technique. This 
method is based on the DIRECT algorithm, a deterministic global optimization scheme that 
performs very well. We have shown that a four-steps scheme using the DIRECT method to 
initialize the parameters and the LM method to refine the first estimation close to the 
minimum leads to a very robust technique. 

\begin{figure*}
\includegraphics[width=\columnwidth]{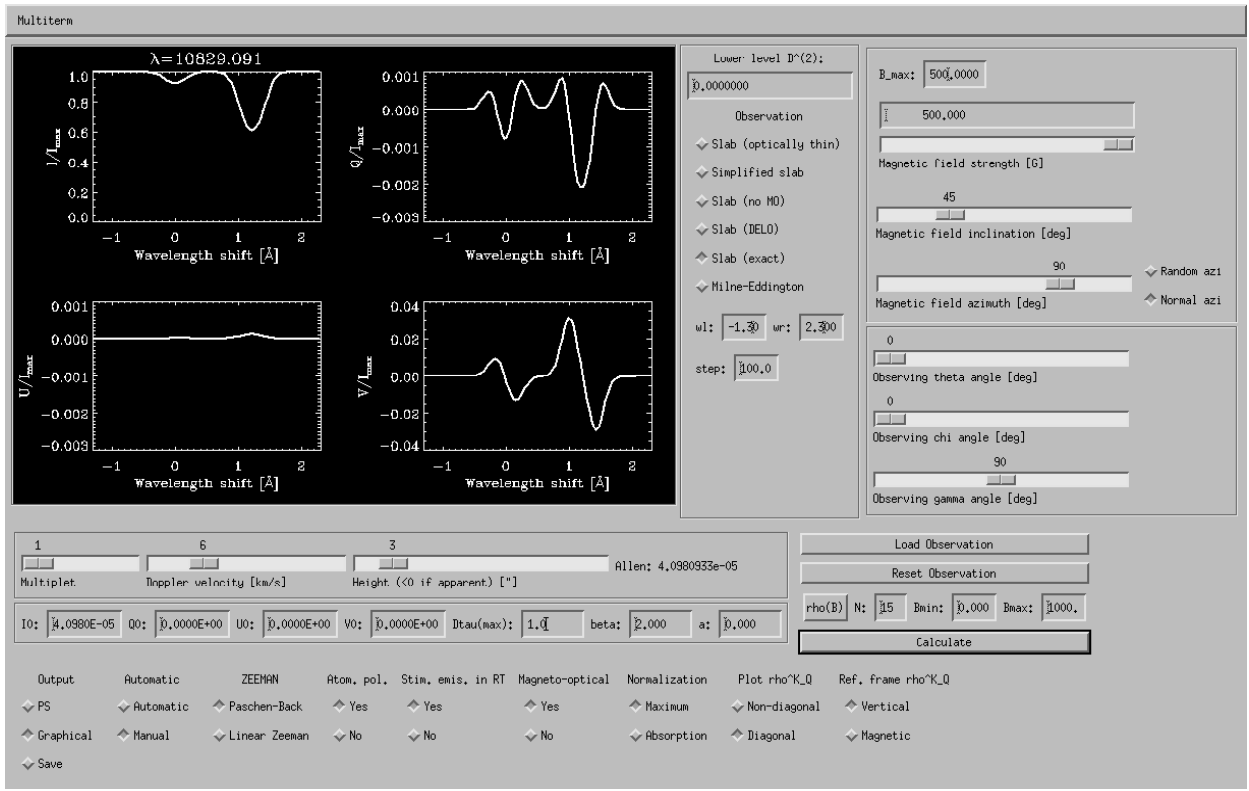}
\caption{Screen dump of the graphical front-end used for the synthesis.
\label{fig:front-end}}
\end{figure*}

\begin{figure}
\plottwo{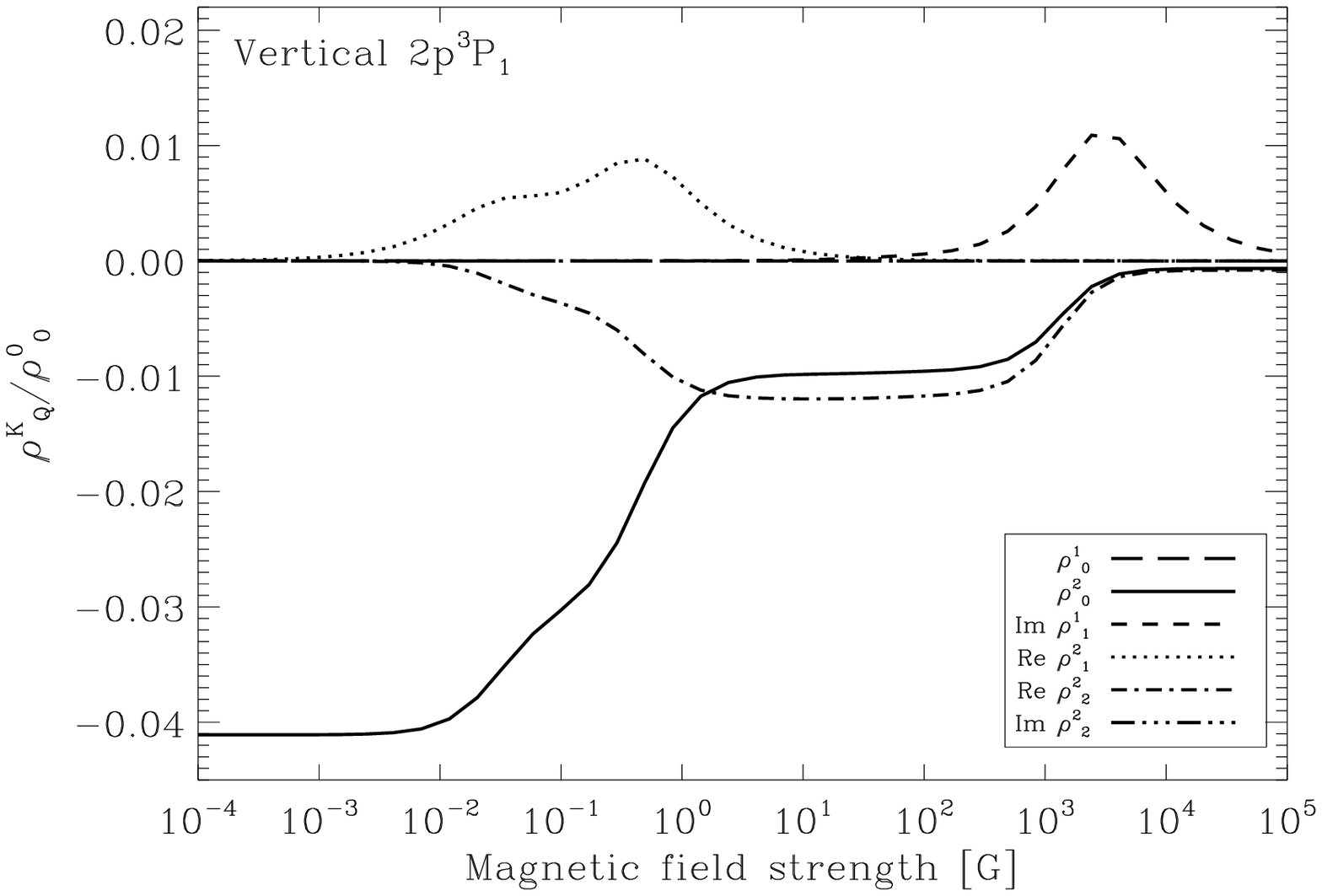}{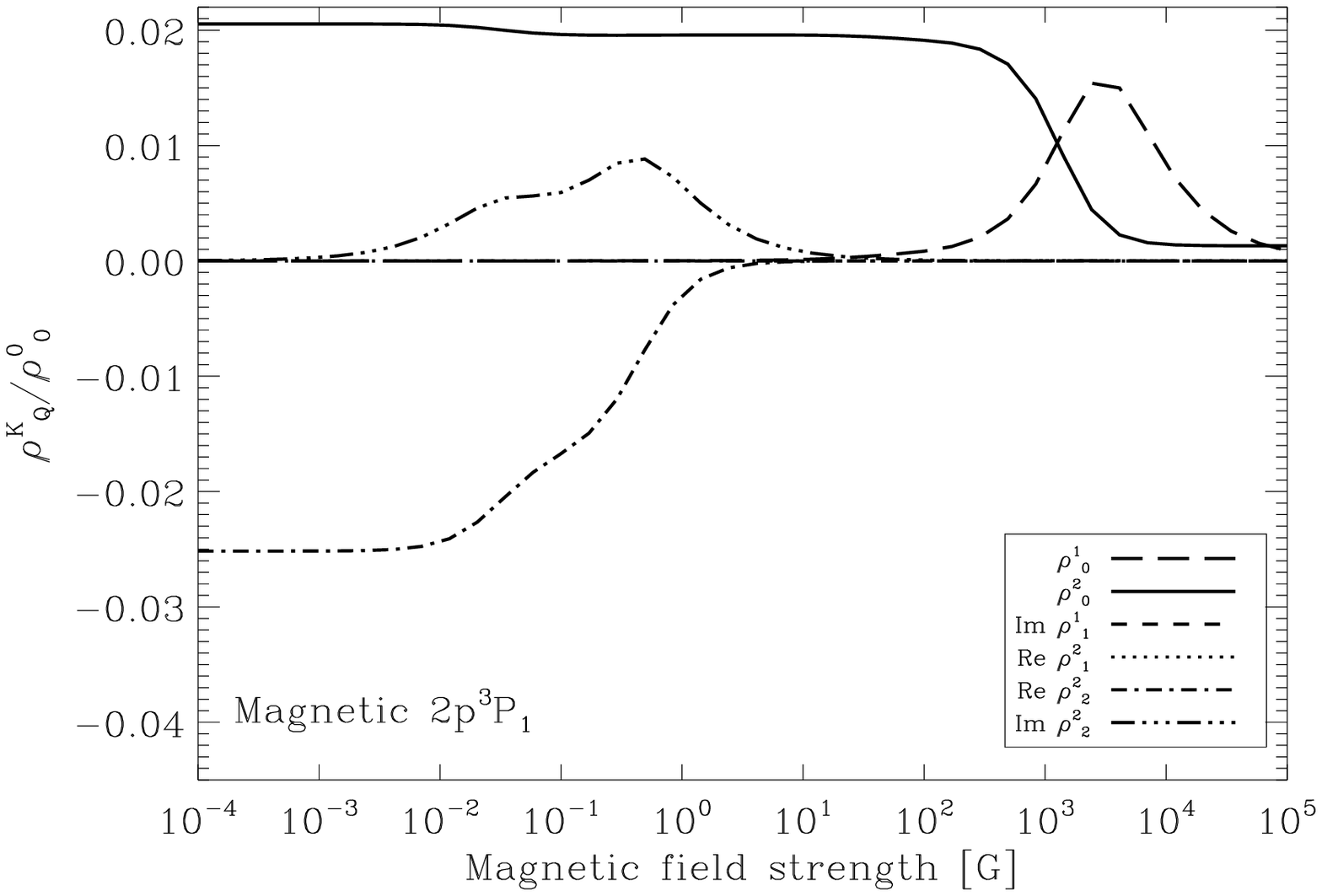}
\plottwo{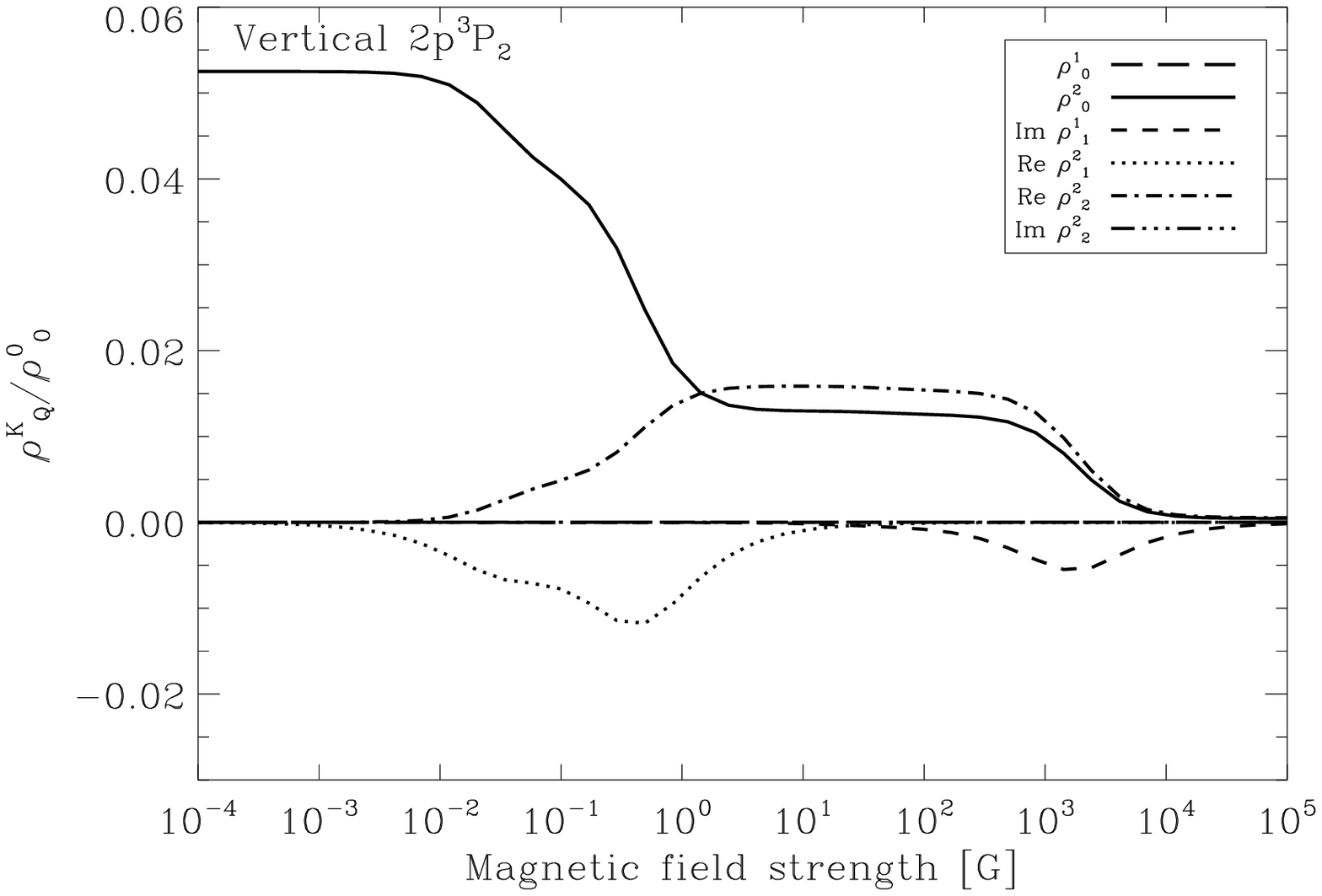}{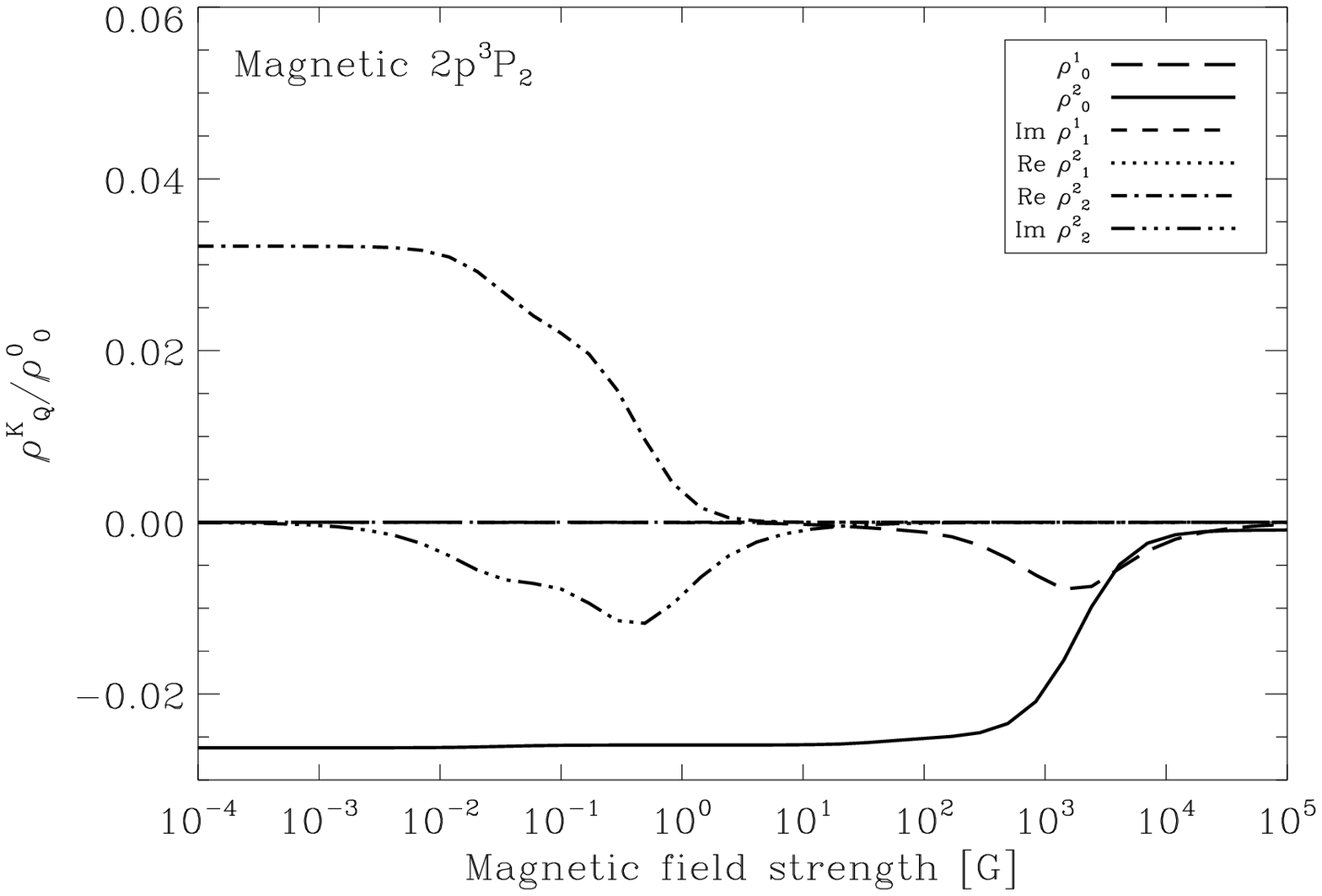}
\plottwo{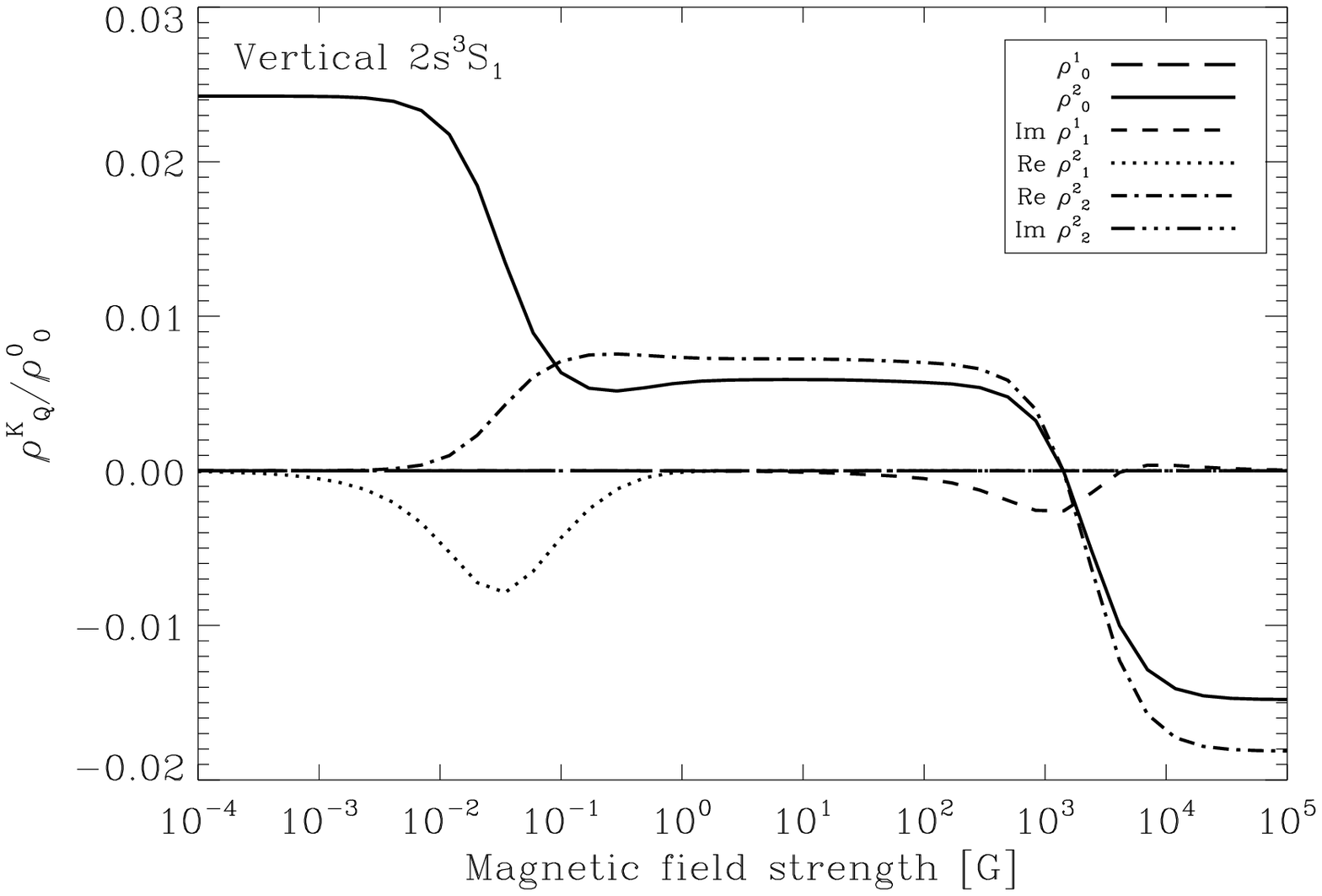}{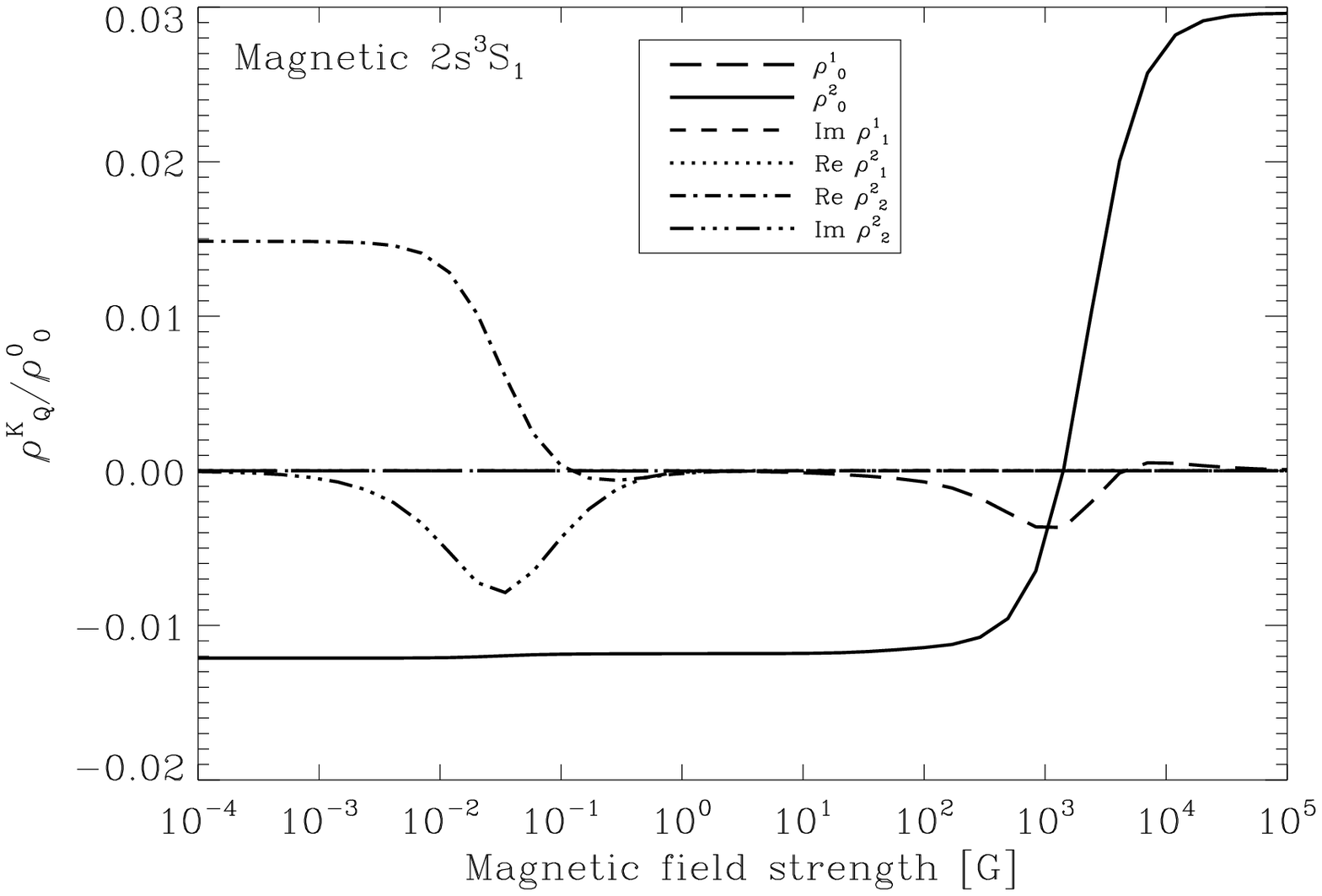}
\caption{Variation of the fractional population imbalances, $\rho^K_0(J)/\rho^0_0(J)$, and
of the non-zero quantum coherences between
magnetic sublevels pertaining to a given $J$-level, $\rho^K_Q(J)/\rho^0_0(J)$,
for different values of the strength of a horizontal magnetic field. We show
these quantities for the $J_u=1$ and $J_u=2$
levels of the upper term of the 10830 \AA\ transition (upper and middle panel,
respectively) and for the
$J_l=1$ level of the lower term of the 10830 \AA\ transition (lower panel). The
left panels
show the results for the ``vertical'' reference frame in which the quantization
axis is chosen
along the symmetry axis of the radiation field, while the right panels show the
results 
for the magnetic field reference frame in which the quantization
axis is chosen along the (horizontal) magnetic field vector. Note that, at
$B{\approx}0$ G, only population imbalances
are present in the vertical reference frame, while both population imbalances
and coherences are present in the 
magnetic reference frame. Note that in the magnetic field reference frame the
quantum coherences are zero and $\rho^2_0(J)$ is constant for
$10<B<100$ G.
\label{fig:coherences}}
\end{figure}

\begin{figure}
\plottwo{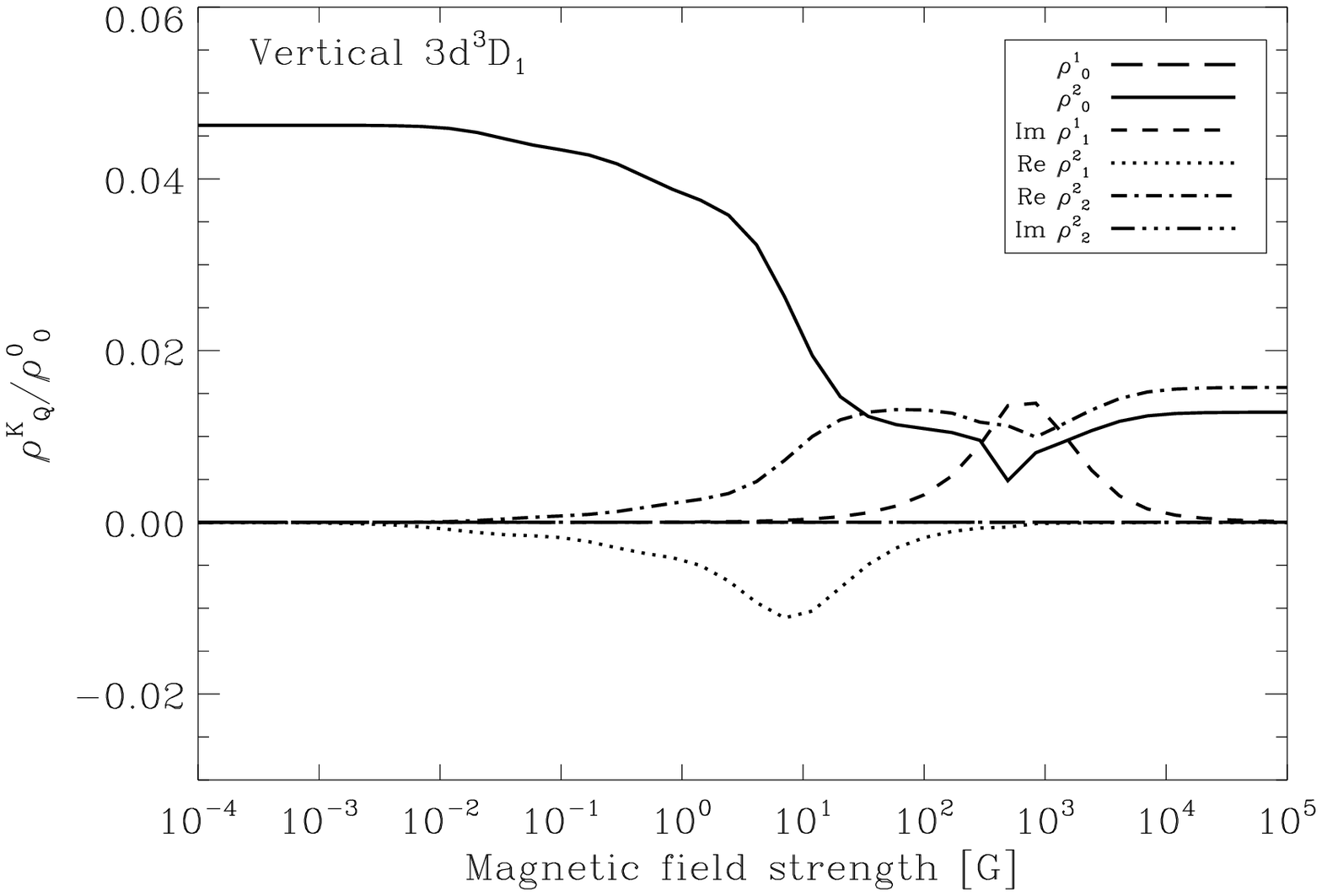}{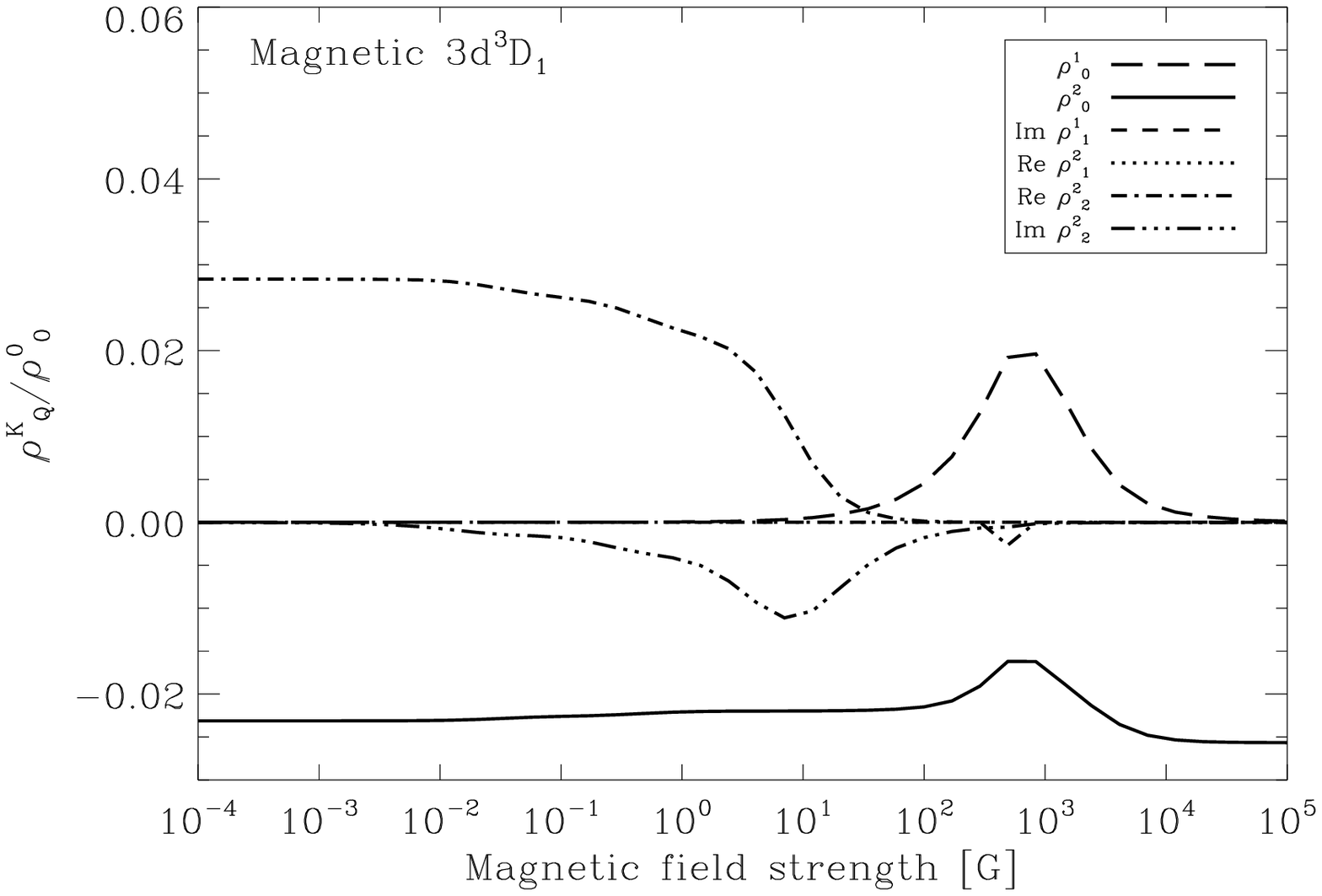}
\plottwo{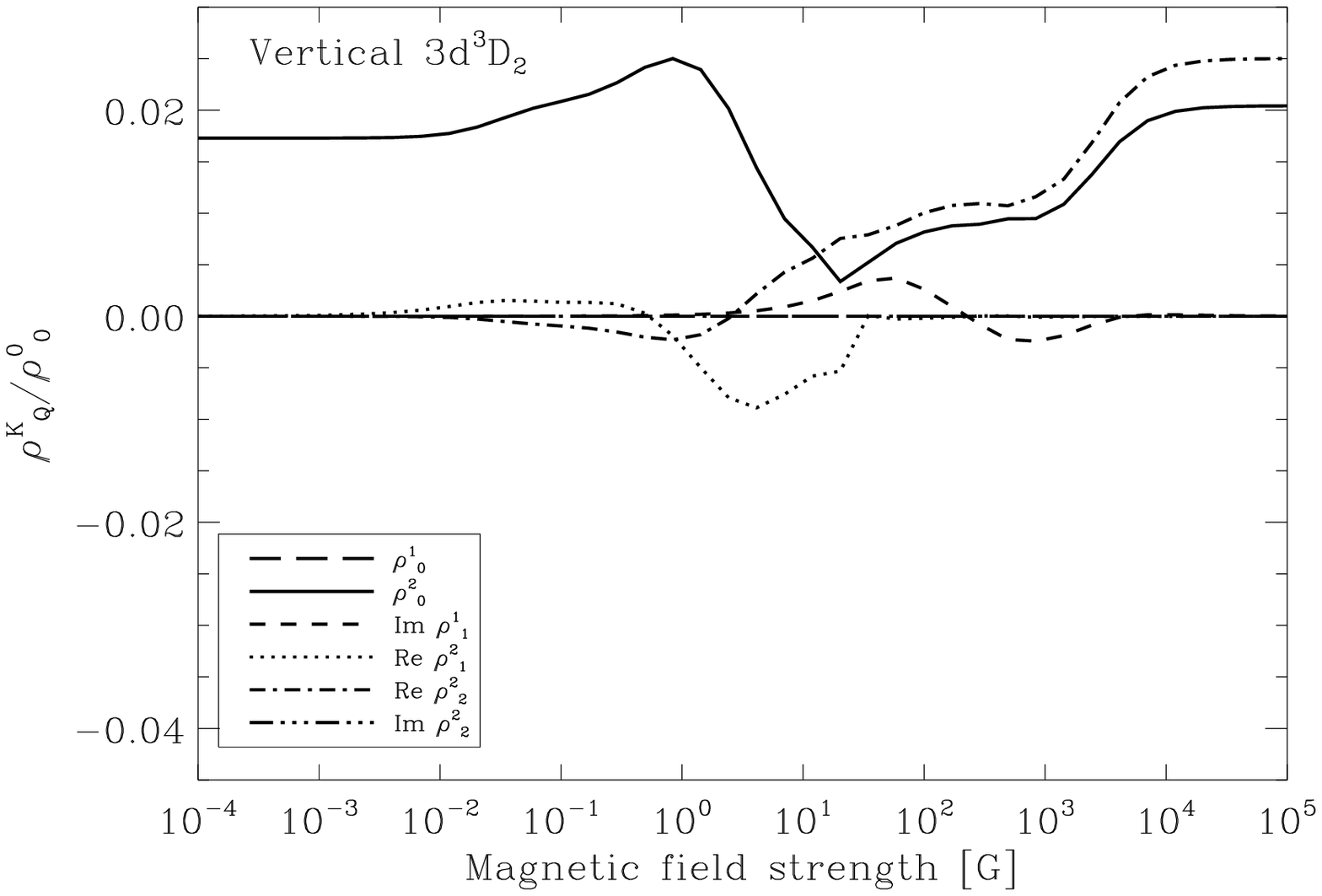}{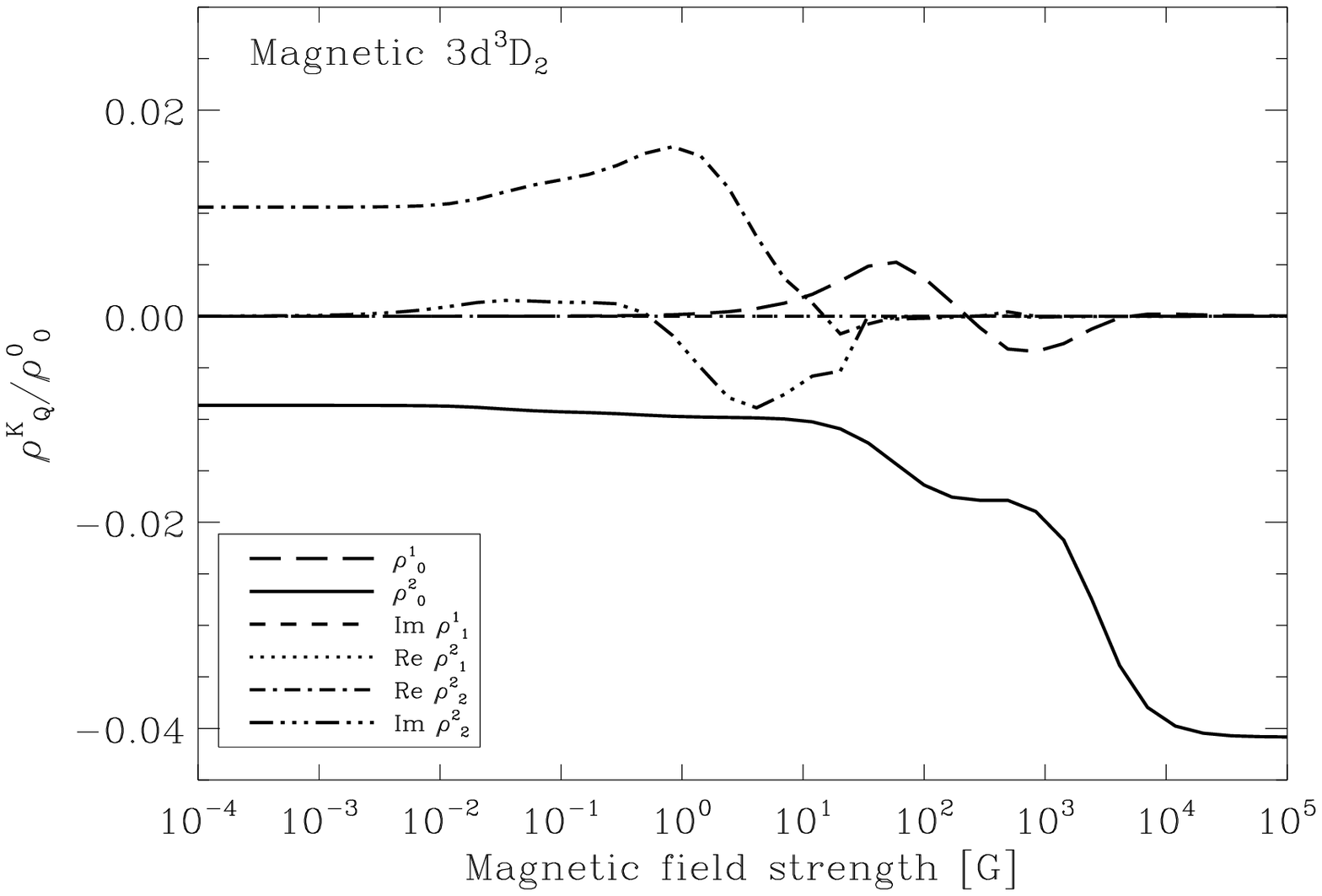}
\plottwo{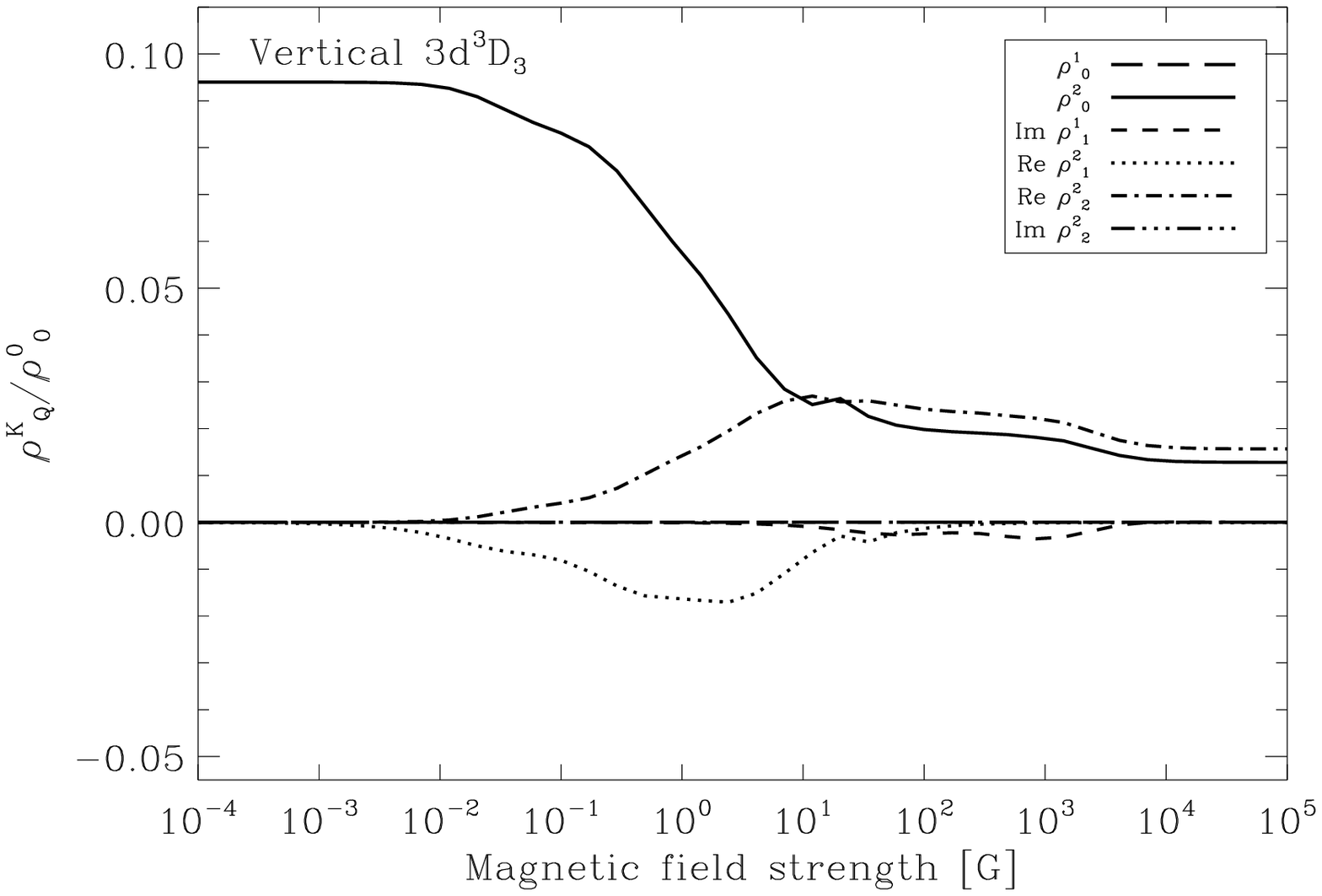}{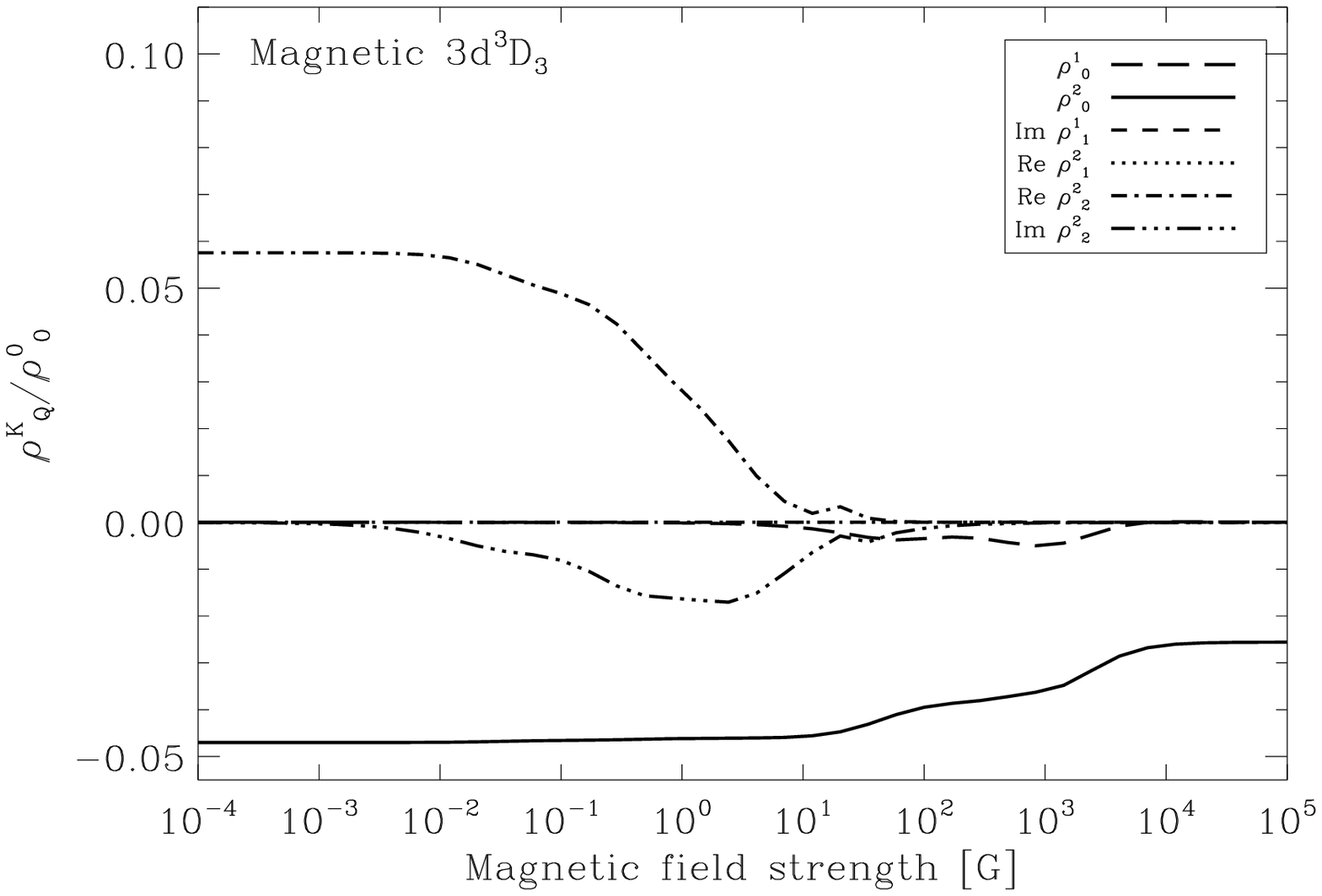}
\caption{Same as Fig. \ref{fig:coherences}, but for the $J$-levels of the upper term of the He {\sc i} D$_3$ multiplet.
\label{fig:coherencesD3}}
\end{figure}

\begin{figure*}
\plottwo{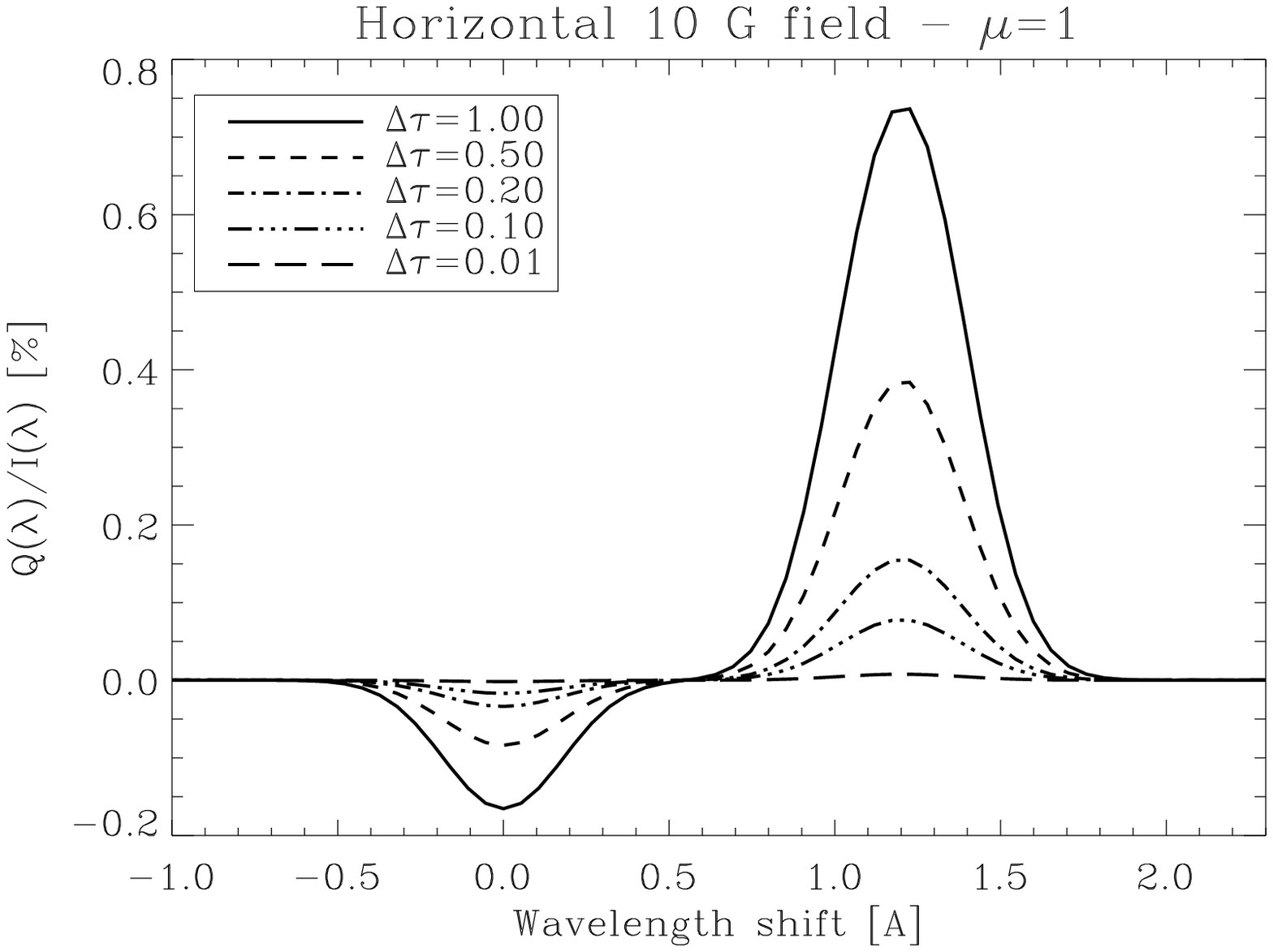}{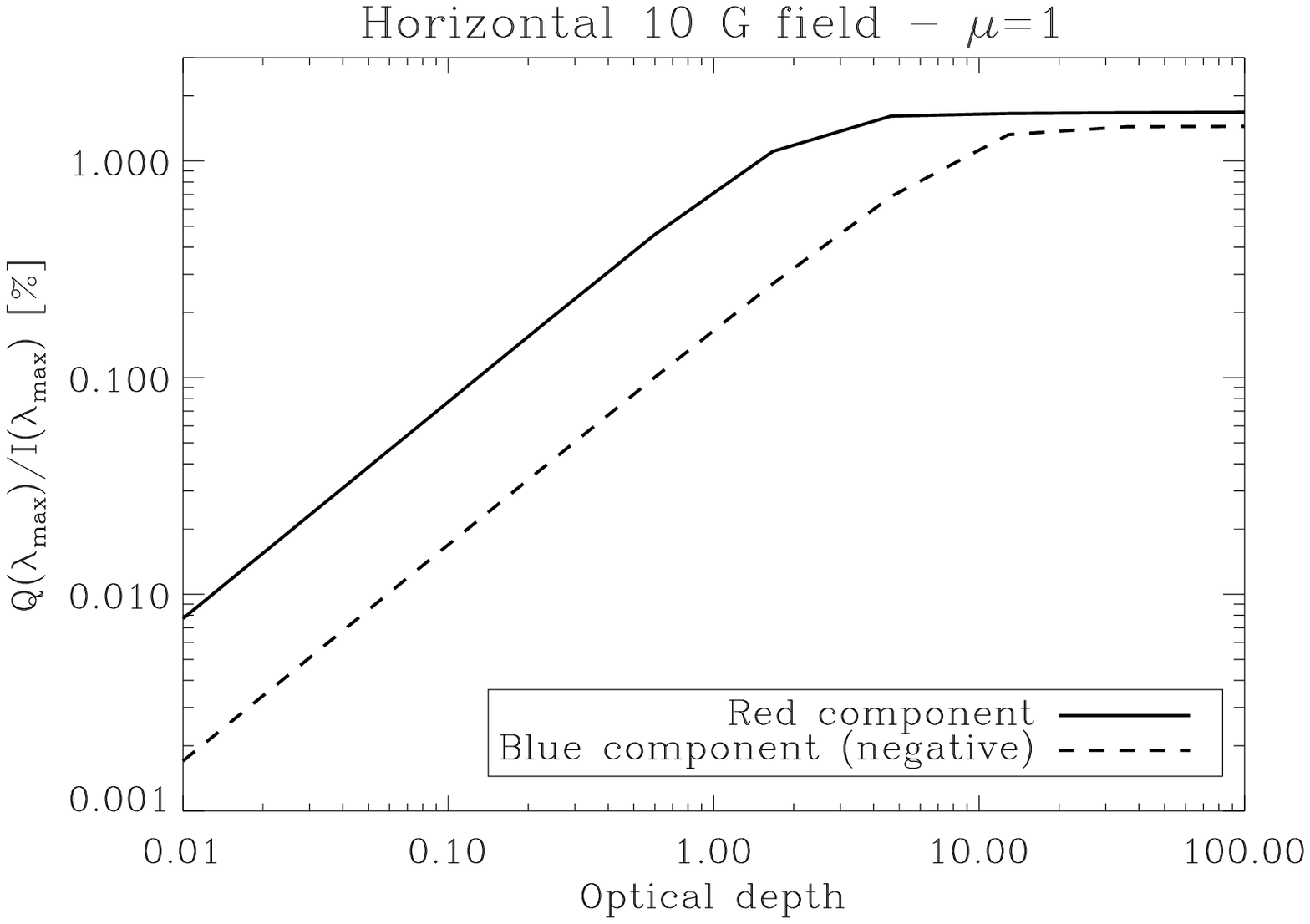}
\caption{Emergent fractional linear polarization at the solar disk
center in the lines of the \ion{He}{1} 10830 \AA\ multiplet, assuming that a
constant-property slab of helium atoms at a height of 3 arcseconds is permeated
by a horizontal magnetic field of 10 G. The various $Q/I$ profiles of the left
panel correspond to the slab's optical thickness indicated in the inset,
calculated at the wavelength of the red blended component. The right panel
indicates that the fractional polarization amplitudes of the red and blue
components increase exponentially with the slab's optical thickness, till  line saturation sets in.
These calculations have been carried out using the exact solution of the 
radiative transfer equation given by Eq. (\ref{eq:slab_peo}). The
positive reference direction for the definition of the Stokes $Q$ parameter lies along the horizontal magnetic field vector.
\label{fig:canopies}}
\end{figure*}

\begin{figure*}
\plottwo{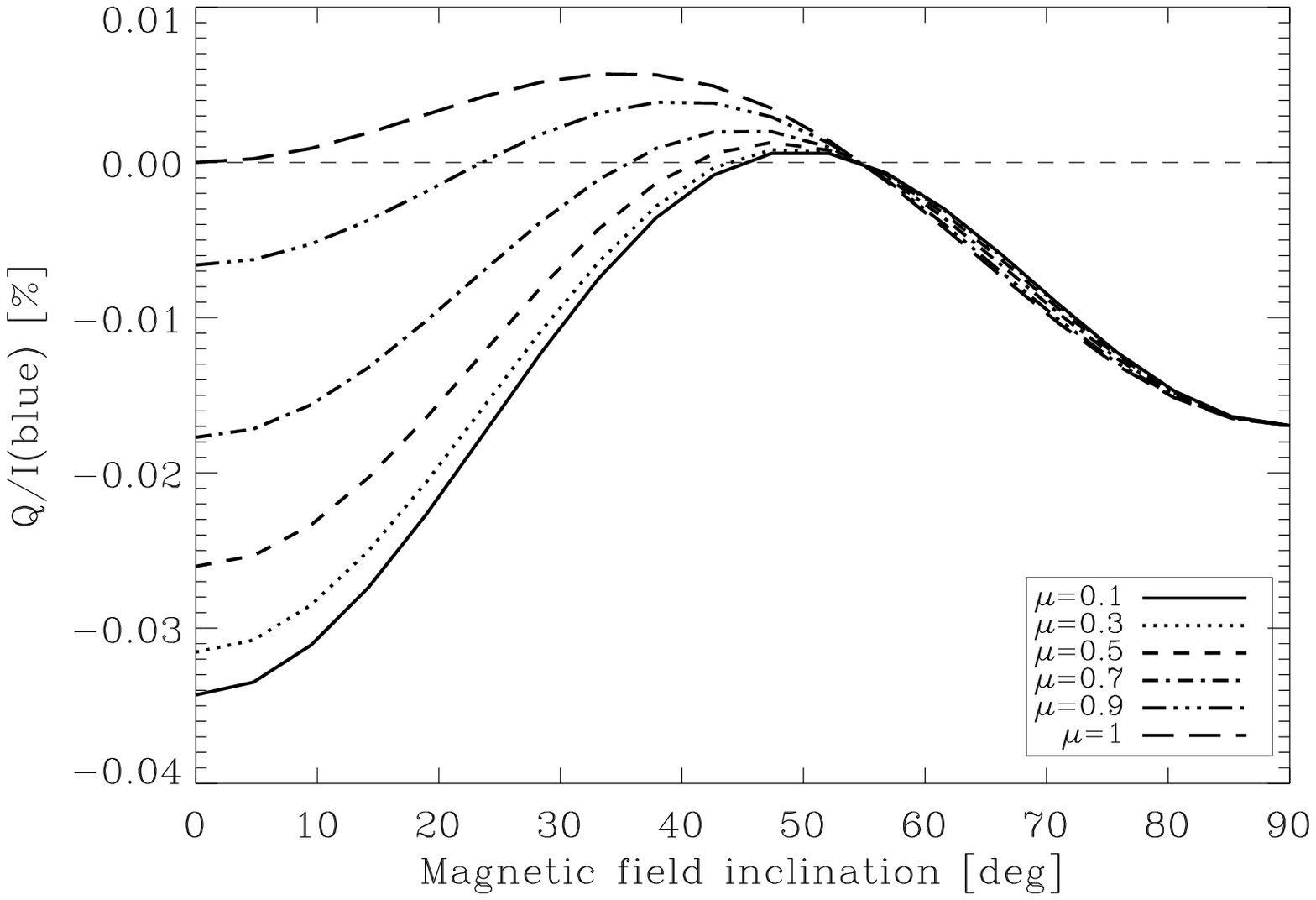}{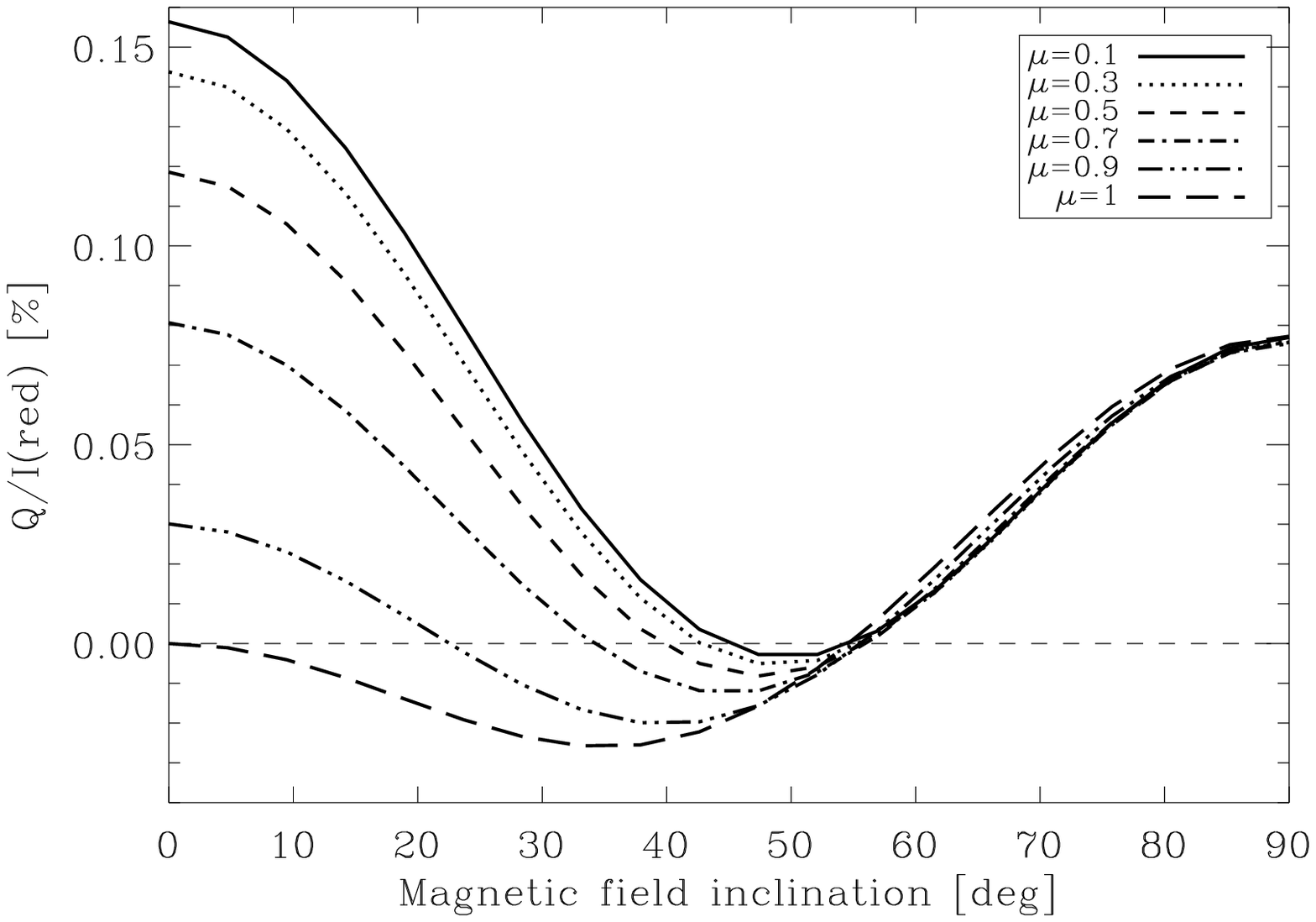}
\plottwo{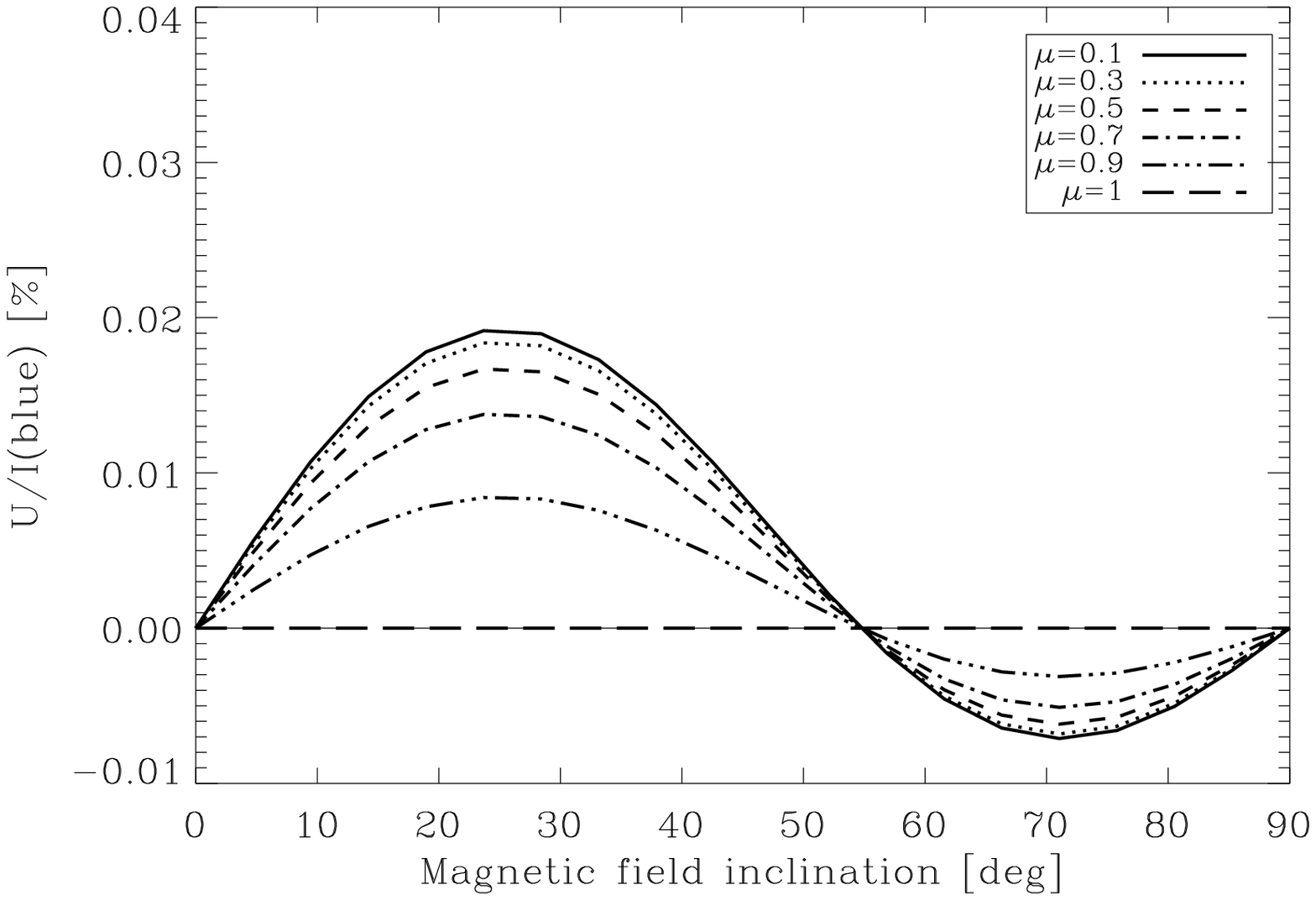}{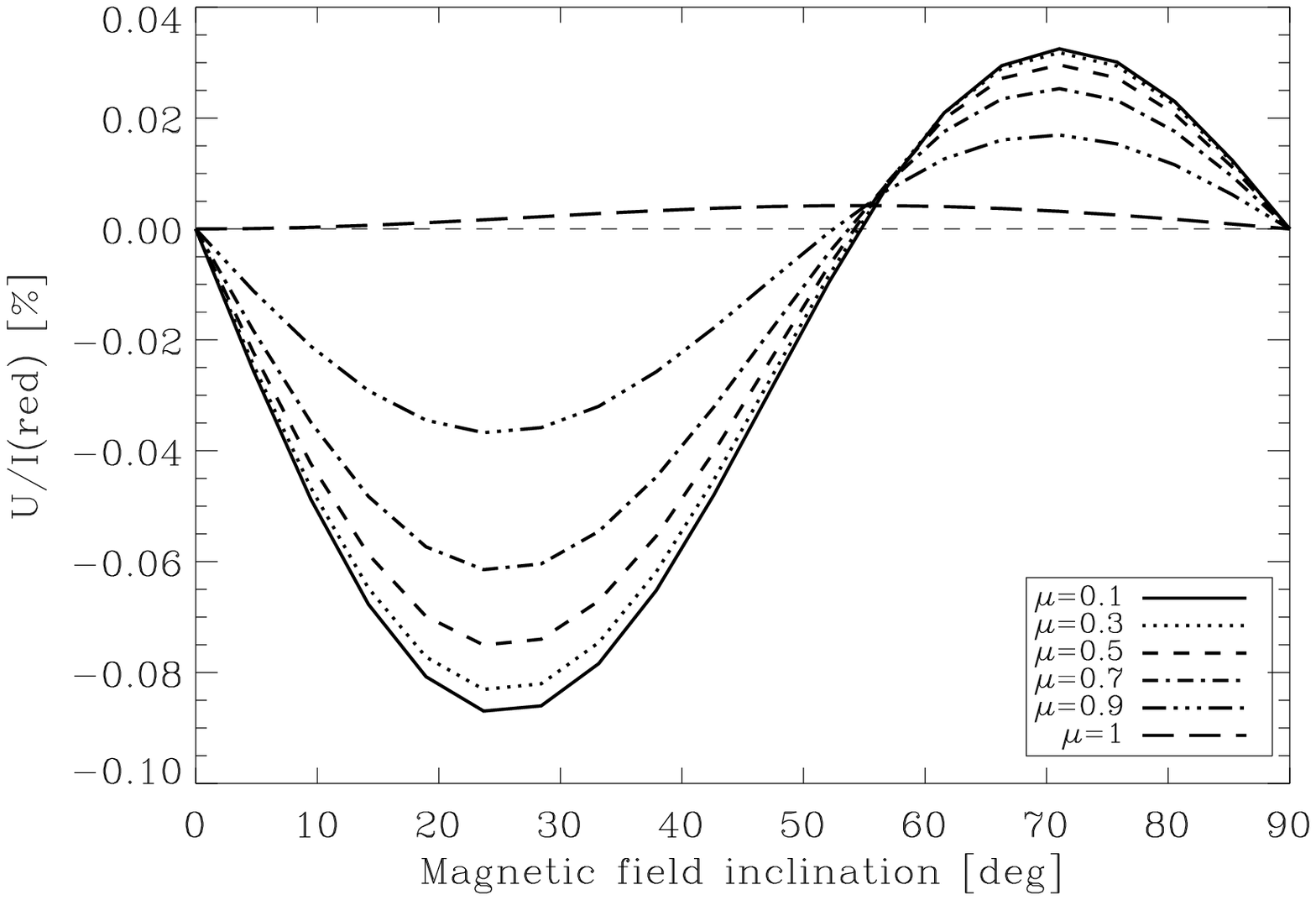}
\caption{Variation of the linear polarization at the line center of the blue
(left panel) and red (right panel)
components of the \ion{He}{1} 10830 \AA\ multiplet for different inclinations of
the magnetic
field vector with respect to the local vertical and for different observing
angles. The upper panels
show Stokes $Q$ while the lower panels show Stokes $U$, normalized to the value
of Stokes $I$ at the 
line center of each component. The calculations have been obtained
assuming that a constant-property slab of helium atoms at a height of 3
arcseconds is permeated
by a magnetic field of 10 G with an inclination $\theta_B$ with respect to the
solar
local vertical direction. The slab's optical thickness at the wavelength of the
red blended component is 
$\Delta \tau_\mathrm{red}=0.1$. 
The positive reference direction for Stokes $Q$ is along the projection of the
magnetic field vector on the
solar surface, which makes an angle of $90^{\circ}$ with any of the considered
line-of-sights.
\label{fig:canopies_qi_peak}}
\end{figure*}

\begin{figure*}
\plotone{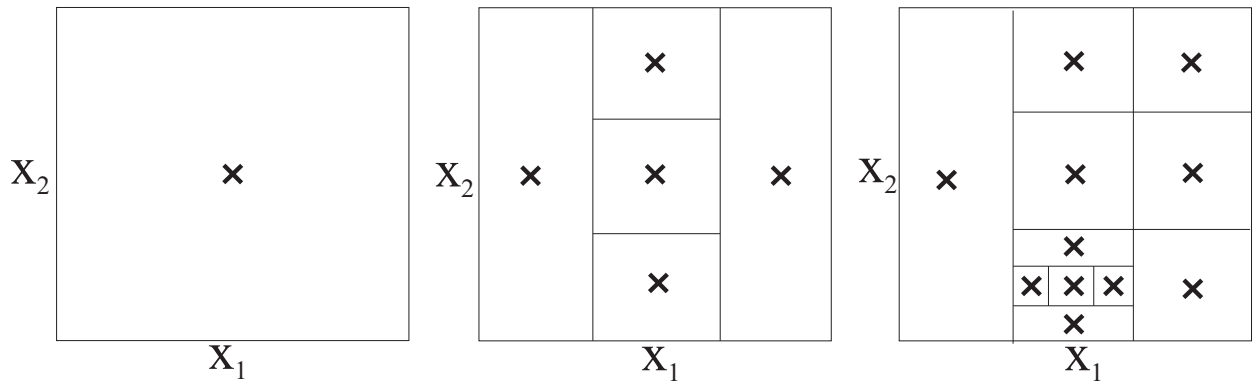}
\caption{This figure illustrates the philosophy of the DIRECT method for
searching 
the region where the global minimum is 
located. In this case, we present an illustrative example in two dimensions.
After the evaluation of 
the merit function at some selected points inside each region, 
the DIRECT algorithm decides, using the Lipschitz condition, which rectangles
should
be further subdivided in case they are potentially optimal (optimal means that a
low value of the merit function has
been found or that the rectangles are large and a finer sampling has to be
carried out). This method rapidly 
finds the region where the minimum is located.
\label{fig:direct_method}}
\end{figure*}

\begin{figure}
\plotone{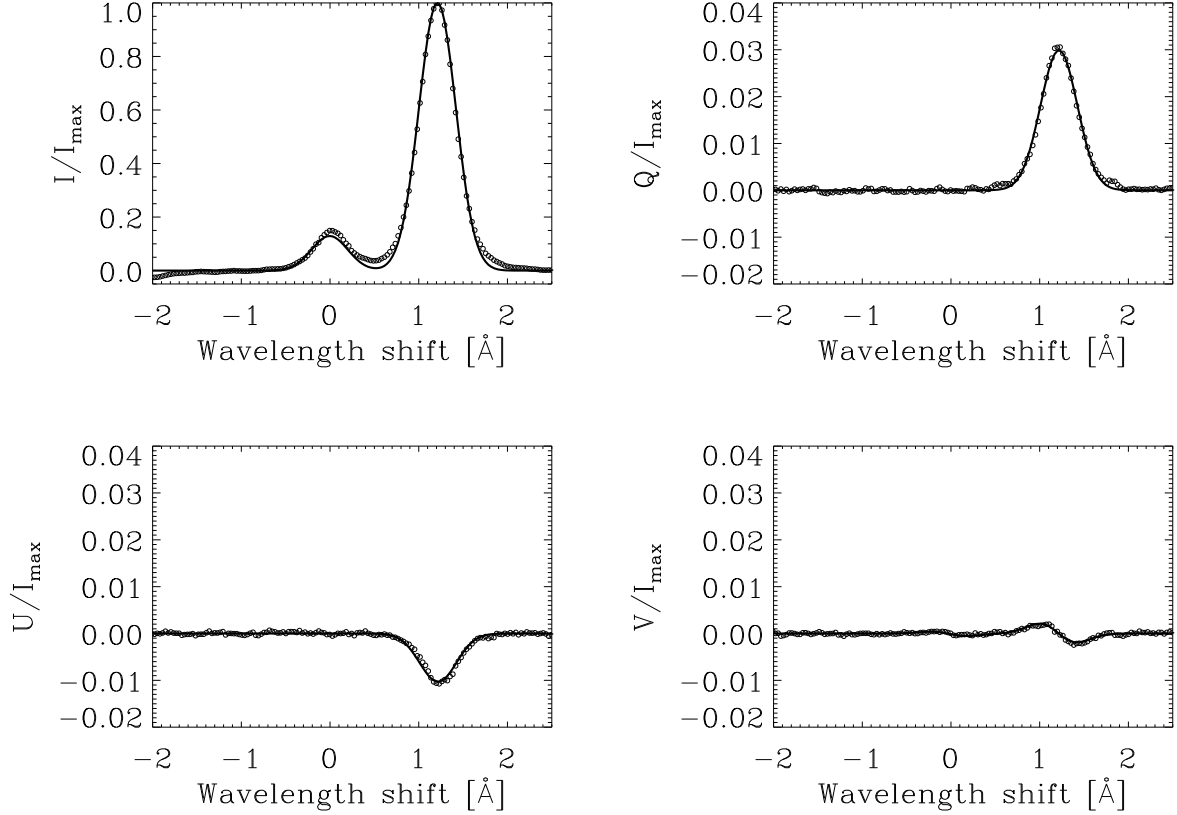}
\caption{The solar prominence case. Example of the best theoretical fit obtained
with the inversion code to some of the 
observed Stokes profiles presented in \cite{merenda06}.
The emergent Stokes profiles have been obtained by assuming an optically thin 
plasma. The inferred magnetic field vector is 
$B=26.8$ G, $\theta_B=25.5^\circ$ and $\chi_B=161^\circ$ and the 
inferred thermal velocity  
$v_\mathrm{th}=7.97$ km s$^{-1}$. All these values are in good agreement with 
the results obtained by \cite{merenda06}.
The positive reference direction for Stokes $Q$ is the parallel to the solar
limb.
\label{fig:prominence}}
\end{figure}

\begin{figure}
\plotone{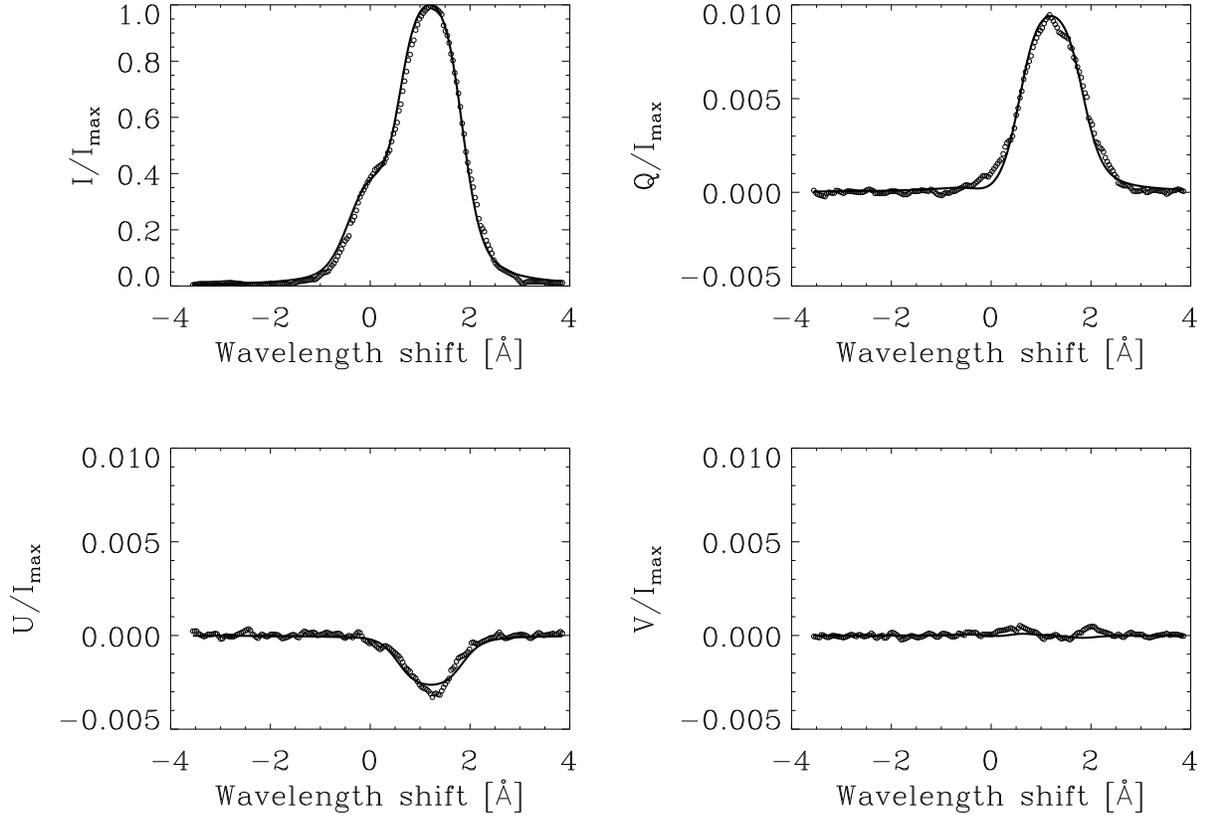}
\caption{The case of solar chromospheric spicules. Example of the best
theoretical fit obtained with the inversion code to the 
observed profiles presented in \cite{trujillo_merenda05}.
The emergent Stokes profiles have been obtained by taking into account 
radiative transfer effects in a constant-property slab model. The inferred magnetic field vector
is 
$B=2.6$ G, $\theta_B=37^\circ$ and $\chi_B=35^\circ$, although an equally good
fit is obtained for other combinations like $B=10$ G, $\theta_B=37^\circ$ and $\chi_B=172^\circ$ 
(see the text for more information). Furthermore, the 
inferred thermal velocity is 
$v_\mathrm{th}=13.9$ km s$^{-1}$, the slab's optical thickness in the red
blended component is $\Delta \tau_{\rm red}=2.54$
and the damping parameter is $a=0.22$. The positive reference direction for
Stokes $Q$ is the parallel to the solar limb.
\label{fig:spicules}}
\end{figure}

\begin{figure*}
\plottwo{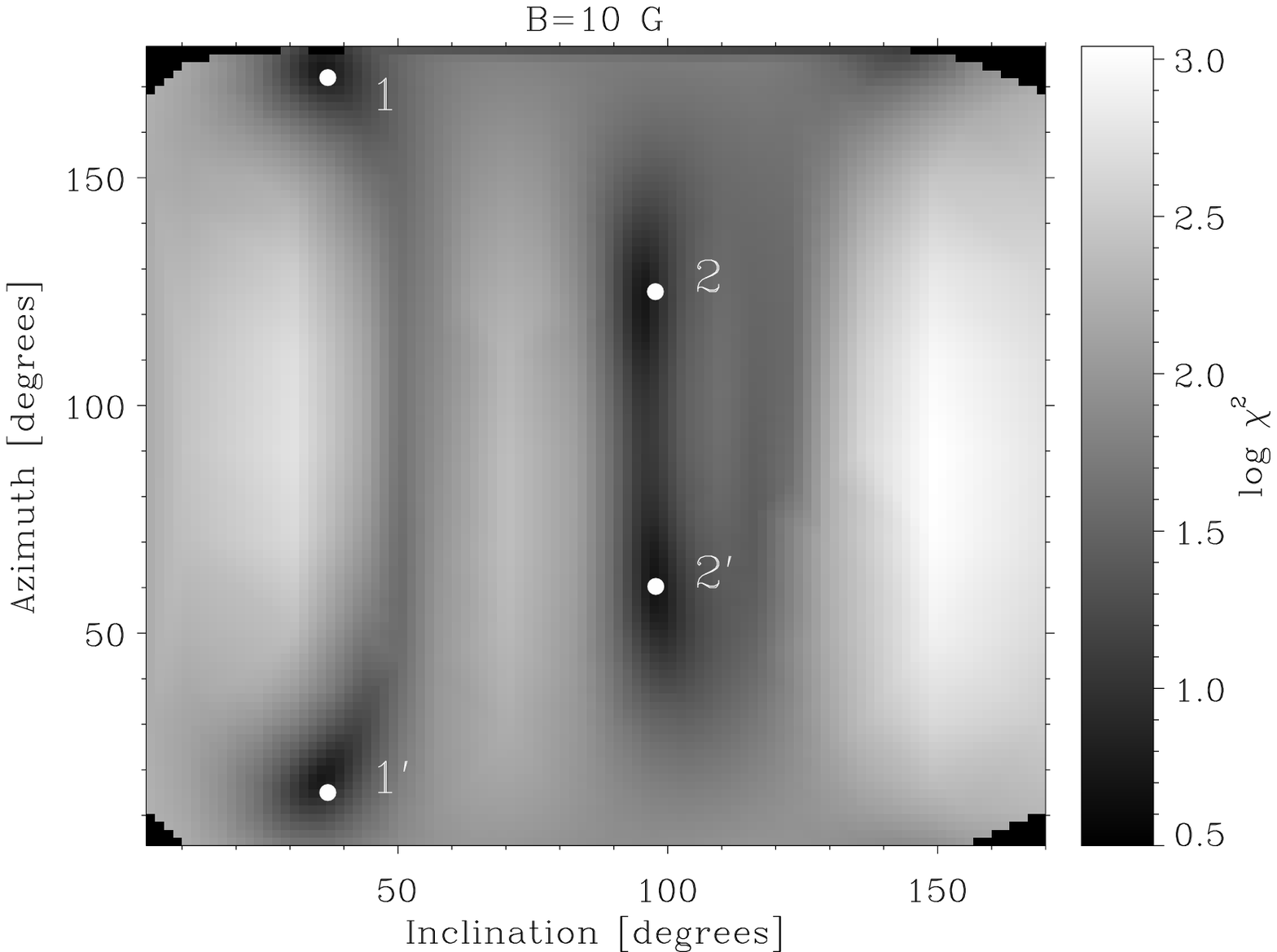}{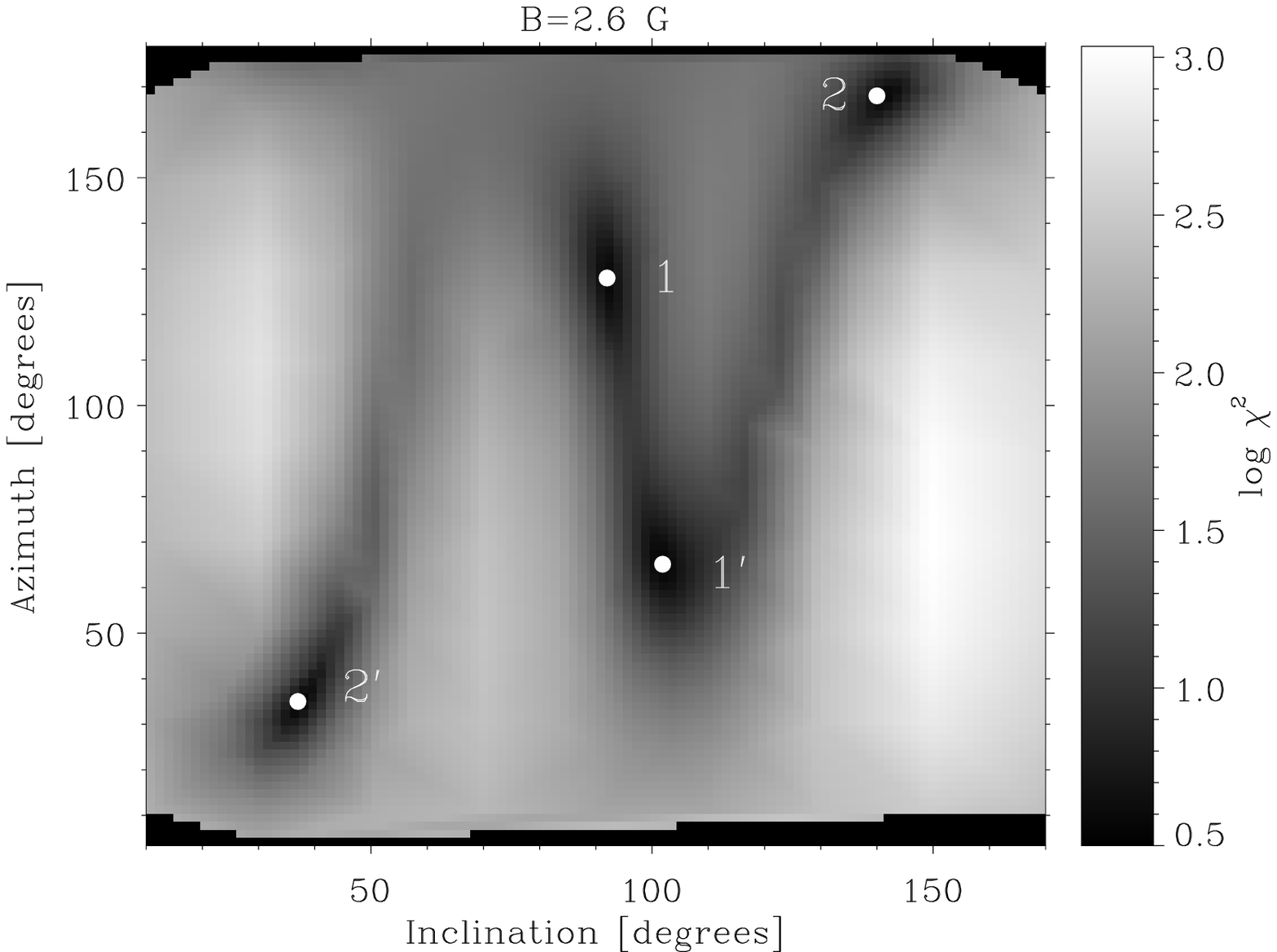}
\caption{Values of the $\chi^2$-function for the possible fits of the Stokes
profiles of the spicular material
observed by \cite{trujillo_merenda05}, after considering several combinations of
the inclination and azimuth of
the magnetic field vector. We fixed the thermal velocity, the optical depth of
the
slab and the damping constant to the following values: $v_\mathrm{th}=13.9$
km$s^{-1}$, $\Delta \tau_{\rm red}=2.54$
and $a=0.22$. The magnetic field strength is 10 G in the left panel while it is
2.6 G in the right panel. Note 
the presence of several local 
minima. The white dots indicate the position of the global minimum of the
$\chi^2$-function. The solutions $1$ and $2$ correspond to the Van-Vleck ambiguity. The same
happens for the solutions $1'$ and $2'$. 
Additionally, the solutions $1$ and $1'$ are equivalent for the code even if they do produce a sign change in the
Stokes $V$ signal because the observed Stokes $V$ profile is at the level of noise. The same happens for the
solutions $2$ and $2'$.
\label{fig:spicules_chi2}}
\end{figure*}

\begin{figure*}
\plotone{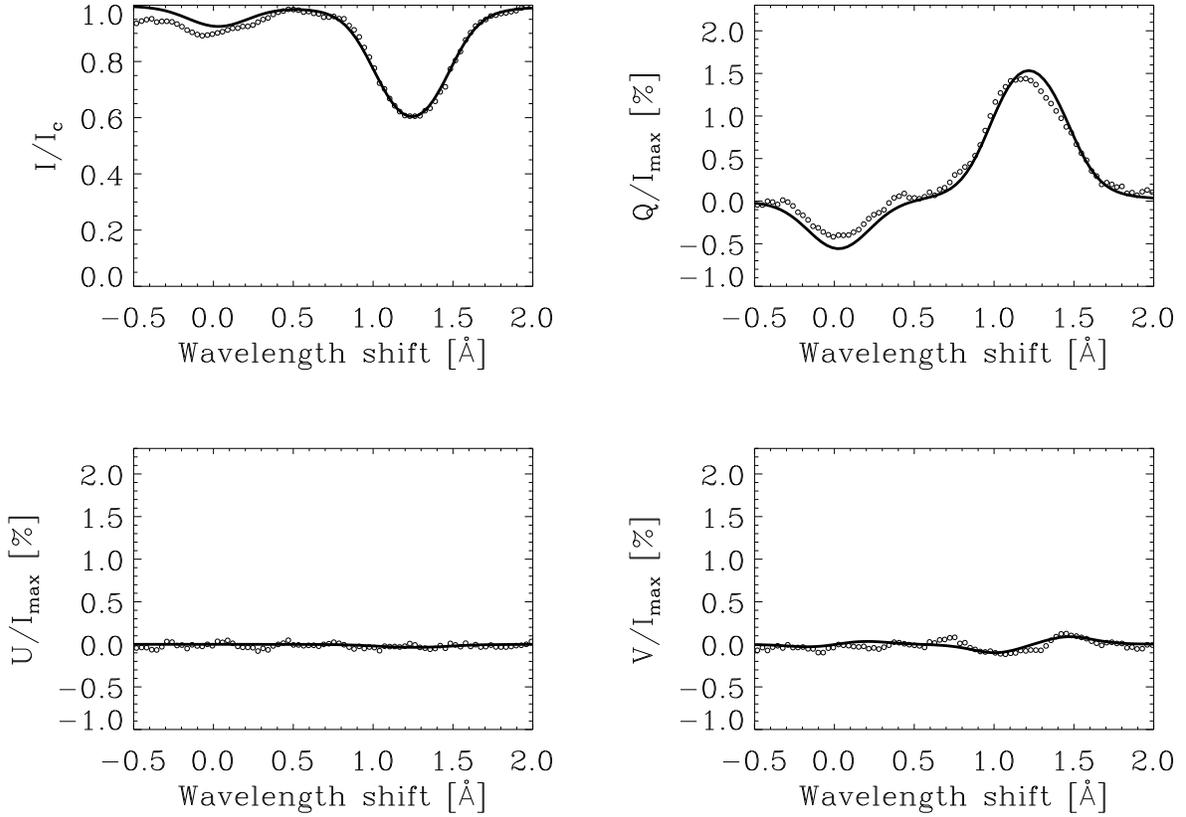}
\caption{The case of a solar coronal filament. Example of the Stokes profiles of
the \ion{He}{1} 10830 \AA\ 
multiplet observed in a coronal filament
at solar disk center. Our results confirm the conclusions of
\cite{trujillo_nature02}, since we obtain
$\Delta \tau=0.86$, $v_\mathrm{th}=6.6$ km s$^{-1}$, $a=0.19$, 
$B=18$ G and $\theta_B=105^\circ$.
The positive direction of Stokes $Q$ is parallel to the projection of the
magnetic field vector on the solar surface. 
\label{fig:filament_observation}}
\end{figure*}

Our computer program has been
developed with the aim of being computationally efficient and 
user-friendly. The relevant equations of the problem result from 
a general and robust theory, so that it is straightforward to
treat limiting cases and include or discard several 
physical effects in a very transparent way. It is appropriate for 
its application to a wide variety of problems, from simple Zeeman-dominated Stokes profiles to more complex situations in which 
the influence of atomic level polarization cannot be neglected.
The code is written in FORTRAN 90, and incorporates a user-friendly
front-end based on IDL\footnote{\texttt{http://www.ittvis.com/idl}} which
facilitates the execution and analysis of the synthesis and/or inversion calculations (see Fig. \ref{fig:front-end}). 

\begin{figure*}
\plotone{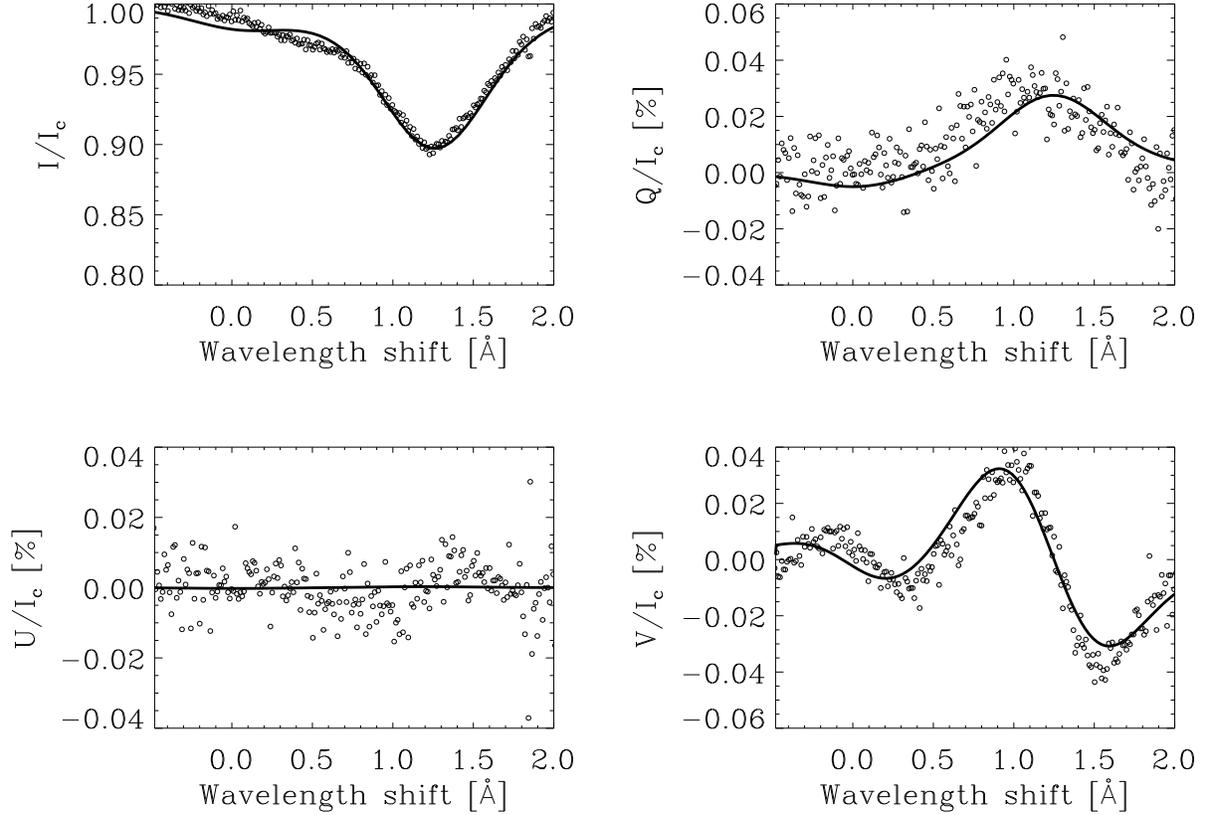}
\caption{The quiet chromosphere case. Example of the Stokes profiles of the
\ion{He}{1} 10830 \AA\ multiplet 
observed in a disk center ($\mu=0.98$) internetwork 
region surrounded by an enhanced network region of almost circular shape. The
observed Stokes profiles selected
for the inversion correspond to a temporal and spatial average within the
internetwork region. The solid 
line presents a
fit to the observations which corresponds to a magnetic field vector with $B=35$
G, $\theta_B=21^\circ$ 
and $\chi_B=0^\circ$. The inferred optical depth in the red blended component is
$\Delta {\tau}_{\rm red} \approx 0.2$, while $v_\mathrm{th}=9.2$ km s$^{-1}$ and $a=0.62$.
An equally good fit is obtained for other field configurations like $B=47$ G,
$\theta_B=47^\circ$ and 
$\chi_B=0^\circ$. 
The fact that with a ${\theta_B}<54.74^{\circ}$ the Stokes $Q$ signal of the 
red component is positive (after our rotation of the reference system to minimize Stokes 
$U$) indicates that the reference direction for Stokes $Q$ lies along the direction perpendicular to the 
projection of the magnetic field vector on the solar disk (see the $\mu=1$ case of Fig. 9).
\label{fig:canopy_observation}}
\end{figure*}

\begin{figure*}
\plotone{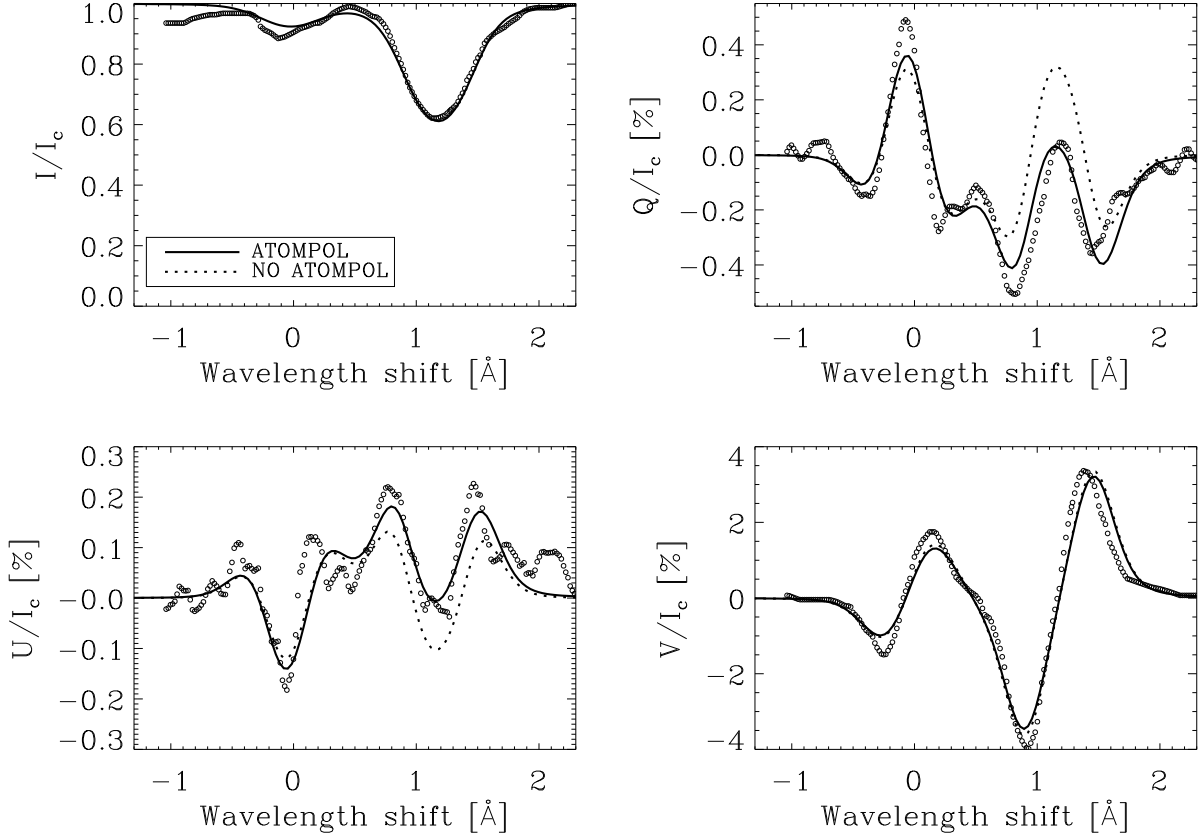}
\caption{The case of an emerging flux region. Example of the Stokes profiles of
the \ion{He}{1} 
10830 \AA\ multiplet in an emerging flux region located at $\mu=0.8$. The
circles present the observed Stokes profiles obtained from Fig. 2 of
\cite{Lagg04}, but after rotating the reference system by $-56^\circ$
to have the reference direction 
for the Stokes $Q$ profile along the direction perpendicular to the straight line joining the 
disk center with the observed region. For this reason the inversion has been carried out 
using $\chi=0^{\circ}$ and $\gamma=90^{\circ}$ (see Fig. 1). The solid line presents 
the best fit obtained with our inversion code, taking into account the effect of atomic 
polarization on the emergent Stokes profiles, in addition to the Zeeman effect treated 
within the framework of the Paschen-Back effect theory. This full solution corresponds 
to a magnetic field vector with $B=1073$ G, $\theta_B=86^\circ$ and $\chi_B=170.7^\circ$, 
while $\Delta \tau_\mathrm{red} = 1.07$, $v_\mathrm{th}=7.24$ km s$^{-1}$ and $a=0.25$. 
The dotted line is the best fit obtained when neglecting atomic polarization. These 
results confirm the conclusion by \citet{trujillo_asensio07} that the presence of 
atomic polarization in the levels of the He {\sc i} 10830 \AA\ multiplet produces 
observable signatures on the emergent Stokes profiles even for fields as large as 
1000 G. We point out that there is a typing error in the caption of figure 3 of 
\cite{trujillo_asensio07}, since the theoretical Stokes profiles in that figure 
were obtained for a LOS with $\mu=\cos \theta=0.8$ (i.e., not for the $\mu=1$ 
forward scattering case) and with $\chi=\gamma=0^{\circ}$ (see Fig. 1).
\label{fig:lagg_emerging}}
\end{figure*}

\begin{figure*}
\plotone{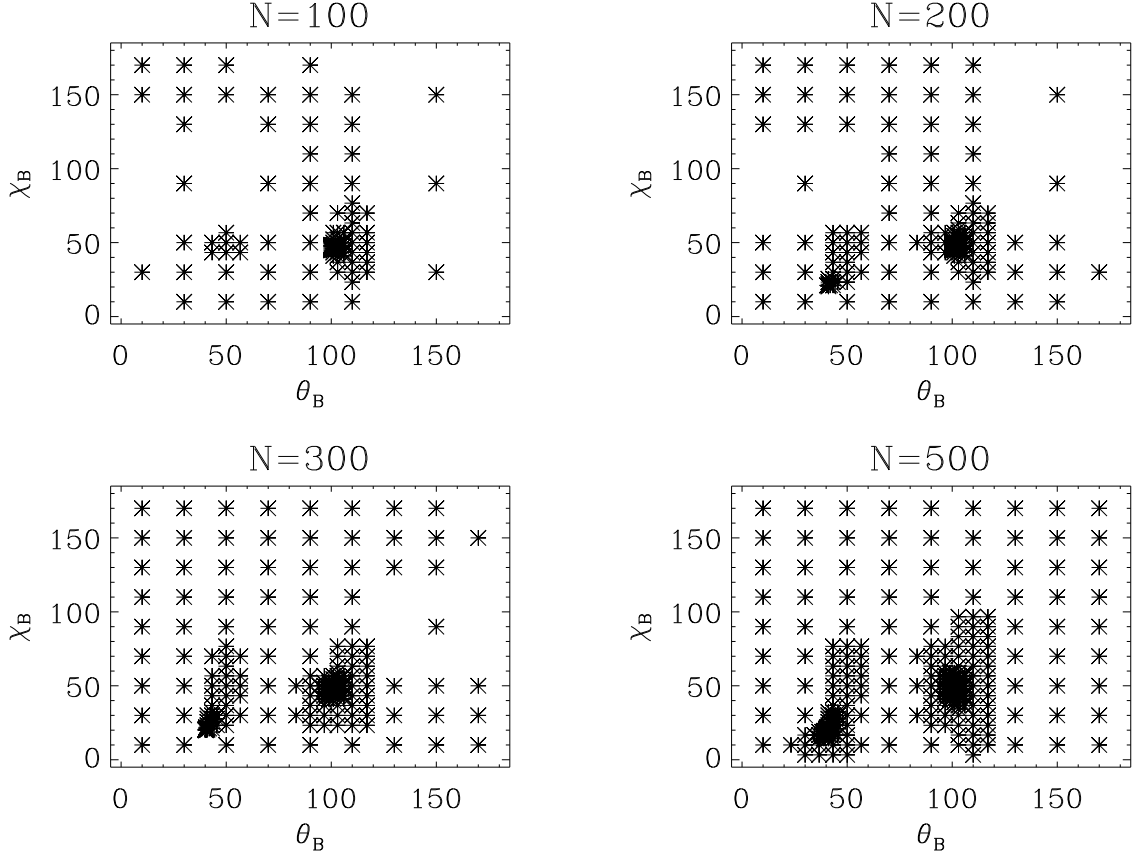}
\caption{Value of $\theta_B$ and $\chi_B$ at which the merit function is
evaluated 
for detecting possible ambiguities using the
DIRECT method. The number $N$ of 
function evaluations is shown on the top of each panel. The two global minima 
seen in the figures result from the Van Vleck ambiguity,
showing two different magnetic field vectors that produce the same emergent 
Stokes profiles. This figure illustrates the power of the
DIRECT method, which allows to recover the two values of the magnetic field
vector 
in this ambiguous case with less than 200
function evaluations. We have shown only half of the space of parameters in
order
to avoid the 180$^\circ$ ambiguity.
\label{fig:vanvleck_ambiguity}}
\end{figure*}

\begin{figure*}
\plottwo{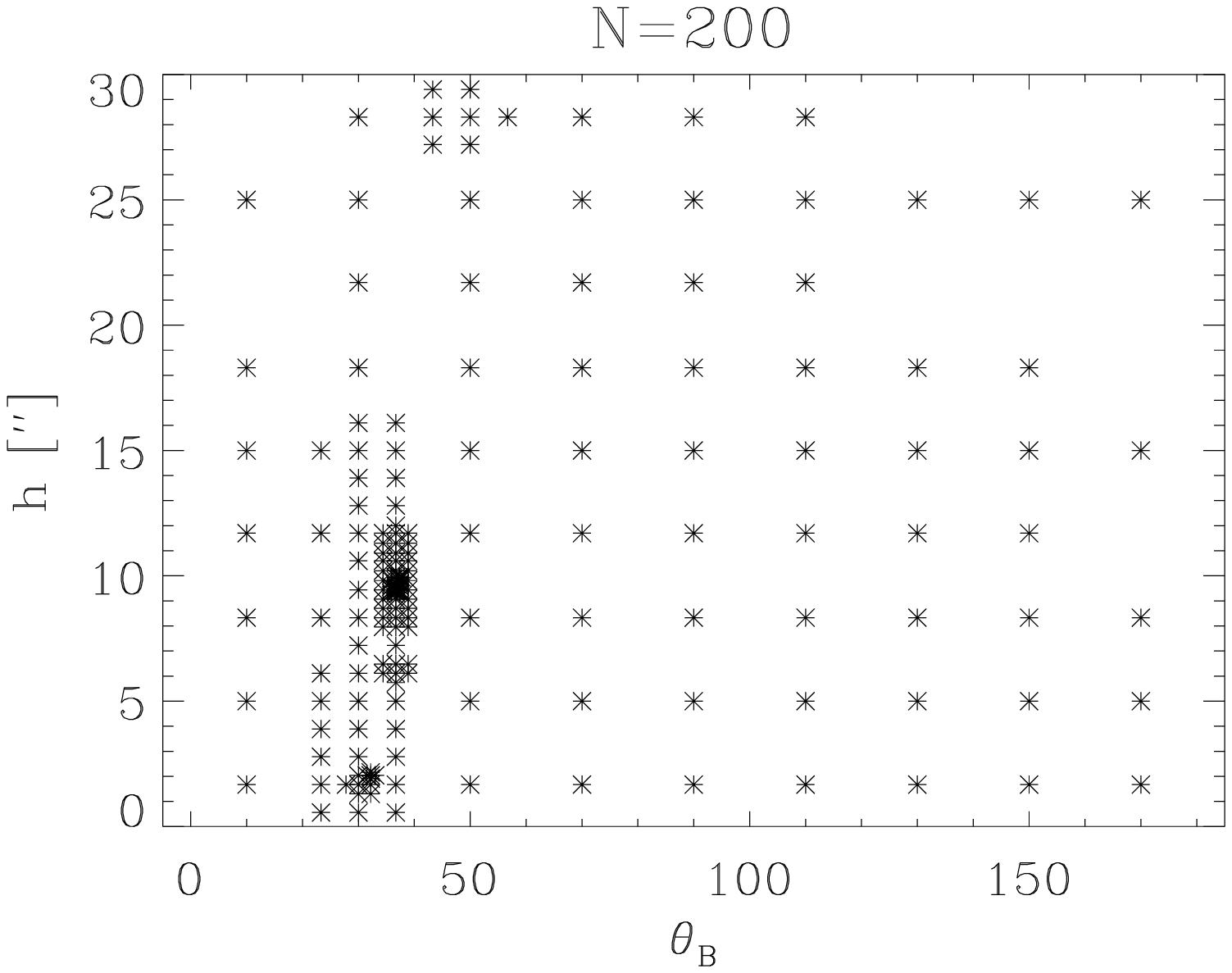}{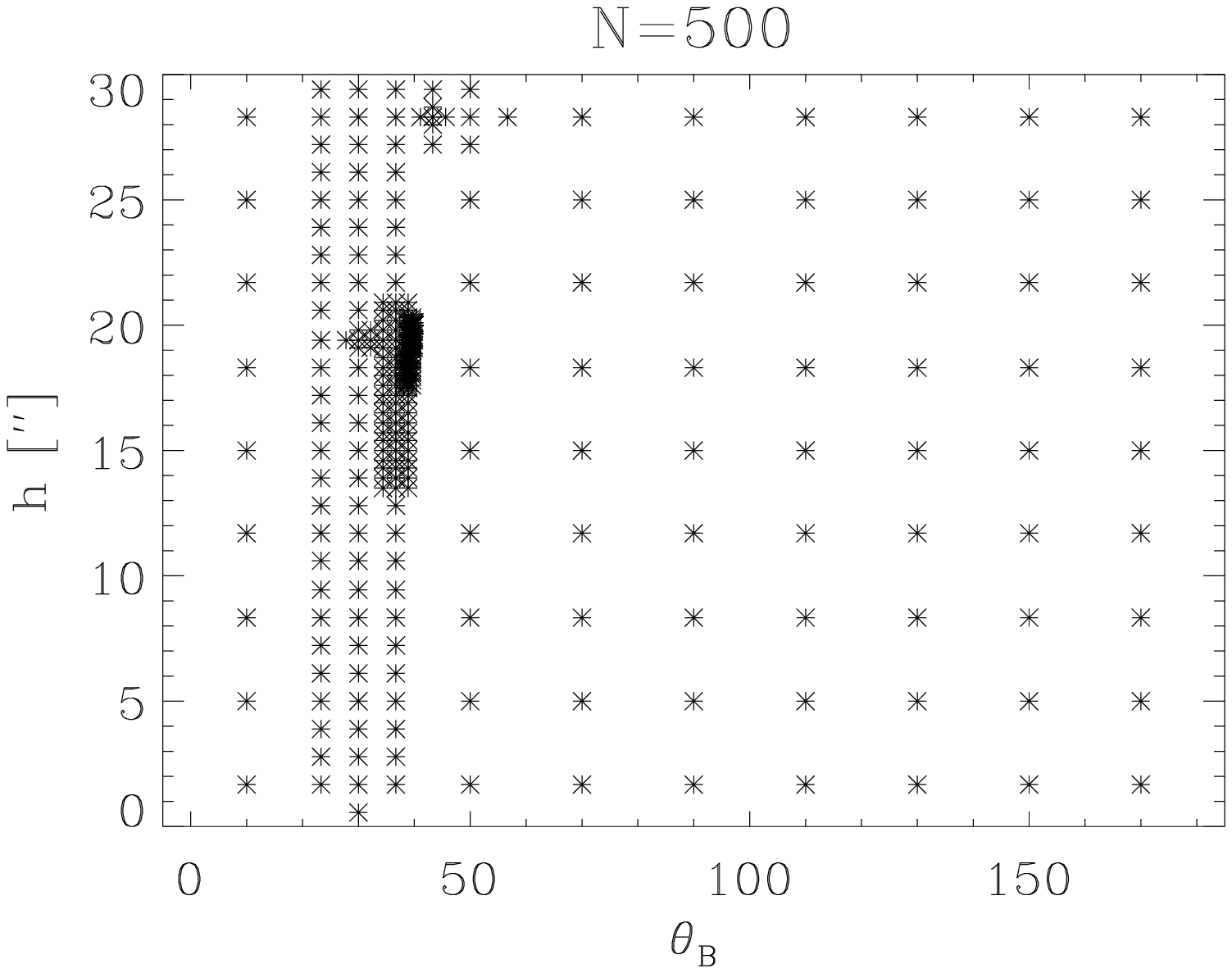}
\plottwo{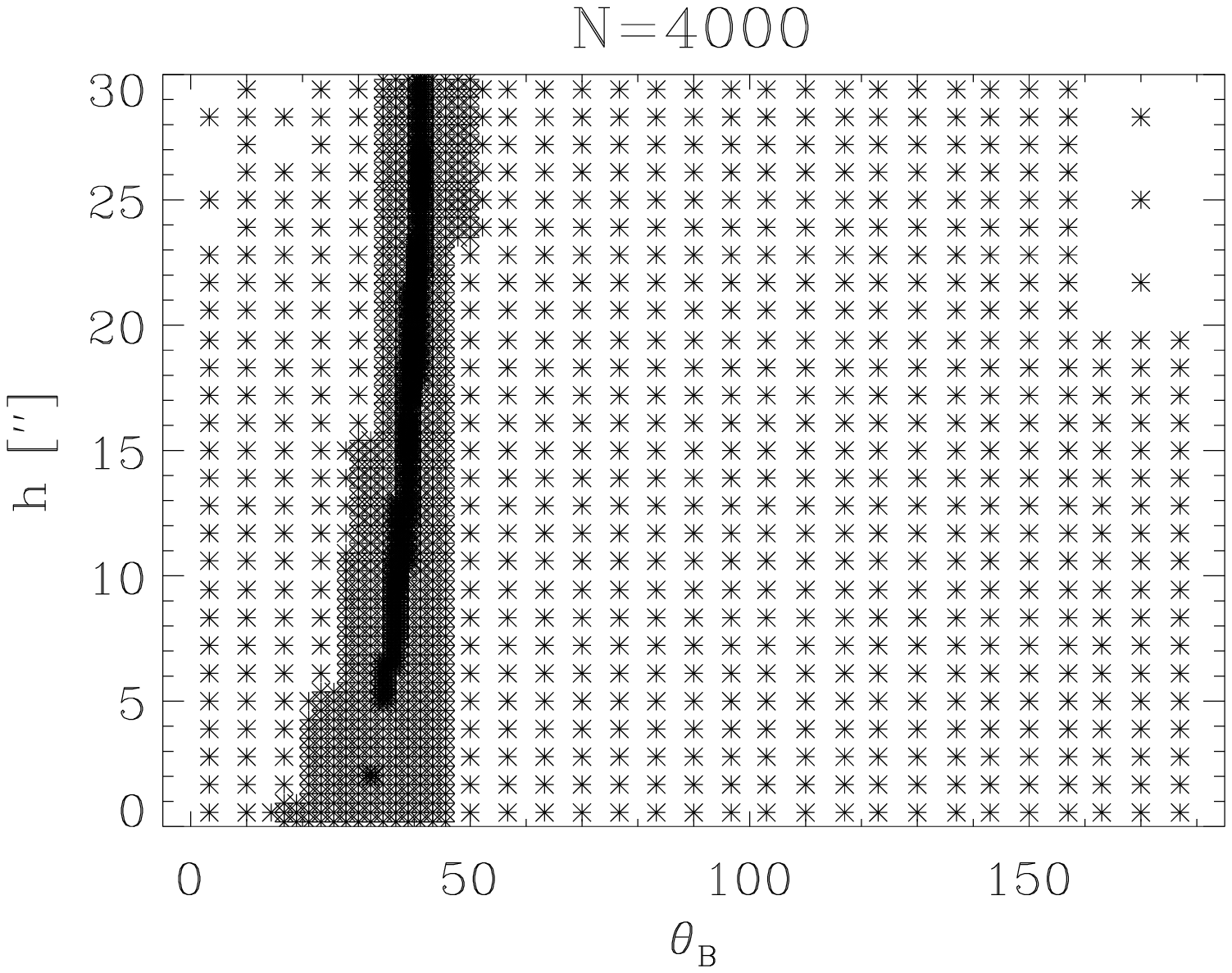}{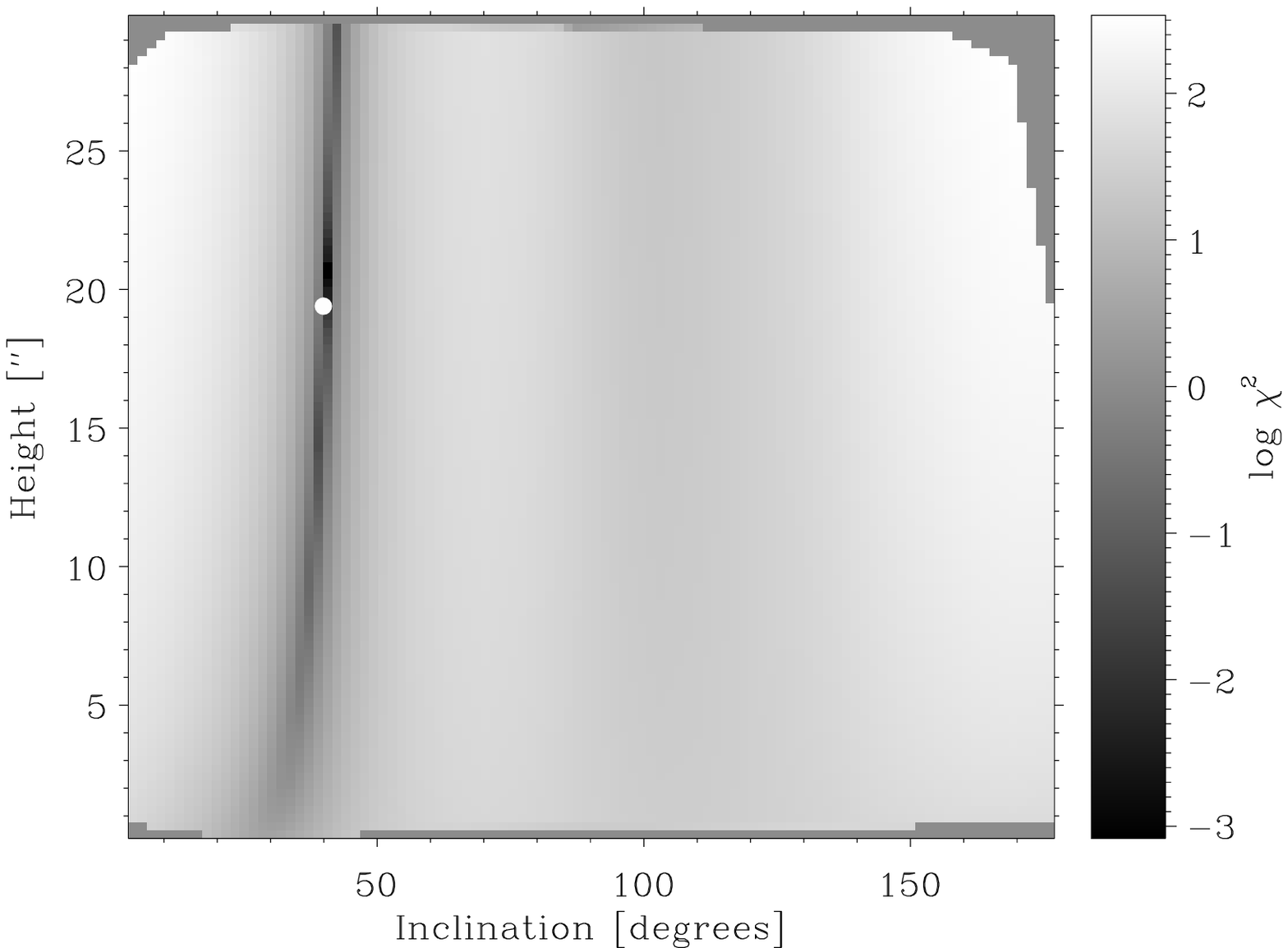}
\caption{Value of $\theta_B$ and $h$ at which the merit function is evaluated
when 
using the
DIRECT method for the inversion of the off-limb Stokes profiles indicated in the
text 
(upper panels and bottom left panel). The number $N$ of 
function evaluations is shown on the top of each panel. The presence of a very
deep and shallow strip where the minimum is located makes it difficult to obtain
it.
The DIRECT method needs at least 1000 function evaluations to find the
region where the minimum is located. 
The bottom right panel shows the value of the $\chi^2$ merit function with the
white dot
indicating the combination of parameters that give the smallest value of the
merit function.
This is an example of a problem that poses severe difficulties to any
gradient-based method like Levenberg-Marquardt.
\label{fig:height_degeneration}}
\end{figure*}

\begin{figure*}
\plottwo{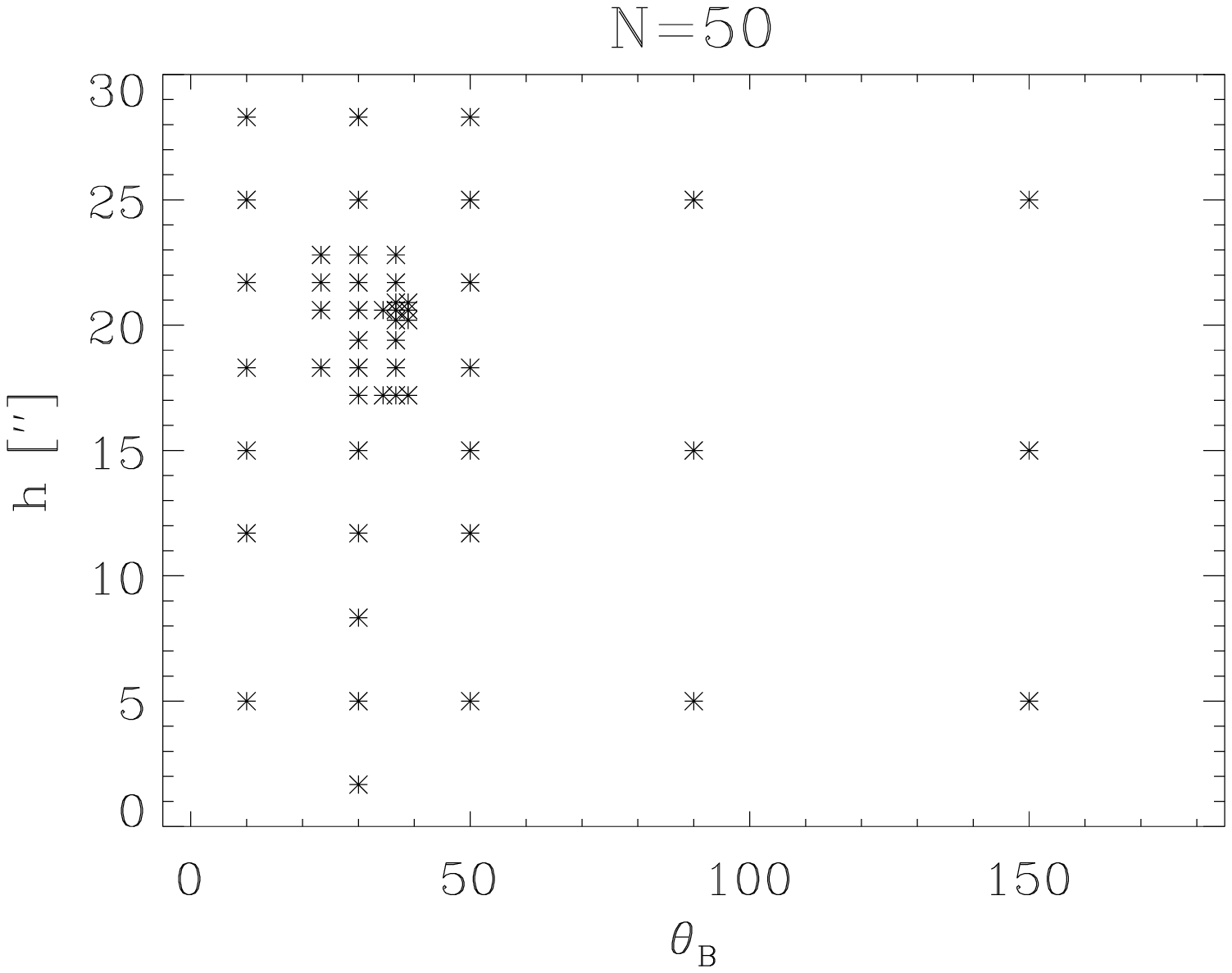}{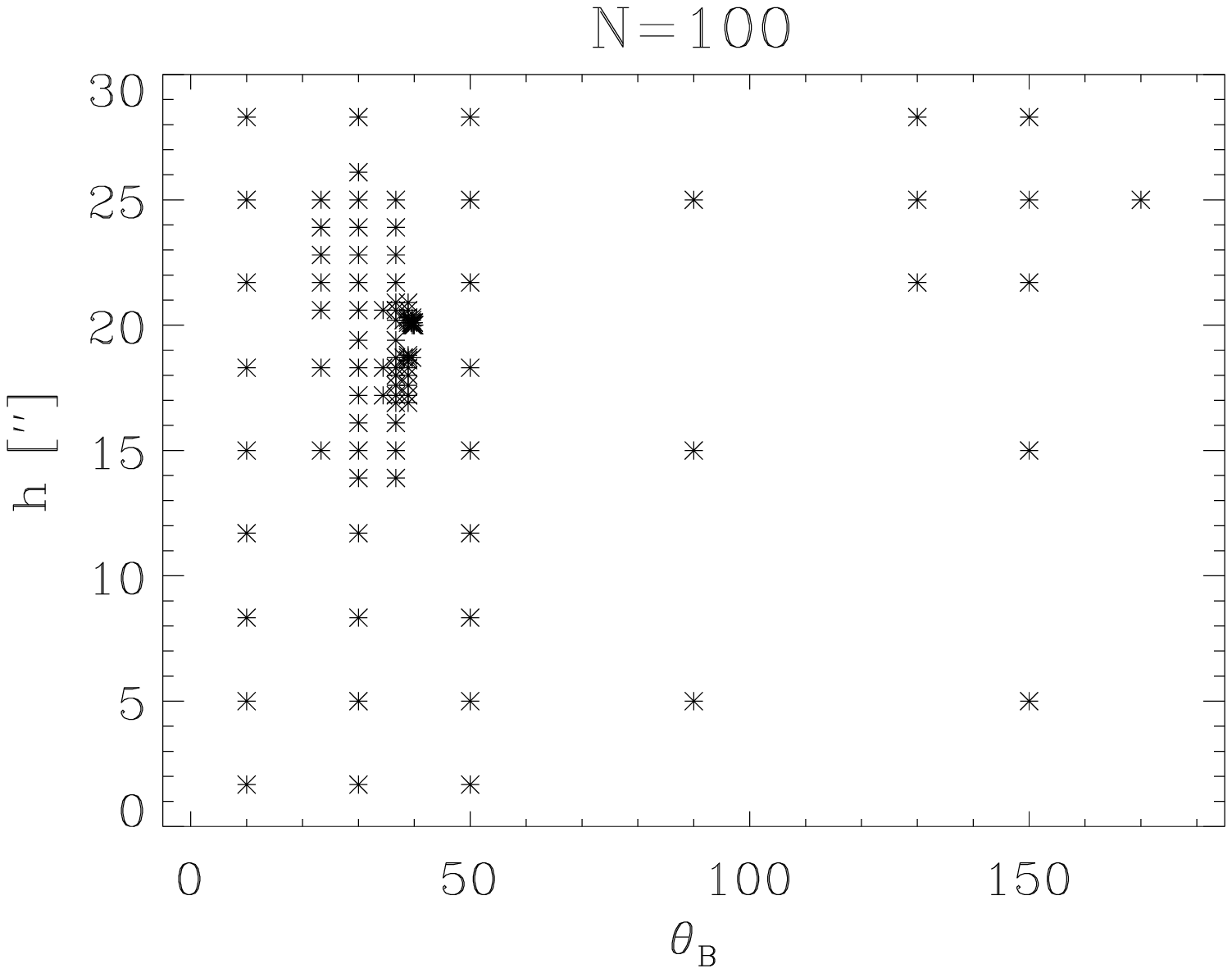}
\plottwo{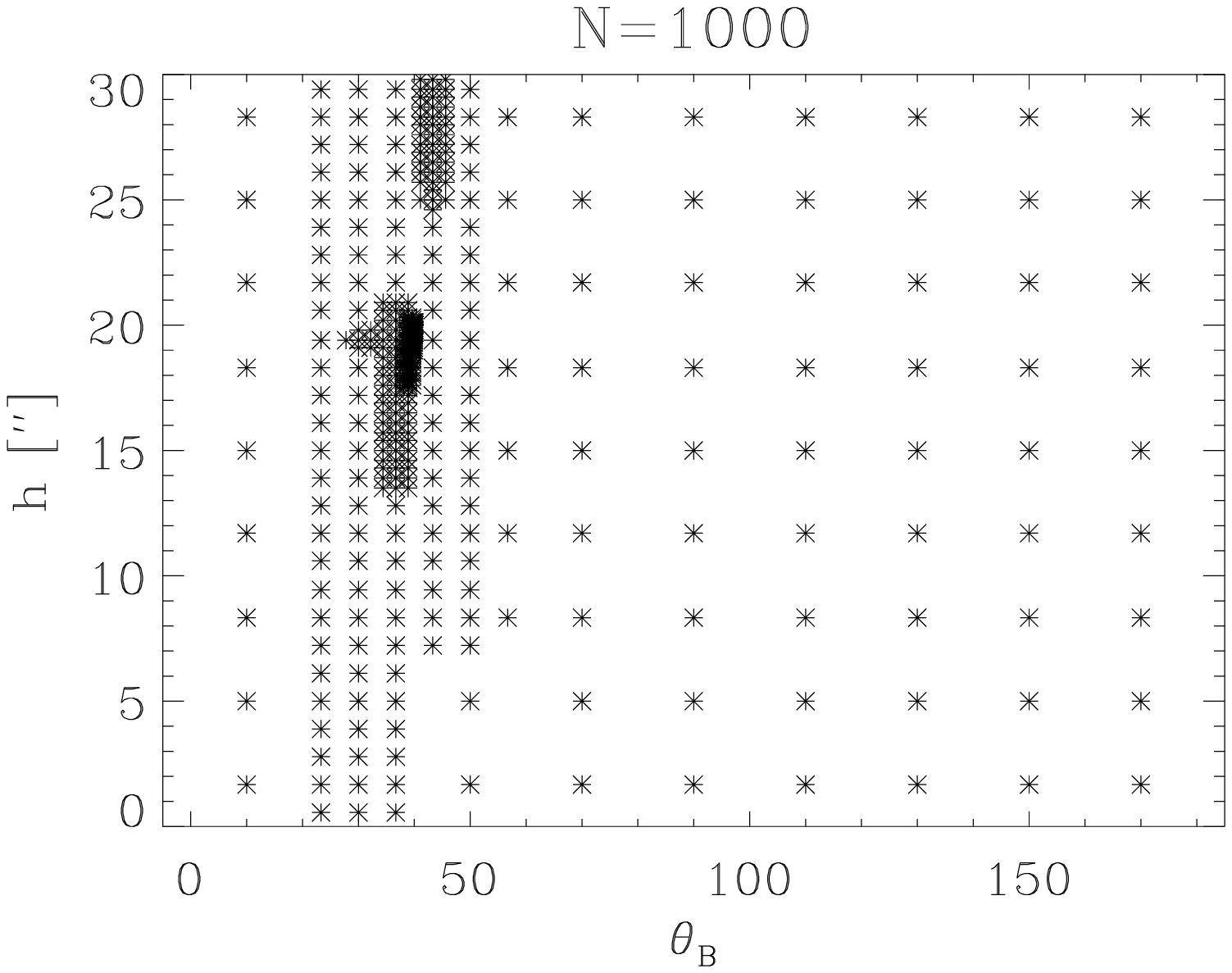}{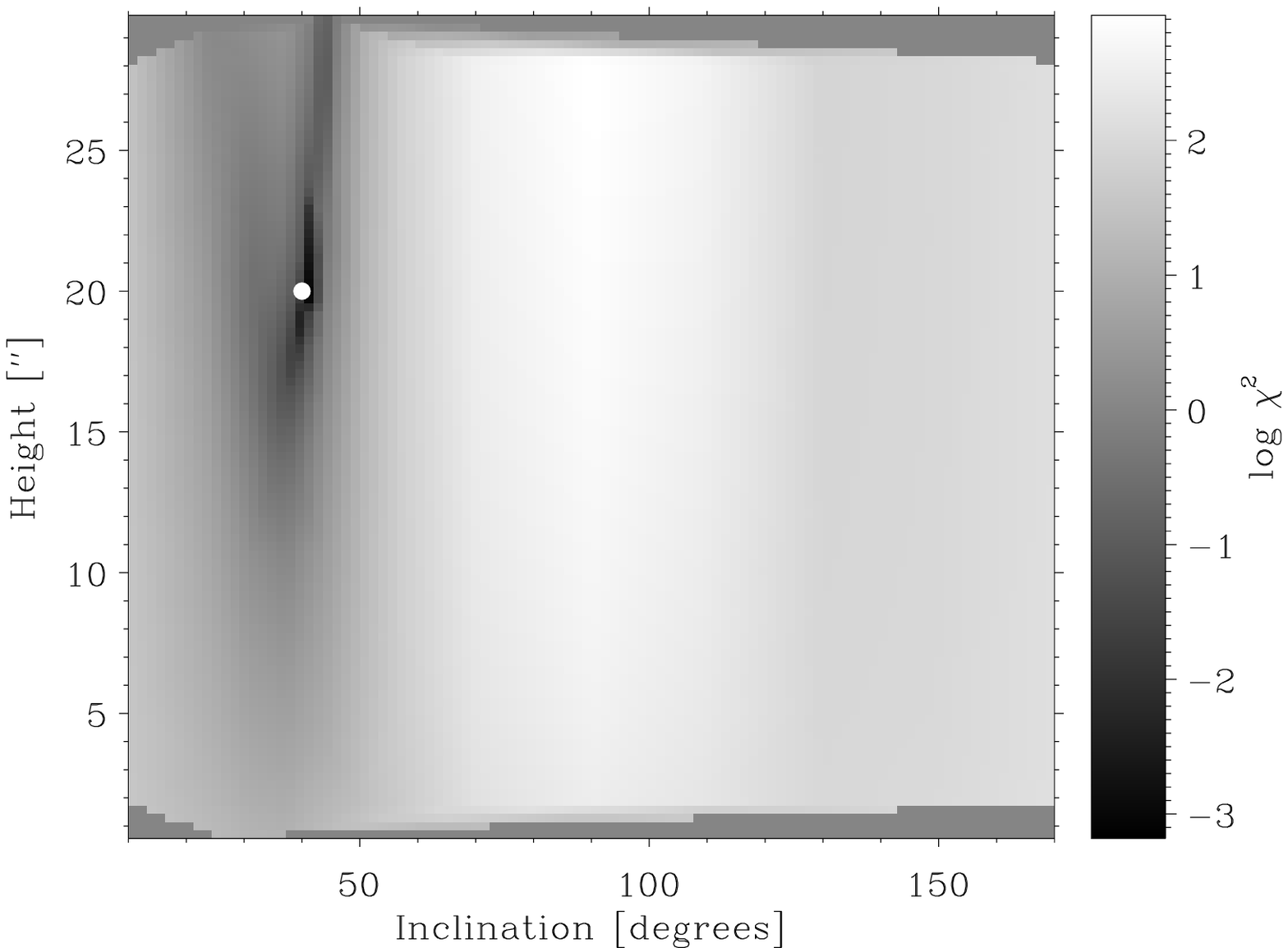}
\caption{Value of $\theta_B$ and $h$ at which the merit function is evaluated
when 
using the
DIRECT method for the inversion of the on-disk Stokes profiles indicated in the
text. The 
number $N$ of 
function evaluations is shown on the top of each panel (upper panels and bottom
left panel). Even 
with only $N=50$, the DIRECT method is able to locate the
region of the global minimum. When the number of evaluations increases, it 
converges towards the values $\theta_B=40^\circ$ and
$h=20"$. The bottom right panel shows the value of the $\chi^2$ merit function
with the white dot
indicating the combination of parameters that give the smallest value of the
merit function.
\label{fig:height_diskcenter}}
\end{figure*}

Obviously, our inversion strategy cannot 
compete in speed with algorithms based on look-up tables, like those applied by 
\cite{casini03} and \cite{merenda06}. At present, with a modern portable computer, 
we need of the order of 1 min. for the inversion of the Stokes profiles shown in 
Fig. \ref{fig:canopy_observation}. The strength of our approach is that it is very 
general and robust, and very suitable also to investigate the impact of the different 
physical mechanisms and parameters on the retrieved models. Concerning future improvements, 
we think that it would be worthwhile to treat the inversion problem within the framework of 
Bayesian inference techniques \citep[see][for a first application of such techniques 
to the inference of Milne-Eddington parameters from Stokes profiles induced 
by the Zeeman effect]{asensio_martinez_rubino07}. The
aim is to sample the joint posterior probability distribution of the parameters of the model 
once the observation has been taken into account, and to carry out marginalizations
to infer the probability distribution of each parameter. One of the main obstacles to overcome is to determine how
to sample efficiently the full posterior probability distribution in the complex physical 
problem that we have investigated in this paper. A possible solution could be to rely on machine learning
techniques for a fast solution of the forward problem, something that could be in perfect synergy
with Markov Chain Montecarlo methods \citep{mackay03}.

The reliability of the developments presented in this paper has been  
demonstrated through several model calculations and applications. Of particular 
interest is the investigation described in Section 3.4, which aimed at clarifying which is the
optimum strategy for determining, from He {\sc i} 10830 \AA\ 
spectropolarimetric observations, whether or not we have magnetic canopies with horizontal fields in the quiet solar chromosphere. The results of an aplication to an observation of 
a disk-center internetwork region can be found in
\S5.4, which suggest the presence of magnetic fields inclined by no more than
$50^{\circ}$ in the observed quiet chromospheric region.

We have also discussed the potential problems that one may
encounter. For example, we have investigated the presence
of degeneracies, paying particular attention to the possibility of determining
the height of the observed plasma structure from the observed Stokes profiles
themselves and to demonstrate that the
DIRECT method is a very efficient technique
for detecting the solutions associated to the Van Vleck ambiguity.

``HAZEL" (an acronym from HAnle and ZEeman Light) is the name we have given to our IAC computer program for the synthesis and inversion of Stokes profiles resulting from the joint action of the Hanle and Zeeman effects in slabs of finite optical thickness. HAZEL will be continuously improved over the years (e.g., with extensions to more complicated radiative transfer models), but is now ready for systematic applications to a variety of spectropolarimetric observations in the spectral lines of the \ion{He}{1} 10830 \AA\ and D$_3$ multiplets. We offer it to the astrophysical community with the hope that it will help researchers to achieve new breakthroughs in solar and stellar physics. To get a copy, it suffices with making an e-mail request to the authors of this paper.

\acknowledgments
{\bf Acknowledgments} 
We thank Roberto Casini (HAO) for carefully reviewing of our paper.
Finantial support by the Spanish Ministry of Education and Science through project AYA2007-63881 and by the European Commission through the SOLAIRE network (MTRN-CT-2006-035484) is gratefully acknowledged.

\clearpage 

\clearpage 

\clearpage

\clearpage 

\clearpage 

\clearpage 

\clearpage

\clearpage  

\clearpage

\clearpage 

\clearpage 

\clearpage 

\clearpage

\clearpage

\clearpage  

\clearpage 

\clearpage 

\clearpage 

\clearpage 

\begin{table}
\begin{center}
\caption{Parameters of the inversion.\label{tab:parameters}}
\begin{tabular}{ccc}
\tableline\tableline
Free parameter & Description & Units \\
\tableline
$B$ & Magnetic field strength & gauss \\
$\theta_B$ & Inclination of the magnetic field vector with respect to the
vertical & degrees \\
$\chi_B$ & Azimuth of the magnetic field vector & degrees \\
$v_\mathrm{th}$ & Thermal velocity affecting the width of the line & km s$^{-1}$
\\
$v_\mathrm{mac}$ & Bulk velocity of the plasma leading to a red/blue-shift & km
s$^{-1}$ \\
$\Delta \tau$ & Optical depth of the line & $-$ \\
$a$ & Line damping parameter & $-$ \\
\tableline
\end{tabular}
\end{center}
\end{table}

\clearpage 

\begin{table}
\begin{center}
\caption{Inversion scheme.\label{tab:inversion}}
\vspace{0.3cm}
\begin{tabular}{cccc}
\tableline\tableline
Step & Method & Free parameters & Stokes profiles \\
\tableline
1 & DIRECT & $v_\mathrm{th}$, $v_\mathrm{mac}$, $\Delta \tau$, $a$ & I \\
2 & LM & $v_\mathrm{th}$, $v_\mathrm{mac}$, $\Delta \tau$, $a$ & I \\
3 & DIRECT & $B$, $\theta_B$, $\chi_B$ & I, Q, U, V \\
4 & LM & $B$, $\theta_B$, $\chi_B$ & I, Q, U, V \\
\tableline
\end{tabular}
\end{center}
\end{table}

\clearpage 

\begin{table}
\begin{center}
\caption{\ion{He}{1} atomic data.\label{tab:tab_einstein}}
\vspace{0.3cm}
\begin{tabular}{cccccc}
\tableline\tableline
Transition ($u \to \ell$) & Air wavelength [\AA] & 
$A(\beta_u L_u S J_u \to \beta_\ell L_\ell S J_l)$ [s$^{-1}$] &
$B_\mathrm{critical}^\mathrm{upper}$ [G]\\
\tableline
2p $^3$P$_0$ -- 2s $^3$S$_1$ & 10829.0911 & 1.022$\times$10$^7$ & undefined \\
2p $^3$P$_1$ -- 2s $^3$S$_1$ & 10830.2501 & 1.022$\times$10$^7$ & 0.77 \\
2p $^3$P$_2$ -- 2s $^3$S$_1$ & 10830.3398 & 1.022$\times$10$^7$ & 0.77 \\
3p $^3$P$_0$ -- 2s $^3$S$_1$ & 3888.6046 & 9.478$\times$10$^6$ & undefined\\
3p $^3$P$_1$ -- 2s $^3$S$_1$ & 3888.6456 & 9.478$\times$10$^6$ & 0.72 \\
3p $^3$P$_2$ -- 2s $^3$S$_1$ & 3888.6489 & 9.478$\times$10$^6$ & 0.72 \\
3s $^3$S$_1$ -- 2p $^3$P$_0$ & 7065.7085 & 3.080$\times$10$^6$ & 0.18 \\
3s $^3$S$_1$ -- 2p $^3$P$_1$ & 7065.2150 & 9.250$\times$10$^6$ & 0.53 \\
3s $^3$S$_1$ -- 2p $^3$P$_2$ & 7065.1769 & 1.540$\times$10$^7$ & 0.88 \\
3d $^3$D$_1$ -- 2p $^3$P$_0$ & 5875.9663 & 3.920$\times$10$^7$ & 8.92 \\
3d $^3$D$_2$ -- 2p $^3$P$_1$ & 5875.6405 & 5.290$\times$10$^7$ & 5.16 \\
3d $^3$D$_1$ -- 2p $^3$P$_1$ & 5875.6251 & 2.940$\times$10$^7$ & 6.69 \\
3d $^3$D$_3$ -- 2p $^3$P$_2$ & 5875.6150 & 7.060$\times$10$^7$ & 6.02 \\
3d $^3$D$_2$ -- 2p $^3$P$_2$ & 5875.6141 & 1.760$\times$10$^7$ & 1.72 \\
3d $^3$D$_1$ -- 2p $^3$P$_2$ & 5875.5987 & 1.960$\times$10$^6$ & 0.45 \\
\tableline
\end{tabular}
\end{center}
\end{table}


\end{document}